\documentclass[fleqn]{book}
\usepackage[x11names]{xcolor}
\usepackage{amsmath}
\usepackage{bclogo}
\newcommand{\underconstruction}{\marginpar{\bcpanchant}}
\colorlet{shadecolor}{lightgray}
\usepackage{framed}
\usepackage{natbib}
\usepackage[plainpages=false, hypertexnames=true, pdfpagelabels=true,
    hyperindex=true, linktocpage, pagebackref=true, pdfa=true]{hyperref}
\usepackage{clr-coloured}
\usepackage{wrapfig}
\usepackage{pgf}
\usepackage{tikz}
\usetikzlibrary{arrows,automata,decorations.pathmorphing,positioning}
\usepackage[thinlines]{easytable}
\usepackage{graphicx}
\usepackage{verbatim}
\usepackage{makeidx}
\makeindex
\usepackage{array}
\usepackage[gen]{eurosym}
\usepackage{mathrsfs}

\title{\vspace*{-2cm}
{\Huge\bf 25 Additional Problems }
\\[5mm]
{\bf - {Extension to  the Book}}\\[2mm] 
{\bf {"125 Problems in Text Algorithms"}}\\[1cm]
\begin{tikzpicture}[scale=.45, node distance=1mm, auto, >=latex', on grid, thick]
\pgftext{\includegraphics[width=14cm]{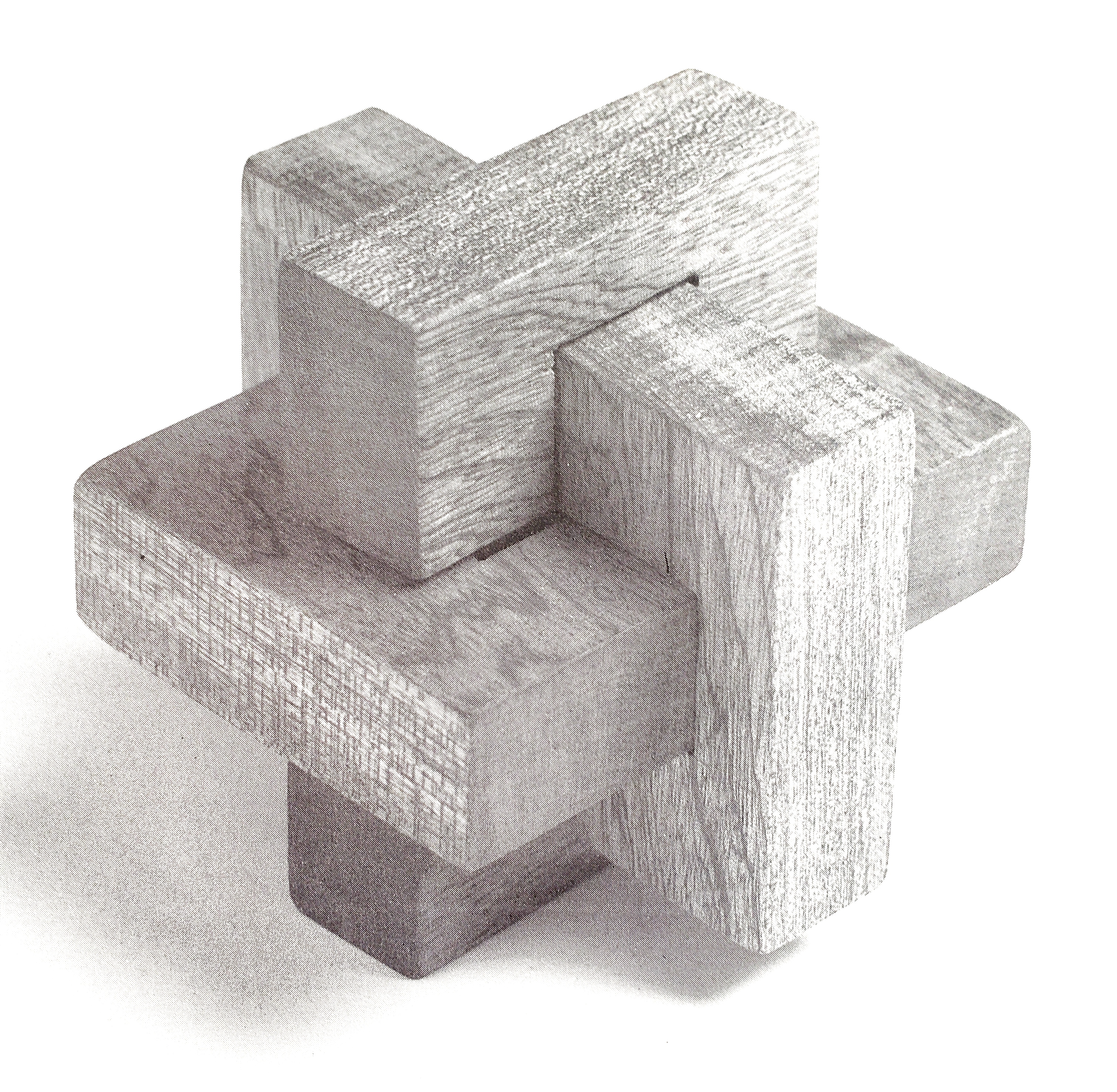}} at (0,0);
\draw[opacity=0.6,fill=red] (-3.5,3.6) -- (1.7,6) -- (3.3,5.2) -- (-1.8,2.8)
 -- (-1.8,-0.5) -- (-3.5,0.6) -- (-3.5,3.6);
 \draw[opacity=0.6,fill=green] (-1.1,-4.1) -- (-6,-0.9) -- (-6,0.9) -- (-1.1,-2.2)
 -- (1.8,-0.8) -- (1.7,-2.7) -- (-1.1,-4.1);
\draw[opacity=0.6,fill=yellow] (2.6,-5.1) -- (2.9,0.8) -- (0,2.5) -- (1.7,3.3)
 -- (4.4,1.7) -- (4.1,-4.2) -- (2.6,-5.1) ;
\end{tikzpicture}
}
\author{%
{\huge \bf Maxime Crochemore}\\[4mm]
{\huge \bf Thierry Lecroq}\\[4mm]
{\huge \bf Wojciech Rytter}\\[4mm]
}
\tikzstyle{dotted}=[dashed,thick,dash pattern=on \pgflinewidth off 2pt]

\begin{document}

\maketitle
\pagenumbering{roman}

\vspace*{1cm}
\noindent
\begin{center}
{\Huge\bf Preface}
\end{center}

\noindent
\textcolor{violet}{\rule{110mm}{.5mm}} 

\medskip

\bigskip\noindent
This very preliminary text is related to ``Algorithms on Texts'', also called ``Algorithmic Stringology''. It is an extension of the book 
{``125 Problems in Text Algorithms''} (see reference \cite{CLR21cup}) providing, in the 
same compact style, more problems with solutions. 

\bigskip
We refer also to the companions to  ``Text algorithms''
available at\\
\url{http://monge.univ-mlv.fr/~mac/CLR/clr1-20.pdf}
and at the web page
\url{http://125-problems.univ-mlv.fr}, where  all 150
problems (including the ones presented here) are briefly announced.

\bigskip\noindent
The selected problems satisfy three
criteria: 

\begin{itemize}
\item challenging, 
\item  having short tricky solutions
\item solvable
with only very basic background in stringology. 
\end{itemize}

\bigskip\noindent
For the basics in stringology we refer to  \cite[Chapter 1]{CLR21cup}
and to\\
\url{http://monge.univ-mlv.fr/~mac/CLR/clr1-20.pdf}.

\vspace{1cm}
\noindent
\textcolor{blue}{\rule{110mm}{.5mm}} 

\vspace{2cm}
{\sf January 2025}

\vspace{2cm}
{ \large  \textit{M. Crochemore, T. Lecroq, W. Rytter}}

\medskip
Paris, Rouen (France), Warsaw (Poland)

\newpage
\vspace{1cm}
\begin{center}{\Huge\bf Table of contents}
\end{center}
\noindent
\textcolor{violet}{\rule{110mm}{.5mm}} 
\bigskip
{\small \tableofcontents}
\vspace{1cm}
\noindent
\textcolor{blue}{\rule{110mm}{.5mm}} 
\pagenumbering{arabic}
\setcounter{page}{0}

\setcounter{EXERCISE}{125}
\begin{EXERCISE}[Subsequence Covers]\label{pb138}
A word $x$ is a subsequence cover (s-cover, in short)
of a word $y$ if each position on $y$ belongs to an
occurrence of $x$ as a subsequence of $y$.

\paragraph{\bf Example}
The word  $x=\sa{010}$ is a (shortest)  s-cover of  $y=\sa{0110110}$ as well as of $y=\sa{000011000}$. However, $x=\sa{010010}$ is not because it is not a subsequence of them, nor $x=\sa{0101}$ because it does not s-cover their last position.

\begin{QUESTION}
Let $y$ be a word in $\{\sa{0},\sa{1},\dots,n-1\}^n$.
Design a linear-time algorithm that checks if a given word $x$ of length
$m<n$ is an s-cover of $y$. 
\end{QUESTION}

\begin{SOLUTION}
Let $y=y[0\dd n-1]$ and $x=x[0\dd m-1]$.
Define the two lists of positions on $y$,
$\mathbf{L}=(p_1,p_2,\dots,p_m)$ and $\mathbf{R}=(q_1,q_2,\dots,q_m),$ as the lexicographically first and last subsequences of positions on $y$, respectively, corresponding to $x$ as a subsequence of $y$. Additionally, it is required that $p_1=0$ and $q_m=m-1$.

Note that $\mathbf{L}$ or $\mathbf{R}$ may not exist (see the above example). This can be tested readily in linear time with a greedy algorithm while computing the lists. Then, we assume up to now that $\mathbf{L}$ and $\mathbf{R}$ exist as defined. By definition, 
$x=y[p_1,p_2,\dots,p_m]=y[q_1,q_2,\dots,q_m]$.

\paragraph{Example}
When $x=\sa{020}$ and $y=\underline{\sa{02}}\sa{1}\underline{\sa{0}}\sa{0020101}\underline{\sa{0}}\sa{1}\underline{\sa{20}}$, many lists of positions on $y$ are associated with $x$ as a subsequence of $y$. Among them, the choosen lists are
$\mathbf{L}=(0,1,3)$ and $\mathbf{R}=(11,13,14)$ (underlined letters).

\paragraph{Observation 1}
A position $i$ on $y$ is s-covered by $x$ (as a subsequence) if
there is a prefix $p_1,p_2,\dots, p_{k-1}$ of $\mathbf{L}$ and a
suffix $q_{m-k+1},q_{m-k+2},\dots, q_m$ of $\mathbf{R}$ for which
$p_{k-1}<i<q_{m-k+1}$ and $x[k]=y[i]$.

\medskip
To implement efficiently the observation, for $i$ position on $y$ define two auxiliary tables
\[\ \ \mathsf{LEFT}[i]=|\{k\in \mathbf{L}\mid k<i\}|,\]
\[\ \ \mathsf{RIGHT}[i]=|\{k\in \mathbf{R}\mid k>i\}|.\]
We  also define, for $0< i <m-1$:
\[\ \ \mathsf{P}[i]= \max\{k \mid k\leq \mathsf{LEFT}[i]+1 \mbox{ and } x[k]=y[i]\} \cup \{0\}.\]

\paragraph{\bf Example}
For $y=\sa{010210201}$ 
and $x=\sa{01201}$ 
we get \[\mathbf{L}=\{0,1,3,5,8\},\ \ \mathbf{R}=\{2,4,6,7,8\}\]

$$\begin{array}{l|ccccccccc}
i & \scriptsize 0 & \scriptsize 1 & \scriptsize 2 & \scriptsize 3 & \scriptsize 4 & \scriptsize 5 & \scriptsize 6 & \scriptsize 7 & \scriptsize 8 \\ \hline
\mathsf{LEFT}  & 0 & 1 & 2 & 2 & 3 & 3 & 4 & 4 & 4 \\
\mathsf{RIGHT} & 5 & 5 & 4 & 4 & 3 & 3 & 2 & 1 & 0 \\
\mathsf{P} & 1 & 2 & 1 & 3 & 2 & 4 & 3 & 4 & 5 \\
\end{array}$$

\noindent Let $\mathbf{\Psi}$ be the predicate
$$\mathbf{\Psi}(x,y)\;\stackrel{def}{\equiv}\; \forall i\in [0\dd m-1]\;\mathsf{P}[i]>0\mbox{ and } \mathsf{P}[i]+\mathsf{RIGHT}[i]\geq |x|,$$
With this terminology Observation 1 restates as follows and leads to Algorithm 
\Algo{s-Cover}.

\paragraph{Observation 2} 
The word $x$ is an s-cover of $y$ if and only if $\mathbf{\Psi}(x,y)$.


 \vspace*{0.3cm}
The algorithm can be written as the following pseudocode.

\begin{algo}{s-Cover }{x,y \mbox{ non-empty words}}

%
\ACT"{compute $\mathsf{LEFT}[i],\mathsf{RIGHT}[i]$ for each $i$}
%

\ACT"{initially $\mathsf{F}[s]=0$ for each letter $s$}

\SET{k}{1};
\SET{\mathsf{F}[\,x[0]\,]}{1}

\COM{\bf Computing the table $\mathsf{P}$}
  \DOFORI{i}{1}{|y|-2}
   \SET{j}{\mathsf{LEFT}[i]}
   \IF{i=p_{j+1}}
   \SET{\mathsf{F}[\,y[i]\,]}{j+1}
   \FI 
   \SET{\ \mathsf{P}[i]}{\mathsf{F}[y[i]]}
\OD
\RETURN{\mathbf{\Psi}(x,y)
}
\vspace*{0.3cm}
\end{algo}

If the tables $\mathsf{P},\, \mathsf{RIGHT}$ are known then $\mathbf{\Psi}(x,y)$ can be computed in linear time.

\medskip \noindent The computation of tables $\mathsf{LEFT},\, 
\mathsf{RIGHT}$ is very simple, we omit  details.
 The table
 $\mathsf{P}$ is computed on-line using the auxiliary table $\mathsf{F}$.
 This table satisfies:
 
 in the moment 
 immediately after we execute "$P[i]=\mathsf{F}(y[i])$",  
  for each symbol $s$
 the value of $F(s)$ is the length of the longest prefix of $x$ which ends with $s$
 and which is a subsequence of $y[0\dots i]$.

  \medskip\noindent
Correctness follows from Observation 1 and Observation 2.
\end{SOLUTION}

\begin{NOTES}
Our algorithm is a version of the one in  \cite{CharalampopoulosPRRWZ22}.

\noindent The notion of an s-cover differs substantially from the notion of\\ a standard cover:
\begin{itemize}
\item
Two shortest s-covers of a same string
can be distinct.
%
\item Computing the length of a shortest s-cover is probably NP-hard.
\item Every binary word of length at least 4 admits a nontrivial
s-cover.
\item In general, if the size $k$ of the alphabet is fixed,
then the length $\gamma(k)$ of the longest word without any nontrivial s-cover is finite, though exponential w.r.t. $k$. It is known that $\gamma(3)=8,\, \gamma(4)=19$.
The exact value of $\gamma(5)$ is unknown.
\end{itemize}
\end{NOTES}
\end{EXERCISE}
\begin{EXERCISE}[String attractors]\label{pb127}
%
The notion of the \MC{string attractor} provides a unifying framework for known
dictionary-based compressors. We say that a subset
$\Att$ of positions on a word $x$ \emph{touches} a factor $u$ of $x$ if
there are positions $i$,
$j$ satisfying $u=x[i\dd j]$, $[i\dd j]\cap \Att \neq \emptyset$.
$\Att$ is a \emph{string attractor} on  $x$ if 
$\Att$ touches each factor $u$ of $x$.
We concentrate on attractors on two special families of  words.

Thue--Morse words are defined by the  recurrence:

\centerline{
$
 \tau_0      =  \sa{a},  \
 \tau_{k+1}  =  \tau_k\overline{\tau_k},\  \mbox{for}\  k > 0,$}
 
 \noindent
 where the bar morphism is defined by
 $\overline{\sa{a}}=\sa{b}$ and $\overline{\sa{b}}=\sa{a}$.
Note the length of the $k$th Thue--Morse word is $|\tau_k|=2^k$ and
$\overline{\tau_{k+1}}  =  \overline{\tau_k}\tau_k$.

\paragraph{Example}
 $\{4,6,8,12\}$ and $\{4,8,10,12\}$ are
attractors on $\tau_4$.

\medskip\noindent
\begin{tabular}{@{}l*{21}{p{1.5mm}}@{}}
$i$&0&1&2&3&\textbf{4}&5&\textbf{6}&7&\textbf{8}&9&\makebox[0mm]{10}
&\makebox[0mm]{11}&\makebox[0mm]{\textbf{12}}&\makebox[0mm]{13}
&\makebox[0mm]{14}&\makebox[0mm]{15}\\
\hline
$\tau_4[i]$&\sa a&\sa b&\sa b&\sa a&\sa b&\sa a&\sa a&\sa b
&\sa b&\sa a&\sa a&\sa b&\sa a&\sa b&\sa b&\sa a
\end{tabular}

\begin{QUESTION}
Construct an attractor of size at most $4$ for Thue-Morse words $\tau_k$, $k\geq 4$.
\end{QUESTION}

%
%
%

\begin{SOLUTION}
\textbf{Properties of Thue-Morse words.}
The clue of the solution is to consider middle positions in words. By the
middle position of a word with even length $2\times m$ we mean position $m$. Indeed
such a position captures many (distinct) factors occurring in the word. 

\medskip For example, position $3$ on $\sa{aaabbb}$ is covered by $12$ factors and
position $m$ on $\sa{a}^m\sa{b}^m$ covered by $m(m+1)$ factors, a quadratic
number with respect to the length of the word. Adding only position $m-1$
gives the attractor $\{m-1,m\}$.

\medskip
Let $\Mid(x)$ be the set of factors of $x$ that have an occurrence in $x$
covering the middle of $x$ and let $\Fact(x)$ be the set of all nonempty
factors of $x$. We have the following fact for $k\geq 4$.

\paragraph{Fact}
{\bf (a)} $\Fact(\tau_k)= \Mid(\tau_k)\cup \Fact(\tau_{k-1})
\cup \Fact(\overline{\tau_{k-1}})$.\\
{\bf (b)} $\Fact(\tau_k)= \Mid(\tau_k) \cup \Mid(\tau_{k-1}) \cup
\Mid(\overline{\tau_{k-1}})\cup \Mid(\overline{\tau_{k-2}}).$

\begin{PREUVE}
Point (a) follows from the recursive description of $\tau_k$. Due to point
(a)  we have
$$\Fact(\tau_k)=\,\bigcup_{i=1}^k\, \Mid(\tau_i)\,\cup \,\bigcup_{i=1}^{k-1}\,
\Mid(\overline{\tau_i})$$ Now the thesis follows from the fact that
$\tau_{k-2}$ is a central part of $\tau_k$, and similarly
$\overline{\tau_{k-2}}$ is a central part of $\overline{\tau_k}$. The same
holds for $\tau_{k-2}$, $\overline{\tau_{k-2}}$ and their central parts.
Hence, in the above equation it is enough to keep these four largest
Thue-Morse words and their barred images.
\end{PREUVE}

\paragraph{\bf Construction of an attractor on $\tau_k$}
Due to Fact 1 it is enough to take middle points in $\tau_k, \tau_{k-1},
\overline{\tau_{k-1}},\overline{\tau_{k-2}}$. However all these words are
parts of $\tau_k$. When $k\geq 4$, from its recursive definition $\tau_k$ can
be written as a composition of $8$ fragments of the same length $A\cdot
B\cdot B\cdot A\cdot B\cdot A\cdot A\cdot B$, where $\tau_{k-1}=ABBA$,
$\overline{\tau_{k-1}}=BAAB$ and $\overline{\tau_{k-2}}=BA$. The 4 sought
middle points are the middle points of the occurrences of $ABBABAAB$, $ABBA$,
$BAAB$ and $BA$ in $\tau_k$. Then
$$\{2^{k-1},\; 2^{k-2},\; 2^{k-1}+2^{k-2},\; 2^{k-2}+2^{k-3}\}$$
is an attractor on $\tau_k$. Note that $2^{k-1}+2^{k-3}$ can be substituted
for $2^{k-2}+2^{k-3}$ because there are two occurrences of $BA$ in
$ABBABAAB$.

\paragraph{Example}
Splitting $\tau_5$ of length $32$ into $8$ equal-length fragments and pointed
positions of its attractor $\{16, 8, 24, 12\}$:

\centerline{$\,\textcolor{white}{\sa{abba}\cdot\sa{baab}\cdot}\bullet
\textcolor{white}{\sa{aab}\cdot}\bullet
\textcolor{white}{\sa{bba}\cdot}\bullet
\textcolor{white}{\sa{aab}\cdot\sa{abba}\cdot}\bullet
\textcolor{white}{\sa{bba}\cdot\sa{baab}}$}

\centerline{$\sa{abba}\cdot\sa{baab}\cdot\sa{baab}\cdot\sa{abba}\cdot
\sa{baab}\cdot\sa{abba}\cdot\sa{abba}\cdot\sa{baab}.$}

\smallskip\noindent
Another attractor on $\tau_5$ is $\{16, 8, 24, 20\}$.

\bigskip
\textbf{Fibonacci words.} For Fibonacci words, using similar arguments as in
the proof of the previous question, $\{|\fib_{k-1}|-1,\, |\fib_{k-2}|-1\}$ is
an attractor of $\fib_k$. It is obviously of the smallest size because at
least two positions have to be indices of two different letters.
\\
We also show a more attractive attractor consisting of two adjacent positions
on $\fib_k$, namely the last two positions of its prefix $\fib_{k-1}$.
We use a well known known property of Fibonacci words that is recalled
first.

\begin{LEMME}
For $k\geq
2$, $\fib_{k}\fib_{k-1}=uw$ and $\fib_{k-1}\fib_{k}=u\overline{w}$, where
$w=\sa{ab}$ if $k$ is even and $w=\sa{ba}$ if $k$ is odd.
\end{LEMME}

\paragraph{Observation}  For $k>3$,  $\fib_{k-2}^2$
is a prefix of $\fib_k$ and $\fib_{k-2}$ is a suffix of $\fib_k$. Also $\fib_{k-2}$ is a string period of $\fib_k$ without the last two symbols.

\paragraph{Example}
Below two adjacent big dots show positions of an attractor $\{6,7\}$ of $\fib_5$.

\centerline{$\textcolor{white}{\sa{a}\,\sa{b}\,\sa{a}\,\sa{a}\,\sa{b}\,\sa{a}}%
\bullet\bullet%
\textcolor{white}{\cdot\,\sa{a}\,\sa{b}\,\sa{a}\,\sa{a}\,\sa{b}}.$}

\centerline{$\;\sa{a}\,\sa{b}\,\sa{a}\,\sa{a}\,\sa{b}\,\sa{a}\,\sa{b}\,\sa{a}%
\cdot\sa{a}\,\sa{b}\,\sa{a}\,\sa{a}\,\sa{b}.$}

\smallskip\noindent
Note that $\{|\fib_{k-1}|,|\fib_{k-1}|+1\}$ is not an attractor of $\fib_k$,
$k\geq 3$. For $\fib_5$, the set $\{8,9\}$ does not capture the factor
$\sa{baba}$.

\paragraph{Proposition}
The set $X_k=\{|\fib_{k-1}|-2,  |\fib_{k-1}|-1\}$ of positions on $\fib_k$ is
an attractor of $\fib_{k}$ for $k>1$.

\begin{PREUVE}
A direct examination shows that the result holds for $\fib_2=\sa{aba}$ and
$\fib_3=\sa{abaab}$. Indeed, $\{0,1\}=\{|\fib_1|-2,|\fib_1|-1\}$ is an
attractor of $\fib_2$ and $\{1,2\}=\{|\fib_2|-2,|\fib_2|-1\}$ is an attractor
of $\fib_3$.

The rest of the proof is by induction on $k$. Let $k>3$ and assume the result
holds for $\fib_{k-2}$.
Let us take the fragment in $\fib_k$ corresponding to the second (internal) occurrence of $\fib_{k-2}$, In the example it is the underlined fragment.

\smallskip
\centerline{$\textcolor{white}{\sa{a}\,\sa{b}\,\sa{a}\ \sa{a}\,\sa{b}\,
\sa{a}}\ \ \bullet\bullet \textcolor{white}{\sa{b}\,\sa{a}
\,\sa{a}\,\sa{b}\ \sa{a}\,\sa{a}\,\sa{b}.}$} 

\centerline{$\sa{a}\,\sa{b}\,\sa{a}\ \sa{a}\,\sa{b}\,
\underline{\textcolor{red}{\sa{a}\,\sa{b}\,\sa{a}
\,\sa{a}\,\sa{b}}}\ \sa{a}\,\sa{a}\,\sa{b}.$}

\medskip\noindent
 By inductive assumption, $X_k$ touches all factors of this fragment.
Also $X_k$ touches all factors to the right of $X_k$, as they are 
factors in $fib_{k-2}$. Consequently $X_k$ touches all factors in the suffix $v$
of $\fib_k$ starting in position $|\fib_2|$. 

Now it is enough to show that $X_k$ touches all factors fully to the left of $X_k$,
that is the factors in the prefix $u=\fib_k[0\dd |\fib_{k-1}-3|]$.
However, due to Observation 1, $u$ is a prefix of $v$. Hence $X_k$ touches all factors
of $u$, since they are also factors of $v$.
\end{PREUVE}

\end{SOLUTION}

\begin{NOTES} 
The relation between string attractors\INDEX{attractor} and text compression
is by Kempa and Prezza \cite{KempaP18}. 
Further results on attractors can be
found in \cite{MantaciRRRS21}. Testing if a set of positions form an
attractor can be done with the algorithm decribed in \cite[Problem
64]{CLR21cup}. Define $k$-attractor as a set of 
positions which touches all $k$-length factors of the word.
Existence of 2-attractor for a given word is an NP-hard problem, see
\cite{Fuchs}.
The explicit description of attractors on  
Thue-Morse words is by Kutsukake et al. \cite{KutsukakeMNIBT20}.   Fibonacci words
are a subset of (so called) standard Sturmian words,
general construction of smallest attractors for standard Sturmian words is given
in Theorem 22 in \cite{MantaciRRRS21}. Testing if a set
is an attractor can be done with the algorithm decribed in \cite[Problem 64]{CLR21cup}.
\end{NOTES}
\end{EXERCISE}

\begin{EXERCISE}[1-Error Correcting Linear Hamming Codes]\label{pb132}
\def\Codes{\mathit{Codes}}
Sending a message through a noisy line may produce errors. The goal of the problem is to present a method for correcting a message in which only one error is assumed to occur.
A message is a word of bits. To allow checking and correcting a possible transmission error in a word $w \in \{\sa{0},\sa{1}\}^*$, a very short word depending on it and easily computable, $f(w)$, is appended to $w$. The complete message to be sent is then $\code(w)=w\cdot f(w)$. 
We assume that the length of a message $w$ is of the form $k=2^r-r-1$, $|f(w)|=r$ and the total length of the code is then $n=k+r=2^r-1$, for an integer $r>2$. Hence, the size $r$ of the additional part of the code is only logarithmic according to length of the message. Such codes are called $(n,k)$-codes.
We consider only binary words and use linear algebra methods. A word $b_1b_2\cdots b_k$ is identified with the vector $[b_1,b_2,\ldots, b_k]$.

The set  of length-$n$ binary words containing at least two occurrences of $\sa{1}$ has size $k=2^r-r-1$. For example, the set of such length-4 words has 11 elements, all $16$ 4-bit words except the $5$ words 0000, 0001, 0010, 0100,1000.


\paragraph{Example} A possible function $f$ for a $(7,4)$-code is
$$f(b_0b_1b_2b_3) = [b_0+b_1+b_2, b_0+b_1+b_3, b_0+b_2+b_3], \mbox{ and}$$
$$\code(b_0b_1b_2b_3) = [b_0,b_1,b_2,b_3, b_0+b_1+b_2,b_0+b_1+b_3,b_0+b_2+b_3],$$
\noindent where $b_i$ are bits and the operation $+$ is $\textit{xor}$.

\medskip
We construct $f(w)$ as $M\times w^T$, multiplication of a $r\times k$ matrix $M$ by the transposed vector associated with $w$. Then,
$$\code^M(w) = [w,f(w)] = [w,M\times w^T].$$

\paragraph{Example (followed)}
The matrix of the above $(7,4)$-code is
$$M =
\left(\begin{array}{cccc}
1 & 1 & 1 & 0 \\
1 & 1 & 0 & 1 \\
1 & 0 & 1 & 1
\end{array}\right)
$$
For example $\code^M(1010) = 1\,0\,1\,0\ 0\,1\,0$, in this case $f(1010)=010$.
\\
The length-$n$ elements of $\Codes^M_n=\{\code_n^M(w):\, w\in \{0,1\}^k\}$ are called codewords and the set $\Codes^M_n$  is called a \MC{Hamming code} if 
$  \min\,\{\ham(u,v)\;:\; u,v\in \Codes^M_n, u\ne v\,\}\geq 3$,
where $\ham(u,v)$ is the Hamming distance (number of mismatches). 


\begin{QUESTION}
Build a matrix $M$ for which $\Codes^M_n$ is a Hamming code.
\end{QUESTION}

\AIDE{Use the observation}

\begin{QUESTION}
Show how to correct the message assuming it contains at most one error.
\end{QUESTION}

\begin{SOLUTION}
Let $I_r$ be the $r\times r$ identity matrix, and let the Parity checking matrix be 
 horizontal concatenation of matrices $M$ and $I_r$.

\paragraph{\bf Observation} The columns of $P$ are all
 nonzero binary words of size $r$.
Following the above example where $r=3$, it is
$$P =
\left(\begin{array}{ccccccc}
1 & 1 & 1 & 0 &1&0&0 \\
1 & 1 & 0 & 1 &0&1&0 \\
1 & 0 & 1 & 1 &0&0&1
\end{array}\right)
$$
The next property of $P$ is a reformulation of the definition of the function $\code$. Observe that operations on matrices are modulo $2$, and that the equality $x=y \bmod{2}$ is equivalent to $x+y=0 \bmod{2}$.
Let us denote by $\bar{0}$ the vector whose components are zeros.

\paragraph{Fact}
$x\in \Codes^M_n$ if and only if $P\times x^T=\bar{0}$.

\smallskip
We are ready to show property $(*)$. Assume, by contradiction, that this property is false and that, for $u\ne v$, $u,v\in \Codes^M_n$, we have $0<\ham(u,v)<3$. Let $x=u-v$ (subtraction modulo 2). Then $x$ has exactly one or two occurrences of $\sa 1$. 
We have also $P\times x^T=\bar{0}$.

If $x$ has a single $\sa 1$, then $P\times x^T$ is a single column of $P$, which cannot be $\bar{0}$ since all columns of $P$ are
nonzero vectors. Furthermore, if $x$ has exactly two $\sa 1$s then $P\times x^T$ is the sum of two columns of $P$. Then again, it cannot be a zero vector since every two distinct columns of $P$ are linearly independent.

Hence, $P\times x^T>\bar{0}$, which contradicts the equality $P\times x^T=\bar{0}$ and completes the proof of property
$(*)$.

\medskip
\paragraph{Larger example}
Consider $r=4$ and the $4\times 11$ matrix for $(15,11)$-code
$$M =
\left(\begin{array}{ccccccccccc}
1 & 1 & 1 & 1 &1 &1 &1 &0 &0 &0 &0\\
0 & 0 & 0 & 1 &1 &1 &1 &1 &1 &1 &0 \\
1 & 1 & 0 & 1 &1 &0 &0 &1 &1 &0 &1\\
1 & 0 & 1 & 1 &0 &1 &0 &1 &0 &1 &1
\end{array}\right)
$$
The function generating additional 4 bits for $11$-bit messages is
$$f(b_0b_1\cdots b_{10}) = M \times [b_0,b_1,\ldots,b_{10}]^T =
[c_0,c_1,c_2,c_4],\mbox{ where}$$
$$c_0=\sum_{i=0}^{6}b_i,\, c_1=\sum_{i=3}^{9}b_i,\ 
c_2=b_0+b_1+b_3+b_4+b_7+b_8+b_{10},$$
$$c_3=b_0+b_2+b_3+b_5+b_7+b_9+b_{10}.$$
(Here, addition is modulo 2.)

\paragraph{Solution to the second question}
Assume there is one error in the received message treated as a vector $y=x+\alpha$, where $x$ is the message without error. The vector $\alpha$ contains exactly one element equals $\sa 1$, say its $i$-th element. To locate the error we have to find $i$. 
We get
$$P\times y^T=P\times x^T + P\times \alpha^T=P\times \alpha^T,$$ since $P\times x^T=\bar{0}$.
Then, $P\times \alpha^T$ is the $i$-th column of $P$ and, since all columns of $P$ are distinct, this uniquely determines the index $i$ of the column as wanted.
\end{SOLUTION}

\begin{NOTES}
Hamming codes have been introduced by R. W. Hamming in \cite{Hamming50}.
\end{NOTES}
\end{EXERCISE}
\begin{EXERCISE}[Computing short distinguishing subsequence]\label{pb126}
\newcommand{\erase}{\mathit{erase}}%
The problem considers \MCa{distinguishing subsequences}{distinguishing
subsequence} between two different binary words (see \cite[Problem
51]{CLR21cup}).
Denoting by $\SMot(x)$ the set of subsequences of a word $x$, a word $z$ is
said to distinguish $x$ and $y$, $x\neq y$, if it is a subsequence of only
one of them, that is, $z\in \SMot(x)\Leftrightarrow z\notin \SMot(y)$.

\begin{QUESTION}
Construct a distinguishing subsequence of length at most $\lceil (n+1)/2
\rceil$ for two distinct binary words of the same length $n$.
\end{QUESTION}

In fact, the above bound is optimal.

\begin{QUESTION}
For each $n>0$, construct two distinct binary words of length $n$ that do not
have a distinguishing subsequence of length smaller than $\lceil (n+1)/2
\rceil$.
\end{QUESTION}

\begin{SOLUTION}
Let $\{\sa{a},\sa{b}\}$ be the alphabet of the different words $x$ and $y$ of
length $n$. First note that if the words have different numbers of
occurrences of $\sa{a}$ (or of $\sa{b}$) then both $\sa{a}^k$ and
$\sa{b}^\ell$ are distinguishing subsequences for some integers $k$ and
$\ell$. Choosing the shorter answers the question. We then assume that the
words have the same number of occurrences of $\sa{a}$ (and then of $\sa{b}$).

For a word $w$ and a natural number $k$, denote by $\pos(w,k)$ the position
on $w$ of the $k$-th occurrence of $\sa{b}$ if it exists. If not (when
$w\in\sa{a}^*$ for example) $\pos(w,k)=|w|$. Let
$$i = \min\{k \mid \pos(x,k)\neq \pos(y,k)\},$$
which is well defined because $x\neq y$ and at least one of them have
occurrences of $\sa{b}$. We later assume w.l.o.g. that $pos(x,i)<pos(y,i)$.

Let $x=x_1\cdot\sa{b}\cdot x_2$, where $$x_1=x[0\dd \pos(x,i)-1] \mbox{ and }
x_2=x[\pos(x,i)+1\dd n-1]$$ and let $z_1$ and $z_2$ be the two sequences
defined by
$$z_1=\erase(x_1,\sa{b})\cdot\sa{ab}\cdot\erase(x_2,\sa{a}),$$
$$z_2 = \erase(x_1,\sa{a})\cdot\sa{b}\cdot\erase(x_2,\sa{b}),$$
where, for a word $w$ and a symbol $c$, $\erase(w,c)$ denotes the word
resulting from $w$ by erasing all the occurrences of letter $c$ in it.

It is clear that $z_1$ and $z_2$ are both distinguishing subsequences for $x$
and $y$. Additionally, since $|z_1|+|z_2| = n+2$, at least one of the two
subsequences is of length at most $\lceil (n+1)/2 \rceil$.

\paragraph{Example 1}
For $x=\sa{ababababab}$ and $y=\sa{ababaababb}$ of length $10$, we have
$i=3$, $\pos(x,3)=5$, $x_1=\sa{ababa}$, $x_2=\sa{abab}$. Eventually, we get
the two distinguishing subsequences $z_1=\sa{aaa}\cdot\sa{ab}\cdot\sa{bb}$
and $z_2=\sa{bb}\cdot\sa{b}\cdot\sa{aa}$. The second has length $5<\lceil
(10+1)/2 \rceil$.

\paragraph{Example 2}
For $x=\sa{abababababa}$ and $y=\sa{ababaaabbba}$ of length $11$, we have
$i=3$, $\pos(x,3)=5$, $x_1=\sa{ababa}$, $x_2=\sa{ababa}$. Eventually, we get
the two distinguishing subsequences $z_1=\sa{aaa}\cdot\sa{ab}\cdot\sa{bb}$
and $z_2=\sa{bb}\cdot\sa{b}\cdot\sa{aaa}$. The second has length $6=\lceil
(11+1)/2 \rceil$.

\paragraph{Optimal bound}
Let $n=2m$, $x=(\sa{ab})^m$ and $y=(\sa{ba})^m$. Then, any binary word of
length $m$ is a subsequence of each of these two different words. Hence, they
have a shortest distinguishing subsequence of length exactly $m+1=\lceil
(n+1)/2 \rceil$, for example, $\sa{a}^m\sa{b}$.

For $n=2m+1$ we can choose $x'=x\cdot\sa{a}$ and $y'=y\cdot\sa{a}$, for which
$\sa{b}\sa{a}^m$ is a shortest distinguishing subsequence of the expected
length.
\end{SOLUTION}

\begin{NOTES}
A standard solution to compute a shortest distinguishing
subsequence\INDEX{distinguishing subsequence} of two words is a by-product of
testing the equivalence of their minimal (deterministic) subsequence automata
(see \cite[Problem 51]{CLR21cup}) as an application of the UNION-FIND data
structure, see \cite{AhoHU74}.

There is a linear-time algorithm computing a shortest distinguishing
subsequence of two different words. Such an algorithm was first announced by
Imre Simon, but it has not been published by him. The first (quite
complicated) published linear-time algorithm for this problem is by
Gawrychowski et al. \cite{DBLP:conf/stacs/GawrychowskiKKMS21}.
\end{NOTES}
\vfill
\begin{center}
\begin{tikzpicture}[scale=.20, node distance=1mm, auto, >=latex', on grid, thick]
\draw[opacity=0.6,fill=red] (-3.5,3.6) -- (1.7,6) -- (3.3,5.2) -- (-1.8,2.8)
 -- (-1.8,-0.5) -- (-3.5,0.6) -- (-3.5,3.6);
\draw[opacity=0.6,fill=green] (-1.1,-4.1) -- (-6,-0.9) -- (-6,0.9) -- (-1.1,-2.2)
 -- (1.8,-0.8) -- (1.7,-2.7) -- (-1.1,-4.1);
\draw[opacity=0.6,fill=yellow] (2.6,-5.1) -- (2.9,0.8) -- (0,2.5) -- (1.7,3.3)
 -- (4.4,1.7) -- (4.1,-4.2) -- (2.6,-5.1) ;

\draw (-1.8,-0.5) -- (0,0.3);
\draw (1.8,-0.8) -- (0,0.3);
\draw (0,2.5) -- (0,0.3);
\end{tikzpicture}
\end{center}
\vfill \mbox{}

\end{EXERCISE}

\begin{EXERCISE}[Local periodicity lemma with one don't care symbol]\label{pb130}
The problem concerns periodicities occurring inside a word that contains one occurrence of a don't care symbol (also called hole or joker). It is a letter, denoted by $*$, that stands for any other letter of the alphabet, that is, it matches any letter including itself.
For a string $x$, two of its letters, $x[i]$ and $x[j]$, are said to $\approx$-match, written $x[i]\approx  x[j]$,  if they are equal or one of them is the don't care symbol.

Further, an integer $p$ is a \MC{local period} of $x$ if for each position $i$ on $x$, $0\leq i<|x|-p$, we have $x[i]\approx x[i+p]$.
Recall the Periodicity lemma for usual words (see 
\cite[Chapter 1]{CLR21cup}).

\begin{LEMME}[\MC{Periodicity lemma}]
Let $x$ be a word (without don't care symbol) and let $p,q$ be periods of $x$ that satisfy $p+q-\pgcd(p,q)\leq|x|$. Then, $\pgcd(p,q)$ is also a period of $x$.
\end{LEMME}

The problem is related to an extension of the lemma to
words in which only one don't care symbol occurs.


\begin{QUESTION} (\MC{Local periodicity lemma})
Let $x$ be a word with one don't care symbol and $p,q$ be two relatively prime local periods of $x$ that satisfy $p+q\leq|x|$. Then, $1$ is also a local period of $x$.
\end{QUESTION}


\begin{QUESTION}
Give an example word $x$ with one don't care symbol having local periods $p=5$ and $q=7$ with $p+q-1=|x|$ but not having $1$ as local period.
\end{QUESTION}

The example in this question shows the inequality in the first question is tight.

\begin{SOLUTION}
Let $n=|x|$ and assume $p+q=n$. The case $p+q<n$ can be easily reduced to this case.

Construct the graph $G(n,p,q)$ whose nodes are $0,1,\dots,n-1$ and whose undirected edges $(i,j)$ are pairs of positions on $x$ with
$|i-j|\in\{p,q\}$. The Periodicity lemma implies that the graph is connected but to get the result we are to prove 
a stronger property in the next lemma, namely the biconnectivity of $G(n,p,q)$.

\begin{LEMME} Assume $p,q$ are relatively prime and the word  $x$ 
has periods $p,q$, where $p+q=n$. If $x$ has only one don't care symbol then $x$ is unary.
\end{LEMME}

\begin{PREUVE}
It is enough to show that the graph $G(n,p,q)$ is biconnected, that is, the removal of any single node, potentially a position of the don't care symbol, does not disconnect the graph.

It is easy to see that each node of $G(n,p,q)$ has degree 2. Hence the graph is a set of cycles. Due to the standard periodicity lemma (no don't care symbol) the graph $G(n-1,p,q)$ is connected.
After removing the node $0$ from $G(n,p,q)$ the remaining graph is isomorphic with $G(n-1,p,q)$, hence it is also connected (its nodes are $1,2,\ldots,n-1$).
Consequently the whole graph $G(n,p,q)$ is a connected graph.

The graph $G(n,p,q)$ does not contain loops and consists of a set of
disjoint simple cycle. Therefore it is just one big cycle, because it is connected.
Hence $G(n,p,q)$ is biconnected, since a single simple cycle is biconnected. This completes the proof.
\end{PREUVE}
\end{SOLUTION}

\paragraph{Solution to the second question}
The word $\texttt{ababaababa}$ of length $10$ has periods $5$ and $7$. Note the Periodicity lemma does not apply to it since $5+7-\pgcd(5,7)=11 > 10$.
The word $x=\texttt{ababaababa}*$ of length $11$ has local periods $5$ and $7$ but obviously not period $1$ as required, despite the equality $5+7-1=|x|$.

\begin{NOTES}
A first proof of the Local periodicity lemma with one don't care symbol was given by Berstel and Boasson in \cite{BerstelB99}, however our solution is different. A version of the Local
periodicity lemma with two don't care symbols is rather nontrivial.

The notion of solid periodicities is thorougly investigated by Kociumaka et al. in \cite{KociumakaRRW22} in conjunction with words containing don't care
symbols. An integer $p$ is a solid period of $x$ if there is a word $z$ without don't cares and with period $p$ for which $x\approx z$. If $p$ is a local period it is not necessarily a solid period, see for example
$x=\texttt{a}*\texttt{b}$ and $p=1$.

Also the Solid periodicity lemmas are different. For example, if $|x|\geq 16$ has two don't cares and has solid periods $5,7$ then it should have $1$
as a solid period. But this is not true for local periods, consider the
example word $x= \texttt{aaaaba}*\texttt{aa}*\texttt{abaaaa}$.

\end{NOTES}
\vfill
\begin{center}
\begin{tikzpicture}[scale=.20, node distance=1mm, auto, >=latex', on grid, thick]
\draw[opacity=0.6,fill=red] (-3.5,3.6) -- (1.7,6) -- (3.3,5.2) -- (-1.8,2.8)
 -- (-1.8,-0.5) -- (-3.5,0.6) -- (-3.5,3.6);
\draw[opacity=0.6,fill=green] (-1.1,-4.1) -- (-6,-0.9) -- (-6,0.9) -- (-1.1,-2.2)
 -- (1.8,-0.8) -- (1.7,-2.7) -- (-1.1,-4.1);
\draw[opacity=0.6,fill=yellow] (2.6,-5.1) -- (2.9,0.8) -- (0,2.5) -- (1.7,3.3)
 -- (4.4,1.7) -- (4.1,-4.2) -- (2.6,-5.1) ;

\draw (-1.8,-0.5) -- (0,0.3);
\draw (1.8,-0.8) -- (0,0.3);
\draw (0,2.5) -- (0,0.3);
\end{tikzpicture}
\end{center}
\vfill \mbox{}

\end{EXERCISE}
\begin{EXERCISE}[Text index for patterns with one don't care symbol]\label{pb150}
\newcommand{\D}{\mathbf{D}}%
\newcommand{\NewTree}{\mbox{\sf NewTree}}%
\newcommand{\Off}{\mbox{\sf OffTrie}}
\newcommand{\MaxChild}{\mathit{MaxChild}}%

For a string $w$ of size $n$ over a (large) integer alphabet 
$\Sigma$ we want to 
create a data structure $\D(w)$, of size $O(n\log n)$, 
called the {\it text index}, which allows to
search for a pattern $P$  in $w$  in $O(|P|)$ time
(usually  $|P|<<n$). The pattern $P$ can  contain
a single occurrence of a special symbol $\theta\notin \Sigma$ called a {\it don't care} 
or a {\it wildcard}, which matches any other symbol in $w$.
In this simplified problem we do not ask about time complexity of constructing  $\D(w)$,
since it is quite technical (see the notes).
Our main aim here is only a small size of $\D(w)$ and fast searching of the pattern.

\paragraph{\bf Combinatorics of trees} \mbox{ \ }\\
 We consider only trees with each internal nodes having at least two children. By a size of a tree we mean the number of its
 leaves.
For each internal node $v$ denote by $T_v$ the subtree rooted at $v$.
An edge $v\rightarrow u$, where $v$ is the {\it parent} of $u$ is called
{\it heavy} if $T_u$ has largest size among subtrees rooted
at children of $v$ (in case of ties we choose a single edge).
Other edges are called {\it light}. 
If $v\rightarrow u$ is heavy then subtrees rooted at other children of $v$ are
called {\it light subtrees}. 

Observe that each path from a given leaf to the root contains only
logarithmically many light edges, 
hence each leaf belongs to logarithmically many light subtrees. Consequently we have the following fact.

\paragraph{Observation 1}
The sum of sizes of all light subtrees is
 $O(|T|\log |T|)$.

\begin{QUESTION} Construct  the text index $\D(w)$ of size $O(n
\log n)$ with searching time  $O(|P|)$.
\end{QUESTION}

\AIDE{Use Observation 1}

\begin{SOLUTION}
We assume a word $w$ ends with special endmarker. Let $ST(w)$ be the suffix tree of $w$.
For a trie $T'$
denote by $strings(T')$ the set of strings corresponding to
paths $root\stackrel{*}{\rightarrow} leaf$ in $T'$.

Denote by $LightStrings(v)$ the set of strings corresponding to paths $v\stackrel{*}{\rightarrow} leaf$ in $ST(w)$ starting with light edges originating at $v$.

Let $\NewTree(v)$  be a compacted trie $T'$ such that 
\[\alpha \in strings(T') \ \mbox{if and only if}\
(\,\exists\;a\in \Sigma\,)\; a\alpha\in LightStrings(v).\]
If $q$ is the total size of light subtrees hanging at $v$ 
then it can be easily seen that $|\NewTree(v)|=O(q)$. 
(we do not ask about time complexity of constructing $\NewTree(v)$),
 and we refer to \cite{DBLP:conf/stoc/ColeGL04}.

\noindent A pseudocode of the  construction of $\D(w)$ is given below, see also the figure. 

\medskip\noindent
\begin{center}
\begin{minipage}{8cm}
\noindent {\bf Algorithm} Construct $\D(w)$
\mbox{ \ }

\medskip\noindent
For each original non-leaf node $v$ of $ST(w)$ do

\smallskip\quad
 $T'=\NewTree(v)$;   $r := root(T')$

\smallskip \quad $r := root(T')$

\smallskip \quad  $next(v)=r$;\ $parent(r)=v$

 
\quad create additional edge $v\stackrel{\Theta}{\rightarrow} r$
 \end{minipage}
 \end{center}
 


\paragraph{\bf Size of $\D(w)$}
The total size of additional (after merging) trees 
 is at most the total size of all light trees.
Hence, due to Observation 1,  $|\D(w)|=O(n\log n)$.

\bigskip
\includegraphics[width=10cm]{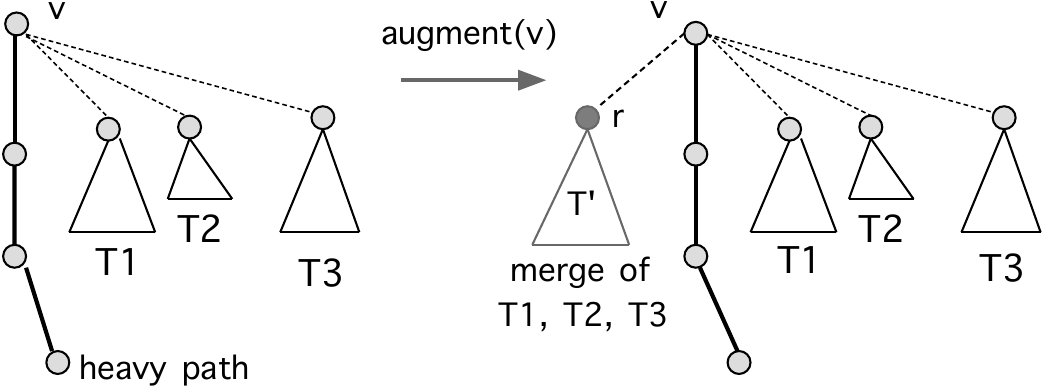}

\paragraph{\bf Searching the pattern $P$}
We scan $P$ and follow the downward path in $T$.
However when we see $\theta$ in $P$ we split the search.
We go to $Next(v)$ and to the next node on the heavy path.
Then we follow two disjoint paths in $\D(w)$.
It still takes $O(|P|)$ time.
\end{SOLUTION}

\begin{NOTES}
In case of $k$ don't care symbols, for $k=O(1)$, one can 
construct the text index $\D(w)$ of size $O(n\log^k n)$ with searching time $O(|P|)$. Initially we proceed similarly as in case one error. Then we recursively process 
 the newly added subtrees (in the merges) with respect to $k-1$ don't cares.
 
In the case of don't cares in the pattern the bottleneck of time complexity of constructing $\D(w)$ is the construction of the tries $\NewTree(v)$, it is technical and we refer to 
\cite{DBLP:conf/stoc/ColeGL04}. 

Our presentation is a version of the simplest case of 
 an approximate text index presented in
 \cite{DBLP:conf/stoc/ColeGL04}, where  
  don't cares and 
 edit operations in the text  were allowed, however
the general case is very technical.
\end{NOTES}
\end{EXERCISE}
\begin{EXERCISE}[Words with  distinct cyclic k-factors]\label{pb128}
\newcommand{\RingWord}{\mathsf{RingWord}}
\newcommand{\Family}{\mathsf{Compl}}
\newcommand{\CC}{\mathsf{ComputeChain}} 
\newcommand{\Glue}{\mathsf{GLUE}}
Assume  the alphabet is $\{0,1\}$.
A binary word $v$  is called  a cyclic factor of a word 
$w$, if $v$
is a (standard) factor of $w^{\infty}$.
A binary (cyclic) word $w$ of length $k\leq n\le 2^k$ is called a 
{\tt $k$-ring word} if
each  cyclic $k$-factor  of $w$ occurs once.
For example, the word
 $w=000101101$ is a binary $4$-ring word.
Observe that
 string of length $k$ is a $k$-ring if and only if it 
is primitive.

We refer to \cite[Problem 18]{CLR21cup}, \cite[Problem 69]{CLR21cup}  for the 
 definition of de Bruijn graph $G_{k}$.
Two edges of $G_{k}$ are loops and in this problem we disregard these two edges.
The   nodes of $G_{k}$ are binary words of length $k-1$, and edges correspond to words of length $k$.
The number of edges of $G_k$ is $2^k$.
The size of these
graphs is $O(n)$ since we can choose minimal $k$ such that $n\le 2^k$.
A {\bf closed chain} (c-chain, in short) is a path $C$ ending and starting in
the same node and containing each of its edges exactly once. 
Denote by  $w=\RingWord(C)$ a word $w$ resulting by spelling labels of 
consecutive edges of $C$.

\begin{QUESTION}
Design a linear time algorithm constructing a binary $k$-ring word,
for given $n,k$, such that $k\le n\le 2^k$.

\smallskip\noindent
{\bf Equivalent formulation:} 
construct a closed chain $C$ of length $n$ in
$G_k$, then $\RingWord(C)$ is a $k$-ring word of length $n$.
\end{QUESTION}
%

\begin{SOLUTION}
Assume the c-chains are represented as cyclic lists
of consecutive nodes. 

\paragraph{Fact 1}
Assume $H$ is a regular subgraph of  $G_k$,  
such that each node of $G_k$ is contained in an edge in $H$.
Then we can compute a  single 
c-chain $\Glue(H)$ in $G_k$  of length $n$.

\begin{PREUVE}
Graphs $G_k$  have 
the following   simple property (\cite[Problem 69]{CLR21cup}).

\smallskip\noindent {\bf Claim.}
Assume we are given 
two node-disjoint c-chains $C_1,C_2$, and an edge $u\rightarrow v$ in $G_k$,
where $ u\in C_1,v\in C_2$.
Then in time $O(1)$ we can create a new c-chain $merge(C_1,C_2)$ of
length $|C_1|+|C_2|$, whose set of nodes
is the union of sets of nodes of $C_1,C_2$.

\medskip
Each connected component of $H$ is  an
Eulerian graph. We can compute c-chains, containing all nodes
of this component, using  Euler algorithm in linear time. If there is one component we are done.
Otherwise there should be two c-chains $C_1,C_2$ satisfying assumption of Fact 1,
since $H$ has no isolated nodes, and because $G_k$ is a connected graph.
Then,  we replace $C_1,C_2$ by $merge(C_1,C_2)$ into a single c-chain.
We iterate this process until we get a single c-chain 
which is a required output.
\end{PREUVE}

We say that a set $X$ of edge-disjoint c-chains is covering $G_k$ if
it  contains each edge of $G_k$.

\paragraph{\bf Fact 2}
We are given   a  c-chain $C$  in $G_{k}$. 
Then we can compute  in time $O(|G_k|)$ a set $\Family(C,k)$ 
of c-chains  such that $\Family(C,k)\cup \{C\}$ is covering $G_k$ (in particular $\Family(C,k)$
has together $2^k-|C|$ edges).

\begin{PREUVE}
A directed graph is called {\it regular} if for each 
node the numbers of its out-going and in-going edges are equal (though can differ for 
distinct nodes).
It is known that  a directed graph has an Euler cycle if and only if it is
regular and connected.
We remove edges of $C$ (but not nodes) and receive the graph $G_k-C$, afterword each connected component contains a directed Eulerian graph.
Then $\Family(C,k)$ consists of Eulerian cycles of these 
components and the cycle $C$. 
\end{PREUVE}
Each edge the form $a_1a_2\cdots a_{k-1}\Arrow{a_k} a_2a_3\cdots a_k,$ in 
$G_k$
corresponds to the node $a_1a_2\cdots a_k$ in $G_{k+1}$.
Each c-chain $C$ of length $n\le 2^{k-1}$
in $G_{k-1}$ corresponds to a simple cycle of length $n$ in
$G_k$, denote this cycle by $\Phi_k(C)$. The edges of $C$ correspond to nodes
of $\Phi_k(C)$.

\paragraph{Example} Below we  show a c-chain $C$ in $G_3$ and $\Phi_4(C)$.
{\small
\[
C\,=\, 01\Arrow{0} 10 \Arrow{1}  01 \Arrow{1} 11 \Arrow{0} 10
\Arrow{0}  00 \Arrow{0}  00 \Arrow{1} 01\]
\[\Phi_4(C)\,=\,010\Arrow{1} 101 \Arrow{1} 011 \Arrow{0} 110 \Arrow{0} 100 \Arrow{0} 000 \Arrow{1} 001\Arrow{0} 010\] 
}

A pseudo-code  of the recursive algorithm computing closed chain of length $n$
in $G_k$, for $n\le 2^k$, is given below.

\begin{center}
\begin{minipage}{10.5cm}
\medskip\noindent 
{\bf Algorithm} $\CC(k,n)$

\smallskip\noindent
\begin{enumerate}
\item 
if $n\le 2^{k-1}$ then

\hspace*{0.5cm} $C$ := $\CC(k-1,n)$;\; 
return $\Phi_k(C)$

\smallskip

\item $H:=G_k$

\smallskip
\item
Let  $n=2^{k-1}+r,\ 0<r\leq 2^{k-1}$
\item 

\smallskip
$C$ := $\CC(k-1,r)$\ \ ({\it recursive call})
\item 

\smallskip
For each $C'\in \Family(C,k-1)$ remove edges of $\Phi_k(C')$ from $H$

\smallskip \hspace*{0.2cm} ($H$ has $2^k-(2^{k-1}-r)$ edges, 
it satisfies assumptions of  Fact 1)

\smallskip


\item 
return $\Glue(H)$
\end{enumerate}

%
\end{minipage}
\end{center}

The time complexity $T(n)$  of the algorithm satisfies
$T(n)= O(T(n/2)+O(n).$
 It implies $T(n)=O(n)$.
\end{SOLUTION}

\begin{NOTES}
Our algorithm is a  version of the algorithms in \cite{Yoeli62,DBLP:journals/iandc/GabricHS22}.
Recently, in \cite{DBLP:journals/corr/abs-2401-14341},
  there were investigated words $w$, called   {\it orientable sequences}, such that all cyclic length-$n$ factors of 
  $w$ and $w^R$ are distinct.
\end{NOTES}
\end{EXERCISE}

\begin{EXERCISE}[Huffman codes vs entropy]\label{pb133}
\newcommand{\huf}{\mathsf{Huffman}}%
\newcommand{\Entropy}{\mathsf{Entropy}}
We consider $n$ items with positive weights probabilities
$p_1, p_2, \dots, p_n$ satisfying $\sum_i\, p_i=1$ and
let $\mathbf{p}=(p_1,p_2,\dots,p_n)$.
\MC{Huffman algorithm} (see \cite[Problem 99]{CLR21cup}) constructs a full binary
tree (each non-leaf node has two children) with items assigned to its leaves. Let $l_i$ be the depth of $p_i$, number of edges from the root to $p_i$.
The average length of the \MC{Huffman coding} of items is
$\huf(\mathbf{p})=\sum_i p_i\cdot l_i$.
An important concept in information theory is \MC{entropy}.
The sequence $\mathbf{p}$ is treated as a source of information and we define
 $\Entropy(\mathbf{p})\,=\, -\sum_i p_i\cdot \log_2 p_i.$

\paragraph{Property A} A useful property for the solution below is: if $p_1,\dots,p_n$, and $q_1,\dots,q_n$ are two sequences of positive integers
with the same sum, then $-\sum_i\,p_i \log_2 q_i\geq -\sum_i\,p_i \log_2 p_i$.
From the inequality, it follows directly: $\log_2 x \le x-1$, for $x>0$.

\begin{QUESTION}\\
Show that $\Entropy(\mathbf{p})\le \huf(\mathbf{p})\le \Entropy(\mathbf{p})+1$.
\end{QUESTION}


\paragraph{Example}\\
\parbox{60mm}{Besides is the Huffman tree\\ corresponding to the sequence\\
$\mathbf{p}=[0.1,0.1,0.3,0.5]$.
Then,\\ 
$\huf(\mathbf{p})$ is\\
$0.1\cdot 3 + 0.1\cdot 3 + 0.3\cdot 2 + 0.5\cdot 1 = 1.7$\\
and
$\Entropy(\mathbf{p}) 
\approx 1.68548$.\\
We have $1.68548 \leq 1.7 \leq 2.68548$}.
\hspace{3mm}
\parbox{50mm}{
\begin{tikzpicture}[scale=0.35,node distance=1.5cm, auto,>=latex', thick]
\path[->] node (q23) at (6.375,9) {\scriptsize 1};
\path[->] node (q24) at (9,6.5) {\scriptsize 0.5}
             (q23) edge node {} (q24);
\path[->] node (q25) at (3.75,6.5) {\scriptsize 0.5}
             (q23) edge node {} (q25);
\path[->] node (q26) at (6,4) {\scriptsize 0.3}
             (q25) edge node {} (q26);
\path[->] node (q27) at (1.5,4) {\scriptsize 0.2}
             (q25) edge node {} (q27);
\path[->] node (q28) at (0,1.5) {\scriptsize 0.1}
             (q27) edge node {} (q28);
\path[->] node (q29) at (3,1.5) {\scriptsize 0.1}
             (q27) edge node {} (q29);
\end{tikzpicture}}

\begin{SOLUTION}
The solution is built in three steps.

\paragraph{Fact 1}\\
(i) In any full  binary tree we have:\
$\sum_i\, 2^{-l_i}= 1$.\\
(ii)   $\sum_i\, 2^{-l_i}\le 1$ implies that
there is a binary  full) tree
with depths of leaves $l_1,l_2,\dots,l_n$.

\begin{PREUVE}
Proof of point (i). Choose two leaves that are children of the same node. They are at the same
depth $l$. After removing these leaves their parent becomes a leaf at depth $l-1$
and the whole sum $\sum_i\, 2^{-l_i}$ does not change.
Eventually, we get a full binary tree with only 2 leaves. 

Proof of point (ii). Assume now $\sum_i\, 2^{-l_i}\le 1$. Take the maximum $l_i$. 
If there is another $l_j$ with $l_j=l_i$, then create a node at depth $l_i-1$ with 2 children.
The depths $l_i,l_j$ are removed and replaced by a single depth $l_i-1$.
If there is no such $l_j$, then create a new node at level $l_i-1$
and having a single child.
It is iterated until the required tree is obtained.
\end{PREUVE}

\paragraph{\bf Fact 2} $\Entropy(\mathbf{p})\le \huf(\mathbf{p})$.

\paragraph{\bf Proof}  Let $q_i=2^{-l_i}$. We have $l_i=-\log_2 2^{-l_i}$,
and due to Fact 1
$\sum_i\,p_i=\sum_i \, q_i$.
 Hence
$$\sum_i\,p_il_i=-\sum_i\,p_i\cdot \log_2 q_i
\stackrel{\rm property A}{\geq}
-\sum_i\,p_i\cdot \log_2 q_i=\Entropy(\mathbf{p}).$$

\paragraph{\bf Fact 3}
$\huf(\mathbf{p})\le \Entropy(\mathbf{p})+1$.

\begin{PREUVE}
Let $l_i=\lceil -\log_2 p_i\rceil$.
Then, $2^{-l_i}\le p_i$.
Hence, $\sum_i\, 2^{-l_i}\le \sum_i\, p_i= 1$, and
due to  Fact 1, there is a binary tree  (not necessarily full)
with depths of leaves $l_1,l_2,\dots,l_n$.
The average path length in such a tree is $$\sum_i\, p_i\cdot l_i
= \sum_i\, p_i\cdot \lceil -\log_2 p_i\rceil
 \le  \sum_i\, p_i\cdot (-\log_2 p_i +1)$$
$$\ = -\sum_i\, p_i\cdot \log_2 p_i
+ \sum_i\,p_i =\Entropy(\mathbf{p})+1.$$
Consequently, there is a tree whose cost is at most $\Entropy(\mathbf{p})+1$.
However, the Huffman tree realises the minimum cost,
which shows that we have also $\huf(\mathbf{p})\le \Entropy(\mathbf{p})+1$
as required.
\end{PREUVE}
\end{SOLUTION}

\begin{NOTES}
The relation between Huffman trees and entropy is from \cite{Shannon}.
\end{NOTES}
\end{EXERCISE}

\begin{EXERCISE}[Compressed pattern matching in Thue-Morse words]\label{pb134}
The {Thue-Morse binary word} on the alphabet $\{\sa{0}, \sa{1}\}$ is produced by iterating infinitely from $\sa{0}$ the \MC{Thue-Morse morphism} $\mu$ from $\{\sa{0}, \sa{1}\}^*$ to itself defined by
$$\mu(\sa{0}) = \sa{01}, \
 \mu(\sa{1}) = \sa{10}.
$$
Eventually, the iteration produces the infinite Thue-Morse word:
$$\mathbf{t}\,=\,
 \sa{01101001100101101001011001101001}\cdots.
$$

For a pattern $x\in \{\sa{0}, \sa{1}\}^*$ of even length, let $\mu^{-1}(x)$ be the word $z$ for which $\mu(z)=x$ if it exists and $\textit{nil}$ otherwise.
In other words $\mu^{-1}(x)\ne nil$ if $x\in \{01,10\}^*.$
We also introduce the set $\mathit{EVEN}=\{\sa{0110}, \sa{1010}, \sa{0101}, \sa{1001}\}$.

Let us denote by  $\first_4(x)$ the prefix of $x$ of length 4, if there is any, 
by $\first(x)$ and $\last(x)$ the first and last letters of $x$, respectively, 
and by $\overline{s}$ the negation of a bit $s$ ($\overline{\sa{0}}=\sa{1}$ and $\overline{\sa{1}}=\sa{0}$).

\smallskip
The following algorithm tests in linear time and in a very simple way
if a finite binary pattern $x$ is a factor of $\mathbf{t}$.

\begin{algo}{Test}{x \mbox{ non-empty word}}
\IF{x=\textit{nil}} 
  \RETURN{\False}
\FI
\IF{|x|< 4}
  \RETURN{(x\ne \sa{111} \textrm{ and } x\neq \sa{000})}
\FI
\IF{\first_4(x)\notin \mathit{EVEN}}
  \SET{x}{\overline{\first(x)} \cdot x}
\FI
\IF{|x| \textrm{ is odd}} 
  \SET{x}{x\cdot \overline{\last(x)}}
\FI
\RETURN{\Algo{Test}({\mu^{-1}(x)})}
\end{algo}


\begin{QUESTION} 
Show why this algorithm correctly tests in linear time if $x$ is a factor
of the infinite Thue-Morse word $\mathbf{t}$.
\end{QUESTION}

\begin{SOLUTION}
First note that if $\mu^{-1}(x) \neq \textit{nil}$ then $|\mu^{-1}(x)|=\frac{1}{2}|x|$, which
implies the linear running time.
The correctness is a consequence of three simple observations.

\begin{description}
\item {(i)} The set $\mathit{EVEN}$ is the set of all length-$4$ words that occur in $\mathbf{t}$ starting at even positions.
\item {(ii)} A nonempty word $x$ of even length starts at an even
position in $\mathbf{t}$ if and only if $\mu^{-1}(x)$ occurs in $\mathbf{t}$.
\item {(iii)} If $|x|<4$ then $x$ is a factor of $\mathbf{t}$
if and only if $x\ne \sa{111} \textrm{ and } x\neq \sa{000}$.
\end{description}

The algorithm checks if $x=ayb$ starts at an even position using (i).
If $x$ starts at odd position we add $\overline{a}$ at the  beginning
of $x$. Now if the length of $x$ becomes odd we add $\overline{b}$ at the end.
In this way, we slightly change $x$ forcing $x$ to occur at an even position and
to have an even length. Then the resulted word $x$ is a factor of $\mathbf{t}$
if an only if $\mu^{-1}(x)$ is.
The correctness follows from the observations above.
\end{SOLUTION}

\begin{NOTES}
Equivalent definitions of the Thue-Morse word can be found in \cite[Chapter 2]{Lothaire02} and \cite[Chapter 1]{CLR21cup}, for example.

The infinite {Fibonacci word} $\mathbf{f}$ is generated by iterating, starting from $\sa{a}$, the morphism $\phi$ from $\{\sa{a}, \sa{b}\}^*$ to itself defined
by $\phi(\sa{a}) = \sa{ab},\phi(\sa{b}) = \sa{a}$.

Following the same strategy as above, we get a very simple algorithm testing if a nonempty binary word is a factor of $\mathbf{f}$.

\begin{algo}{Test-Fib}{x \mbox{ non-empty word}}
\ACT"{{\bf if} $x=\textit{nil}$ {\bf then return} {\False}}
\ACT"{{\bf if} $|x|=1 $ {\bf then return} {\True}}
\ACT"{{\bf if} $x=\sa{b}y $ {\bf then } $x\leftarrow \sa{ab}y $}
\ACT"{{\bf if} $x=y\sa{ba}$ {\bf then } $x\leftarrow y\sa{b}$}
\ACT"{{\bf if} $x=y\sa{aa}$ {\bf then } $x\leftarrow x\sa{b}$}
\RETURN {\Algo{Test-Fib}(\phi^{-1}(x))}
\end{algo}

\noindent For example:

\noindent
$\Algo{Test-Fib}(\sa{baa})
=\Algo{Test-Fib}(\phi^{-1}(\sa{abaab}))
=\Algo{Test-Fib}(\sa{aba})\\
=\Algo{Test-Fib}(\phi^{-1}(\sa{ab}))
=\Algo{Test-Fib}(\sa{a})
=\True$.

\noindent However

\noindent
$\Algo{Test-Fib}(\sa{baaa})
=\Algo{Test-Fib}(\phi^{-1}(\sa{abaaab}))
=\Algo{Test-Fib}(\sa{abba})\\
=\Algo{Test-Fib}(\phi^{-1}(\sa{abb}))
=\Algo{Test-Fib}(\textit{nil})=\False$.

\smallskip
Thue-Morse and Fibonacci words are examples of morphic words. Testing a pattern in any morphic word is the subject of \cite[Problem 68]{CHL07cup}.
We showed special cases of pattern-matching in compressed texts.
The fastest algorithm for general case, using {\it recompression technique}, was presented in  \cite{Jez}.
\end{NOTES}
\vfill
\begin{center}
\begin{tikzpicture}[scale=.20, node distance=1mm, auto, >=latex', on grid, thick]
\draw[opacity=0.6,fill=red] (-3.5,3.6) -- (1.7,6) -- (3.3,5.2) -- (-1.8,2.8)
 -- (-1.8,-0.5) -- (-3.5,0.6) -- (-3.5,3.6);
\draw[opacity=0.6,fill=green] (-1.1,-4.1) -- (-6,-0.9) -- (-6,0.9) -- (-1.1,-2.2)
 -- (1.8,-0.8) -- (1.7,-2.7) -- (-1.1,-4.1);
\draw[opacity=0.6,fill=yellow] (2.6,-5.1) -- (2.9,0.8) -- (0,2.5) -- (1.7,3.3)
 -- (4.4,1.7) -- (4.1,-4.2) -- (2.6,-5.1) ;

\draw (-1.8,-0.5) -- (0,0.3);
\draw (1.8,-0.8) -- (0,0.3);
\draw (0,2.5) -- (0,0.3);
\end{tikzpicture}
\end{center}
\vfill \mbox{}

\end{EXERCISE}

\begin{EXERCISE}[Compressed strings of combinatorial generations]\label{pb135}
\newcommand{\BS}{\mathbf{R}}
There are many interesting strings related to combinatorial generations,
their characteristic feature is usually {\it high compressibility}.
Here we discuss  permutation generations.
Usually they produce
each successive permutation by applying some  kind of a
{\it basic operation}.
The sequence of this operations, corresponding to the permutation ordering, is called
the {\it generating sequence}. It is treated as a word over the alphabet
consisting of names of basic operations. The most interesting are
cases when this alphabet is small. We present in detail  five very similar generation sequences, each time
the $n$-th sequence is expressed recursively in terms of the $(n-1)$-th sequence
using stringologic operations. The corresponding recurrences  have similar structure. We use
operations of concatenation,  morphisms and reversals.

Assume the permutation of
numbers $\{1,2,\ldots,n\}$ is stored
in a table with positions numbered from zero. 


\medskip
 We consider compression in terms of straight-line programs.
A straight-line program, briefly SLP, is a context-free grammar that produces a single
word $w$ over a given alphabet $\Sigma$.

An SLP  can be also defined as a
sequence of recurrences (equations), using operations of concatenation of words.
Compression by straight-line programs is also
called Grammar-Based Compression.

\begin{QUESTION}
Construct a generating sequence $\mathbf{Z}_n$ which 
generates all $n$-permutations using operations of prefix reversals 
and can be described by an
SLP of size $O(n\log n)$. 
Show also that each SLP of any generating sequence for $n$-permutations
is of $\Omega(n\log n)$ size.
\end{QUESTION}

\begin{SOLUTION}
In this generation the  alphabet of basic
operations is $\Sigma_n=\{1,2,\ldots,n-1\}$. The symbol $i$ corresponds to
the basic operation:
reverse the prefix $\pi[0\dd i]$ of permutation $\pi$. 

We define  the word $\mathbf{Z}_n$ by recurrences
$$
\mathbf{Z}_2 = 1;\ \fbox{$\mathbf{Z}_{n+1}\,=\,\mathbf{Z}_n\cdot (\, n\cdot \mathbf{Z}_n\,)^n\,$}
\ \mbox{for}\ n>2.
$$
\medskip\noindent
For example 
$\mathbf{Z}_3\,=\,1\,2\,1\,2\,1, \ \ \mathbf{Z}_4\,=\, 
1\,2\,1\,2\,1\,3\,1\,2\,1\,2\,1\, 3\, 1\,2\,1\,2\,1\,3\,
1\,2\,1\,2\,1$

\medskip
$\mathbf{Z}_n$ is a generating sequence: starting from the id-permutation
$\pi=(\pi_0,\pi_1,\pi_2,\ldots,\pi_{n-1})\,=\, (1,2,\ldots,n),$ and
consecutively applying operations from $\mathbf{Z}_n$ all $n$-permutations
are generated, each exactly once.

\paragraph{\bf Example} Recall that we number positions in
permutations starting from 0, but the $n$-permutations consist
of numbers $1,2,\ldots,n$.\\
We have $\mathbf{Z}_3=12121$ and the generation of $\{1,2,3\}$
is:
{\small
$$123\Arrow{1} 213 \Arrow{2}  312\Arrow{1} 132\Arrow{2} 231 \Arrow{1}
321
$$}
The generation of all 24 permutations
of $\{1,2,3,4\}$ has the following structure.
{\small
$$1234\Arrow{\mathbf{Z}_3} 3214\Arrow{3}4123 \Arrow{\mathbf{Z}_3} 2143\Arrow{3} 3412 \Arrow{\mathbf{Z}_3}
1432
\Arrow{3}  2341 \Arrow{\mathbf{Z}_3} 4321$$
}
\smallskip\noindent
It is not exactly an SLP because of exponents.
However $X^{n}$ can be
rewritten as $O(\log n)$-legth  SLP in a strict sence. Hence
the total size of all expressions defining $\mathbf{Z}_n$, using only concatenation, is $O(n \log n)$.


\paragraph{\bf Fact 1}
If we start with $x_0x_1\cdots x_{n-1}$ then $\mathbf{Z}_n$ generates all permutations of
$\{x_0,\ldots,x_{n-1}\}$ and ends in $x_{n-1}x_{n-2}\cdots x_0$.

\begin{PREUVE}
Assume it is true for $n$. We show it holds for $n+1$.
First notice that, due to inductive assumption,
$$x_0x_1\cdots x_{n-1}x_n \Arrow{\mathbf{Z}_{n}}
x_{n-1}x_{n-2}\cdots x_0x_n\Arrow{n} x_nx_0x_1\cdots x_{n-1}$$
Hence each $\mathbf{Z}_n\cdot n$ produces the left cyclic shift
and  $\mathbf{Z}_{n+1}$ works like $n$ left cyclic shifts, followed by reversing the
prefix of size $n$. After $n$ left cyclic shits we get
$x_1\cdots x_{n-1}x_{n}x_0$. Then the last $\mathbf{Z}_n$ reverses $x_1\cdots x_{n-1}x_{n}$ and
we get $x_n x_{n-1}x_{n-2}\cdots x_0$. This completes the proof.
\end{PREUVE}

An SLP of size $k$ can generate only words of single
exponential size $N=O(2^k)$, consequently $k=\Omega(\log N)$.
We have $|\mathbf{Z}_n|=n!-1$, hence in this case $k=\Omega(\log n!)$,
which is $\Omega(n\log  n)$.
\end{SOLUTION}

%

  \paragraph{Modified prefix reversals}
  Now our basic operation $\BS(k)$ ($k$, in short) consists in reversing a prefix of size $k$ and moving it to the end of the word. 
In
other words, if $x=uv,\, |u|=k$, then $\BS(k)(x)\,=\, vu^R$.\\
For example \[(\underline{1,2,3},4,5,6)\,\stackrel{3}{\rightarrow}\, 
(4,5,6,\underline{3,2,1}).\]

\begin{QUESTION} 
Write a compact representation of the  generator  
using modified prefix reversals (reversed prefix is moved
to the end).
\end{QUESTION}

\begin{SOLUTION}
A permutation generator using operations $\BS(k)$
corresponds to another compactly described generating sequences.
An iterative generation with function $\BS(k)$ is exceptionally simple.

\medskip\noindent The following algorithm   is a version of the  iterative algorithm C, 
which Knuth in his 4-th volume of
``The art of computer programming'' (page 56) called ``the simplest 
permutation generator of all''.
We refer to Knuth's book for correctness of the algorithm C.

\begin{minipage}{11cm}
\medskip\noindent 
{\bf Algorithm} $NEXT(x)$

$x$ is a permutation of $\{1,2,\dd n\}$

let $u$ be
the shortest
prefix of $x$ which is not a prefix of $n,\,n-1,\, n-2,\,\dd 2,\, 1$

if $|u|=n$ the STOP

let $x\,=\, uv$

 return $vu^R$.
 \end{minipage}
 
\medskip\noindent
We construct the generation sequence $\mathsf{M}_n\,=\,
(k_1,k_2,k_3,\cdots,k_m)$  of identifiers of
 actions $\BS(k)$. 
 
\noindent   
The word $\mathsf{M}_{n}$ can be defined  by a recurrence,
\begin{equation}
\mathsf{M}_2=1,\  \fbox{$\mathsf{M}_{n+1}\,=\, 1^n\,\prod_{i=1}^m\, (\,(a_i+ 1) \cdot 1^n\,)$}
\end{equation}
{where} $a_1a_2\cdots a_m=\mathsf{M}_n$.

\paragraph{\bf Example}
\noindent We have,
\[\mathsf{M}_2=1,\ \ \mathsf{M}_3\;=\;  1  1  \  2\   1  1, \quad  
\mathsf{M}_4\;=\;1   1    1   \ \mathbf{2} \ 1    1    1   \ \mathbf{2}  \
1    1    1  \  \mathbf{3}\    1    1 
  1   \  \mathbf{2} \    1    1    1   \  \mathbf{2}\     1   1    1.\]
 For $n=3$ the output (sequence of generated permutations) is
\[\hspace*{5mm}
\fbox{123}\stackrel{1}{\rightarrow}\fbox{231}\stackrel{1}{\rightarrow} \fbox{312}\stackrel{2}{\rightarrow} \fbox{213}\stackrel{1}{\rightarrow} \fbox{132}\stackrel{1}{\rightarrow}  \fbox{321}.\]
For $n=4$, the generation is:

\medskip\noindent
$\fbox{123}\,4  \stackrel{1}{\rightarrow}  2341 \stackrel{1}{\rightarrow}
3412 \stackrel{1}{\rightarrow}  4123 \stackrel{2}{\rightarrow}
\fbox{231}\,4 \stackrel{1}{\rightarrow}  3142 \stackrel{1}{\rightarrow}
1423 \stackrel{1}{\rightarrow}
4231 \stackrel{2}{\rightarrow}  \fbox{312}\,4 \stackrel{1}{\rightarrow}
1243 \stackrel{1}{\rightarrow}  2431 \stackrel{1}{\rightarrow}  4312
\stackrel{3}{\rightarrow}  \fbox{213}\,4 \stackrel{1}{\rightarrow}
1342 \stackrel{1}{\rightarrow}  3421 \stackrel{1}{\rightarrow}  4213
\stackrel{2}{\rightarrow}   \fbox{132}\,4 \stackrel{1}{\rightarrow}\,  3241
\stackrel{1}{\rightarrow}\, 2413 \stackrel{1}{\rightarrow}\,  4132
\stackrel{2}{\rightarrow}\, \fbox{321}\,4\, \stackrel{1}{\rightarrow}\,
2143 \stackrel{1}{\rightarrow} \, 1432 \stackrel{1}{\rightarrow} \, 4321 $

\medskip\noindent Observe how the sequence for $n=4$ results 
from the sequence for $n=3$ of boxed fragments in a recursive way.
For each $n$-permutation $\pi$ we replace it by $\pi'\,:=\, \pi\cdot (n+1)$ and 
generate all cyclic shifts of $\pi'$. For example, in case $n=3$,  $312$ is replaced by the sequence $3124,\, 1243,\, 2431\, 3412$.

If we replace $\mathbf{Z}$ by $\beta$ then Fact 1 remains true and correctness proof for the sequence $\beta$ is very similar to that for $\mathbf{Z}$.

\end{SOLUTION}

\paragraph{Generating by transpositions}
Let $\langle i,j\rangle $ represent a  transposition $(x[i],x[j]):=)x[j],x[i])$.

\begin{QUESTION} (Heap's algorithm) 
    Construct a compactly represented sequences $\mathsf{H}_n$ of transpositions 
    which are permutations generators, such that $\mathsf{H}_n$ is
    a prefix of $\mathsf{H}_{n+1}$.
\end{QUESTION}

\begin{SOLUTION}
 The permutations are stored as $x[0,1,\dd n-1]$, where $n$ is the number of elements. Define the words $w_n$, each of size $n-1$, 
for $0\le i \le n-2$, 
as:
\[w_n[i] = \begin{cases}
\langle 0,n-1\rangle & \mbox{ if } n \mbox{ is odd}\\[2mm]
\langle i,n-1\rangle & \mbox{ if } n \mbox{ is even}\\[2mm]
\end{cases}
\]
Then  Heap's algorithm \cite{DBLP:journals/cj/Heap63}
 corresponds
to the sequence $\mathsf{H}_n$ of basic operations defined as follows
\begin{equation}
\mathsf{H}_2 = \langle 0,1\rangle;\ \fbox{$\mathsf{H}_{n+1}\,=\,\mathsf{H}_n\cdot 
\prod_{i=0}^{n-1}\;(\,w_n[i]\cdot \mathsf{H}_n\,)$} \ \mbox{for}\ n>2.
\end{equation}
%

\paragraph{Example}
\noindent We have:
{\small
$\mathsf{H}_3=\mathsf{H}_2\, (\langle 0,2\rangle \, \mathsf{H}_2)^2
\,=\, \langle 0,1\rangle\, \langle 0,2\rangle\, \langle 0,1\rangle\, \langle 0,2\rangle\, \langle 0,1\rangle.$
\[\mathsf{H}_4=
\mathsf{H}_3\, \langle 0,3\rangle \, \mathsf{H}_3\, \langle 1,3\rangle ,\,\mathsf{H}_3\, \langle 2,3\rangle \, \mathsf{H}_3. \
\mathsf{H}_5=
\mathsf{H}_4\, (\langle 0,4\rangle \, \mathsf{H}_4)^4. \]
}

Starting with  (0, 1, 2, 3, 4, 5), Heap's algorithm would produce (3, 4, 1, 2, 5, 0) as last permutation; starting with (0, 1, 2, 3, 4, 5, 6, 7), it would produce
(5, 6, 1, 2, 3, 4, 7, 0) as last permutation.
Note that starting with (0, 1, 2, 3, 4), it would produce (4, 1, 2, 3, 0) as last permutation; starting with (0, 1, 2, 3, 4, 5, 6), it would produce
(6, 1, 2, 3, 4, 5, 0) as last permutation. 

\medskip
\noindent Correctness of $\mathsf{H}_n$ follows  from the general property:
\begin{itemize}
    \item Assume $n> 3$. After performing $\mathsf{H}_n$, starting with the permutation 
    $1,2,\dd n$, we generate each permutation exactly once, and
    finish with $n,2,3,4,\dd n-1,1$, if $n$ is odd, and 
\end{itemize}
\end{SOLUTION}

\begin{NOTES}
Our presentation of reversing-prefixes algorithm 
follows the Zaks algorithm \cite{Zaks84} for permutation generation.
The recurrences  were given originally in  terms of suffixes, but
it is essentially equivalent to taking prefixes.
In \cite{SawadaW16} the same permutation generating sequence $\mathbf{Z}_n$
 was described by a greedy
algorithm. 
Each sequence $\mathbf{Z}_n$ as a prefix of size $n!-1$ of the  sequence
$\rho\,=\,(\rho_1,\rho_2,\rho_3,\ldots )$, where
$$\rho_k= \max \lbrace j : j! \ \mbox{is a divisor of}\ k \rbrace, \ \mbox{for}\ k\geq 1.$$
We have $\rho\,=\, 1\,2\,1\,2\,1\,3\,1\,2\,1\,2\,1\,3\, 1\,2\, \ldots$.
$\rho_n$ s the sequence of values of so called {\it factorial ruler} function.
$\rho_n$ can be also generated on-line using extra  memory of size
$O(n)$ in the following way. 

%
\paragraph{Ehrlich algorithm}
Gideon Ehrlich devised in \cite{Ehrlich73} a tricky version of Zaks algorithm,
this time
the operation $i$ corresponds to the transposition $x_0 \leftrightarrow x_i$,
also called ``star transposition''. 
We cite D. Knuth, as he has written in his 4-th volume of ``The Art of Computer Programming'', fascicle 2, page 57:
``The most amazing thing
about this algorithm ... is that it works.''
Knuth in his book also gives a sketch of  correctness proof  (exercise 55).
We present here our stringologic version of Ehrlich algorithm showing
its remarkable similarity to Zaks algorithm.
\\
Assume $\odot$ is the composition of functions from left to right, and
$Shift_k(x)$ moves the $k$'th element of $x$  to the beginning of $x$.
The sequence $\mathsf{E}_n$ of {\it star transpositions}
generating $n$-permutations is
compactly represented, using morphisms $h_n$, as

\begin{equation}
 \fbox{$\mathsf{E}_{n+1}=\mathsf{E}_n\,\prod_{i=1}^n\,(\,{ n}\,h_n^i(\mathsf{E}_{n})\,),\ \ h_{n+1}=h_n^{n+1}\odot Shift_n,$}
 \end{equation}
 for $n\geq 2$, where
$\mathsf{E}_2=1,\ \ h_2=\mbox{Identity}$.

\paragraph{Example}
\[\mathsf{E}_3 = 1\,2\,1\,2\,1. \ \
\mathsf{E}_4\,=\, \mathsf{E}_3\,\mathbf{3}\, h_3(\mathsf{E}_3)\, \mathbf{3}\,
 h_3^2(\mathsf{E}_3)\, \mathbf{3}\,  h_3^3(\mathsf{E}_3)\]
 \[=\;1\,2\,1\,2\,1\,\mathbf{3}\,2\,1\,2\,1\,2\,\mathbf{3}\,1\,2\,1\,2\,1\,\mathbf{3}\,2\,1\,2\,1\,2.\]
\[
 \mathsf{E}_5\;=\; \mathsf{E}_4 \  \mathbf{4}\  h_4(\mathsf{E}_4)
 \ \mathbf{4}\  h_4^2(\mathsf{E}_4)\  \mathbf{4} \   h_4^3(\mathsf{E}_4) \  \mathbf{4} \ h_4^4(\mathsf{E}_4) \]
\[
=\;1\,2\,1\,2\,1\,\mathbf{3}\,2\,1\,2\,1\,2\,\mathbf{3}\,1\,2\,1\,2\,1\,\mathbf{3}\,2\,1\,2\,1\,2\, \mathbf{4}\, 3\,1\,3\,1\,3\,\mathbf{2}\,
1\,3\,1\,3\,1\,\mathbf{2}.... \]
\noindent
 The morphism $h_n$ involves  
only letters $1,2,\dd n-1$.
We have:

\smallskip
  $h_2 = [1],\  h_3 = [2, 1], $\
  $h_4 = [3, 1, 2], $
  \
  $h_5 = [4, 2, 3, 1,], $
  
\smallskip
  $h_6 = [5, 1, 2, 3, 4],$
  \
  $h_7 = [6, 4, 5, 1, 2, 3],$\ $h_8 = [7, 3, 1, 2, 6, 4, 5].$

\smallskip
 $ h_9 = [8, 5, 1, 7, 3, 4, 2, 6, 9...]$.
 
\paragraph{\bf Steinhaus-Trotter-Johnson algorithm}
In this algorithm we generate recursively all $(n-1)$-permutations of
$0\,1\,2\dd n-2$, 
then for each of them we insert the element $n-1$ in all possible places,
traversing the $(n-1)$-permutation  alternately right-to-left or left-to-right.

Assume each permutation is a sequence $x[0],x[1]\dd x[n-1]$.
The alphabet of basic actions is $\{0,1,2,\dd n-2\}$.
In this case the $i$-th action is "exchange $x[i]$ with $x[i+1]$".
Denote by $\mathsf{S}_n$ the corresponding sequence of basic
operations generating $n$-permutations.
We have $\mathsf{S}_2=0$.

If $n>2$ and $\mathsf{S}_{n-1}=a_1a_2\dd a_N$   then the sequence $\mathsf{S}_n$ of actions
is a word of length $n!-1$  defined as
\begin{equation}
    \fbox{$\mathsf{S}_n\,=\, w_n^R\, \mathbf{b_1}\, w_n\, \mathbf{b_2}\, w_n^R\, 
 \mathbf{b_3}\, w_n\, \mathbf{b_4}\,\dd\,  \mathbf{b_{N-1}}\, w_n^R\,
\mathbf{b_N}\,w_n.$}
\end{equation}
where 

${b_1b_2b_3 \dd b_N}\,=\,  (a_1+1)\, a_2\, (a_3+1)\, a_4\,
(a_5+1)\,a_6\,\dd (a_N+1), 
w_n\, =\,0,1,2,\dd (n-2).$

\paragraph{Example}
 \[\mathsf{S}_2\,=\, 0,\ \ \mathsf{S}_3\,=\, (10)\,\mathbf{1}\, (01)\,=\,  1\,0\,1\,0\,1,\]
\[
\mathsf{S}_4\,=\, (210)\,  \mathbf{2}\, (012)\,  \mathbf{0}\, (210)\,  \mathbf{2}\, (012)\,  \mathbf{0}\, (210)\,  \mathbf{2}\, (012)\]
\[\ =\ 210\,  \mathbf{2}\, 012\,  \mathbf{0}\, 210\,  \mathbf{2}\, 012\,  \mathbf{0}\, 210\,  \mathbf{2}\, 012\,=\, (21020120)^2\, 2102012.\]

\end{NOTES}
\end{EXERCISE}

\begin{EXERCISE}[Algorithm for 2-Anticovers]\label{pb136}
A 2-anticover of a word $x$ is a set of pairwise distinct factors of $x$ of
length 2 that cover the whole word. The notion is dual of the notion of a cover,
for which a unique factor (or a finite number of them) covers the whole word.
The duality is similar to that of powers and antipowers, where the word is a concatenation of the same factor or of distinct factors.
Instead, for anticovers or covers the occurrences factors can overlap or just be adjacent.

\paragraph{Example}
The set \{\texttt{ab},\texttt{aa},\texttt{ac},\texttt{ba},\texttt{cc},\texttt{ca}\} is a  2-anticover of the word $\texttt{abaacbacca}$.

\smallskip\noindent
{\begin{picture}(110,20)(0,0)
\put( 0,11){\makebox(10,10)[b]{\sa a}}
\put(10,11){\makebox(10,10)[b]{\sa b}}
\put(20,11){\makebox(10,10)[b]{\sa a}}
\put(30,11){\makebox(10,10)[b]{\sa a}}
\put(40,11){\makebox(10,10)[b]{\sa c}}
\put(50,11){\makebox(10,10)[b]{\sa b}}
\put(60,11){\makebox(10,10)[b]{\sa a}}
\put(70,11){\makebox(10,10)[b]{\sa c}}
\put(80,11){\makebox(10,10)[b]{\sa c}}
\put(90,11){\makebox(10,10)[b]{\sa a}}
\put( 1,6){\line(0,1){4}}\put( 1,6){\line(1,0){18}}\put(19,6){\line(0,1){4}}
\put(21,6){\line(0,1){4}}\put(21,6){\line(1,0){18}}\put(39,6){\line(0,1){4}}
\put(31,0){\line(0,1){4}}\put(31,0){\line(1,0){18}}\put(49,0){\line(0,1){4}}
\put(51,6){\line(0,1){4}}\put(51,6){\line(1,0){18}}\put(69,6){\line(0,1){4}}
\put(71,0){\line(0,1){4}}\put(71,0){\line(1,0){18}}\put(89,0){\line(0,1){4}}
\put(81,6){\line(0,1){4}}\put(81,6){\line(1,0){18}}\put(99,6){\line(0,1){4}}
\end{picture}}

\smallskip\noindent
Note the word $\texttt{abaababbaab}$ has no 2-anticover because $\texttt{ab}$ is both a prefix and a suffix of it.

\medskip
The notion generalises obviously to $k$-anticover and, for example, the word $\texttt{abaababbaa}$ admits the 3-anticover $\{\texttt{aba},\texttt{aab},\texttt{bab},\texttt{baa}\}$: 

\smallskip\noindent
{\begin{picture}(110,20)(0,0)
\put( 0,11){\makebox(10,10)[b]{\sa a}}
\put(10,11){\makebox(10,10)[b]{\sa b}}
\put(20,11){\makebox(10,10)[b]{\sa a}}
\put(30,11){\makebox(10,10)[b]{\sa a}}
\put(40,11){\makebox(10,10)[b]{\sa b}}
\put(50,11){\makebox(10,10)[b]{\sa a}}
\put(60,11){\makebox(10,10)[b]{\sa b}}
\put(70,11){\makebox(10,10)[b]{\sa b}}
\put(80,11){\makebox(10,10)[b]{\sa a}}
\put(90,11){\makebox(10,10)[b]{\sa a}}
\put( 1,6){\line(0,1){4}}\put( 1,6){\line(1,0){28}}\put(29,6){\line(0,1){4}}
\put(21,0){\line(0,1){4}}\put(21,0){\line(1,0){28}}\put(49,0){\line(0,1){4}}
\put(41,6){\line(0,1){4}}\put(41,6){\line(1,0){28}}\put(69,6){\line(0,1){4}}
\put(71,6){\line(0,1){4}}\put(71,6){\line(1,0){28}}\put(99,6){\line(0,1){4}}
\end{picture}}

\medskip
On an alphabet of size $\sigma$, since the number of words of length $k$ is $\sigma^k$, no word of length larger than $k\sigma^k$ admits a $k$-anticover. This is why it is appropriate to consider an integer alphabet that is potentially infinite.

\begin{QUESTION}
Design a linear-time algorithm testing if the word $x$ admits a 2-anticover,
assume its alphabet is sortable in linear time (integer alphabet).
\end{QUESTION}

\AIDE{
Use a linear-time  algorithm for the satisfiability of 2CNF formulas (CNF = conjunctive normal form).
Each 2CNF formula  is a conjunction of two-variable ``clauses'' (alternatives
of variables and their negations).
An example of a 2CNF formula is: $(v_2\lor v_4) \land (v_1 \lor \lnot v_3)$}

\begin{SOLUTION} As  clauses we use also formulas of the type $(a\rightarrow  b)$
because it is equivalent to $(\lnot a \lor b)$.

For a set of Boolean variables $V=\{v_1,v_2,\dots,v_m\}$, we 
introduce the predicate $$\Delta(V)\equiv |\{i\mid v_i=1\,\}|\le 1$$ 
and use the following fact.

\paragraph{Fact 1} The predicate $\Delta(V)$ can be written as an equivalent
2CNF formula of size $O(m)$ for $V=\{v_1,v_2,\dots,v_m\}$.

\begin{PREUVE}
We introduce variables $\alpha_i$ and $\beta_i$ that are to be interpreted as
$$\alpha_i \equiv (\forall t\le i\;\; v_i=\False) \mbox{ and } \beta_i \equiv (\forall t\geq i\;\; v_i=\False).$$
Then, $\Delta$ can be written as the following conjunction of implications:\\[1mm]
$\forall\, i<m\;\; (v_i \rightarrow \beta_{i+1}) \land (\beta_i \rightarrow \beta_{i+1})$ \\[1mm]
$\land\ \forall\, i>1\;\; (v_i \rightarrow \alpha_{i-1}) \land (\alpha_i \rightarrow \alpha_{i-1})$ \\[1mm]
$\land\ 
 \forall\, 1\le i\le m\;\; (\alpha_i \rightarrow \lnot v_i) \land (\beta_i \rightarrow \lnot v_i).$ \\[1mm]
Consequently, $\Delta(v_1,\dots, v_m)$ is equivalent to a conjunction of $O(m)$ implications.
\end{PREUVE}

\paragraph{Construction of a 2-anticover}
Let $x=a_1a_2\cdots a_n$ ($a_i$ letters) and $\Fact_2(x)$ be the set of factors of length 2 of $x$.
Let $\mathit{Occ}(v,x)$ denote the set of starting positions of occurrences of $v$ in $x$.

We consider the Boolean variables $x_i$ whose value is \True\ 
iff $a_ia_{i+1}$ is an element of our anti-cover.
Now the problem reduces to the satisfiability of the 2CNF formula:\\[1mm]
$\forall\;1<i<n\;\; (x_i\ \lor\ x_{i-1}) \land (x_1  \land x_{n-1})$\\[1mm]
$\land \ \forall\;v\in \Fact_2(w)\;\; \Delta(\{x_i \mid i\in Occ(v,x)\}.$\\[1mm]
Then, for each $i$, $1<i<n$ and $x_i=\True$, we choose $a_ia_{i+1}$ as an element of the 2-anticover. Otherwise we choose $a_{i-1}a_i$. 
We have also to take $a_1a_2$ and $a_{n-1}a_n$ (hence $x_1 \land x_{n-1}$).%

The second part of the formula says that each factor of length 2 is chosen at most once as a
fragment in the 2-anticover.

To conclude, the word admits a 2-anticover if and only if the formula is true, which answers the question.
\end{SOLUTION}

\begin{NOTES}
We briefly sketch a linear-time algorithm testing 2CNF satisfiability. Let $V$ be the set of variables and their negations. We change the 
problem to a set of implications of type $A\rightarrow B$, where $A,B\in V$.
Each implication can be viewed as a directed edge in the graph $G$ whose $V$
is the set of nodes.
Then the formula is not satisfiable  if and only if, for some variable $v$, both $v$ and $\lnot v$
are in the same strongly connected component. The strongly connected components of a graph can 
be computed in linear time.

The $k$-anticover was introduced in \cite{AlzamelCDGIKW20}, where the above result is proved and it is shown that the 3-anticover problem is {\sf NP}-complete.

Some algorithms related to covers are the subject of Problems 20 and 45 in \cite{CLR21cup}. Problem 90 deals with antipowers, see also \cite{BadkobehFP18}.
\end{NOTES}
\end{EXERCISE}\begin{EXERCISE}[Short Supersequence of Shapes of Permutations ]\label{pb137}
\newcommand{\group}{\mathit{group}}%
\newcommand{\jump}{\mathit{jump}}%
\newcommand{\Jumps}{\mathsf{Jumps}}%
\newcommand{\BothEven}{\mathsf{BothEven}}%
\newcommand{\BothOdd}{\mathsf{BothOdd}}%
\newcommand{\Dif}{\mathsf{Dif}}%
\newcommand{\Asc}{\mathbf{A}}%
\newcommand{\Desc}{\mathbf{D}}%
\newcommand{\even}{\mathit{even}}%
\newcommand{\odd}{\mathit{odd}}%
An $n$-permutation is a length-$n$ sequence (or word) of $n$ 
distinct elements from $\{\sa{1},\sa{2},\dots, n\}$.
The aim of the problem is to build a short word ${\mathbf S}_n$,
called a \MC{superpattern} (supersequence of shapes), such that each $n$-permutation is order-equivalent to a subsequence of ${\mathbf S}_n$. The question is similar to finding a short supersequence but the order-preserving feature reduces drastically the length of the searched word. Indeed, the superpattern defined below has length $|{\mathbf S}_n|=(n^2+n)/2$, which is almost half the length $n^2-2n+4$ of the supersequence constructed in \cite[Problem 15]{CLR21cup}.

\smallskip
The word ${\mathbf S}_n$ is drawn from the alphabet $\{\sa{1},\sa{2},\dots,n+1\}$ as follows. Let $\alpha_n$ be the increasing sequence of all odd letters and $\beta_n$ be the decreasing sequence of all even letters of the alphabet ($\alpha_n$ is an ``ascending group'' and $\beta_n$ is a 
``descending group''). Alternation between ascending and descending groups is the main trick of the solution. Then, define,

\medskip
${\mathbf S}_n=
\begin{cases}
(\alpha_n\,\beta_n)^{n/2} & \mbox{if } n \mbox{ is even},\cr
(\alpha_n\,\beta_n)^{\lfloor n/2 \rfloor}\,\alpha_n &  \mbox{otherwise}.
\end{cases}$

\paragraph{Example} With $n=8$,
$\alpha_8 = \sa{1\,3\,5\,7\,9}$, $\beta_8 = \sa{8\,6\,4\,2}$ and\\[1mm]
${\mathbf S}_8= {\sa{1\,3\,5\,7\,9\ 8\,6\,4\,2\ 1\,3\,5\,7\,9\ 8\,6\,4\,2\ 1\,3\,5\,7\,9\ 8\,6\,4\,2\ 1\,3\,5\,7\,9\ 8\,6\,4\,2}.}$\\[1mm]
With $n=7$,
$\alpha_7 = \sa{1\,3\,5\,7}$, $\beta_7 = \sa{8\,6\,4\,2}$ and\\[1mm]
${\mathbf S}_7\,= {\sa{1\,3\,5\,7\ 8\,6\,4\,2\ 1\,3\,5\,7\ 8\,6\,4\,2\ 1\,3\,5\,7\ 8\,6\,4\,2\ 1\,3\,5\,7}.}$

\medskip
For a permutation $\pi=(\pi_1,\pi_2,\dots,\pi_{n})$ of $\{\sa{1},\sa{2},\dots, n\}$, 
let $\pi^+$ denote $(\pi_1+1,\pi_2+1,\dots,\pi_{n}+1)$,
permutation of $\{\sa{2},\sa{3},\dots, n+1\}$.

An embedding of $\pi$ in a word ${\mathbf S}$ is an increasing sequence of positions $(p_1,p_2,\dots,p_n)$ on ${\mathbf S}$ that satisfies
$\pi = {\mathbf S}[p_1]\,{\mathbf S}[p_2] \cdots {\mathbf S}[p_n]$.

\begin{QUESTION}
Show how to compute in linear time an order-preserving embedding of a given $n$-permutation $\pi$ into ${\mathbf S}_n$.
\end{QUESTION}

\AIDE{Show that $\pi$ or $\pi^+$ is a (standard) subsequence of ${\mathbf S}_n$
and can be found by a 
greedy algorithm. Note that $\pi^+$  is order equivalent to $\pi$}

\begin{SOLUTION}
Following the hint, we show that $\pi$ or $\pi^+$ is a subsequence of ${\mathbf S}_n$. 
To do it, we proceed indirectly as follows. We show that $\pi$ and $\pi^+$ are 
subsequences of prefixes of lengths $m_1$ and $m_2$,
respectively, of an infinite word ${\mathbf S}$,  where $m_1+m_2\leq 2|{\mathbf S}_n|$.
Then, since ${\mathbf S}_n$ is a prefix of ${\mathbf S}$, one of $\pi$ or $\pi^+$ is a subsequence of ${\mathbf S}_n$.

Let ${\mathbf S}=(\alpha_n\,\beta_n)^{\infty}$. 
The positions on ${\mathbf S}$ are numbered from 1 and are partitioned into
consecutive disjoint intervals of alternative sizes $|\alpha_n|$ and $|\beta_n|$, called a groups.
${\mathbf S}_n$ is the  prefix of ${\mathbf S}$ whose indices consist of $n$ groups.

Let $\group(j)$ be the number of the group containing $j$ if $j>0$, and set $\group(0)=0$.


\medskip
From an $n$-permutation $\pi$, Algorithm \Algo{Greedy} computes in a greedy manner an embedding $(p_1,p_2,\dots,p_n)$ of $\pi$ in ${\mathbf S}$.

\medskip
\begin{algo}{Greedy}{\pi=(\pi_1,\pi_2,\dots,\pi_{n}),\ \mbox{length-$n$ permutation}}
  \SET{p_0}{0}
  \DOFORI{i}{1}{n}
    \SET{p_i}{\min\{j>p_{i-1}\mid \pi_i={\mathbf S}[j]\}}
    \SET{\jump_{\pi}(i-1)}{\group(p_i)-\group(p_{i-1})} \label{alg137l4}
  \OD
\RETURN{(p_1,p_2,\dots,p_n)}
\end{algo}

\paragraph{Observation} Variable $\jump$ and the instruction at line~\ref{alg137l4} Note that $\jump_{\pi}(i-1) \in \{0,1,2\}$ (see figure).

\begin{figure}
\centering

\begin{tikzpicture}[scale=0.635]
\foreach \a in {0.5,2.6,4.7,6.8,8.9,11,13.1,15.2}{
\draw[shift={(\a,0)}] (0,0) rectangle (2,3.5);
}
\foreach \a in {0.5,4.7,8.9,13.1}{
\draw[shift={(\a,0)}] (.25,.25) node {\sa{1}};
\draw[shift={(\a,0)}] (.625,1) node {\sa{3}};
\draw[shift={(\a,0)}] (1,1.75) node {\sa{5}};
\draw[shift={(\a,0)}] (1.375,2.5) node {\sa{7}};
\draw[shift={(\a,0)}] (1.75,3.25) node {\sa{9}};
}
\foreach \a in {2.6,6.8,11,15.2}{
\draw[shift={(\a,0)}] (1.75,.25) node {\sa{2}};
\draw[shift={(\a,0)}] (1.375,1) node {\sa{4}};
\draw[shift={(\a,0)}] (1,1.75) node {\sa{6}};
\draw[shift={(\a,0)}] (.625,2.5) node {\sa{8}};
}
\draw node (q0) at (0.10,0.40) {};
\draw node (q1) at (1.5,1.75) {\textbf{\sa{5}}};
\draw node (q2) at (4.95,.25) {\textbf{\sa{1}}};
\draw node (q3) at (7.425,2.5) {\textbf{\sa{8}}};
\draw node (q3p) at (7.675,2.5) {};
\draw node (q4) at (8.175,1) {\textbf{\sa{4}}};
\draw node (q4p) at (8.425,1) {};
\draw node (q5) at (9.525,1) {\textbf{\sa{3}}};
\draw node (q5p) at (9.275,1) {};
\draw node (q6) at (10.275,2.5) {\textbf{\sa{7}}};
\draw node (q6p) at (10.025,2.5) {};
\draw node (q7) at (12.75,.25) {\textbf{\sa{2}}};
\draw node (q8) at (16.2,1.75) {\textbf{\sa{6}}};
\draw[dashed] (q0) -- (q1) -- (q2) -- (q3);
\draw[dashed] (q3p) -- (q4p);
\draw[dashed] (q4) -- (q5);
\draw[dashed] (q5p) -- (q6p);
\draw[dashed] (q6) -- (q7) -- (q8);
\end{tikzpicture}

\caption{The {route} showing how Algorithm \Algo{Greedy} processes the permutation $\pi=(\sa{5,1,8,4,3,7,2,6})$,
by successive jumps to the first next appropriate group.  The sequence of jumps is
$1,2,1,0,1,0,1,2$, for a total $\Jumps(\pi)=8$. The output is $(p_1,p_2,\dots, p_8)=(3,10,15,20,22,27,34)$
 } \label{fig-137}
\end{figure}
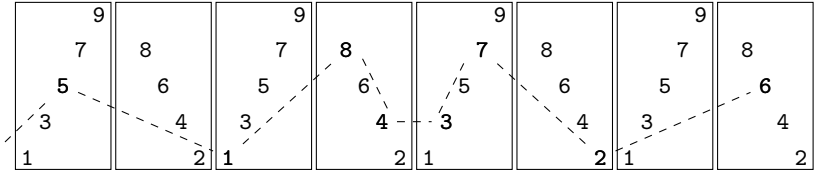

\smallskip
The algorithm is said to be {\it successful} if $p_n\leq |{\mathbf S}_n|$, which means that $\pi = {\mathbf S}_n[p_1 \dd p_n]$ because ${\mathbf S}_n$ is a prefix of ${\mathbf S}$. 

\paragraph{Example (followed)} For $n=8$ and $\pi=(\sa{2,4,6,8,4,3,2,1})$ the algorithm is unsuccessful, but is not for $\pi^+=(\sa{3,5,7,9,5,4,3,2})$ since it returns 
$(2,3,4,5,12,17,20,27)$ and $27 \leq |{\mathbf S}_n| = 36$.

\smallskip
The property has an equivalent formulation in terms of jumps.
Let $\Jumps(\pi)$ be the sum of jumps: $\jump_{\pi}(0)+\jump_{\pi}(1)+\cdots +\jump_{\pi}(n-1)$. Then, we get the following fact to characterise a success.

\paragraph{Fact 1} 
$p_n \leq |\mathbf{S}_n|\Leftrightarrow \Jumps(\pi)\le n.$


\smallskip\noindent
Since $\pi$ and $\pi^+$ are obviously order-equivalent, it is enough to prove that Algorithm \Algo{Greedy} is successful for at least one of $\pi$ and $\pi^+$. This amounts to show that $\pi$ or $\pi^+$ is a (standard) subsequence of ${\mathbf S}_n$ as computed by \Algo{Greedy}. 

\paragraph{Example} Let $n=8$. For $\pi=(\sa{1,2,3,4,5,6,7,8})$, 
$\Jumps(\pi)=8$ and $\Jumps(\pi^+)=9$.
For $\pi=(\sa{2,4,6,8,7,5,3,1})$, $\Jumps(\pi)=15$ and $\Jumps(\pi^+)=2$.
In both cases $\Jumps(\pi)+\Jumps(\pi^+)=2n+1$, which is not accidental, and is a key point to  correctness.

\begin{algo}{Embedding}{\pi=(\pi_1,\pi_2,\dots,\pi_{n}) \mbox{ $n$-permutation}}
  \SET{(p_1,p_2,\dots,p_n)}{\Call{Greedy}{\pi}}
  \IF{p_n > |{\mathbf S}_n|}
    \RETURN{\Call{Greedy}{\pi^+}}
  \FI
  \RETURN{(p_1,p_2,\dots,p_n)}
\end{algo}

The proof of correctness of Algorithm \Algo{Embedding} reduces to the following statement whose proof is after the observation.

\paragraph{Fact 2}
 $\Jumps(\pi)+\Jumps(\pi^+)=2n+1$.
Hence, for an $n$-permutation $\pi$,
either $\Jumps(\pi)\le n$ or $\Jumps(\pi^+)\le n$ and Algorithm \Algo{Greedy} is successful for $\pi$ or for $\pi^+$.

\paragraph{Observation}
When $\pi_i$ and $\pi_{i+1}$ are both even, $p_i,p_{i+1}$ 
belong to  descending groups. In this case, 
if $\pi_i>\pi_{i+1}$ then $\jump(i)=0$ else $\jump(i)=2$. 
Symmetrically, when they are both odd, they belong to the same ascending group and if $\pi_i>\pi_{i+1}$ then $\jump(i)=2$ else $\jump(i)=0$.
When $\pi_i$ and $\pi_{i+1}$ are of distinct parities, $\jump(i)=1$.

\begin{PREUVE}
Let $\BothEven$, $\BothOdd$, $\Dif$ be the set of $i<n$ for which respectively
both $\pi_i,\pi_{i+1}$ are even, both are odd and they are 
of different parities.
We introduce the sets:\\
$\Asc_\even=\{0<i<n\;:\; \pi_i<\pi_{i+1} \mbox{ and } i\in \BothEven\}$,\\[.2mm]
$\Desc_\even=\{0<i<n\;:\; \pi_i>\pi_{i+1} \mbox{ and } i\in \BothEven\}$,\\[.2mm]
$\Asc_{\odd}=\{0<i<n\;:\; \pi_i<\pi_{i+1} \mbox{ and } i\in \BothOdd\}$,\\[.2mm]
$\Desc_{\odd}=\{0<i<n\;:\; \pi_i>\pi_{i+1} \mbox{ and } i\in \BothOdd\}$.

\medskip\noindent 
For $0<i<n$, we have

\smallskip
$(\jump_{\pi}(i),\, \jump_{\pi^+}(i)) =
\begin{cases}
  (0,2) & \mbox{if } i\in \Asc_{\odd}\cup \Desc_{\even},\cr
  (2,0) & \mbox{if } i\in \Asc_{\even}\cup \Desc_{\odd},\cr
  (1,1) & \mbox{if } i\in \Dif.
\end{cases}$

%
%

\medskip\noindent 
Hence, for $0<i<n$, 
$\jump_{\pi}(i)+\jump_{\pi^+}(i)=2.$
This, together with equation $\jump_{\pi}(0)+\jump_{\pi^+}(0)=3$, implies 
\[\Jumps(\pi)+\Jumps(\pi^+)= 2(n-1)+3=2n+1,\]
which completes the proof.
\end{PREUVE}
\end{SOLUTION}

\begin{NOTES}
The present construction is adapted from the version in \cite{Miller09}. 
If the conjecture that the shortest superpattern has length $\frac{1}{2}n^2(1+o(n))$ held, this would imply that our construction is asymptotically optimal.



\end{NOTES}
\end{EXERCISE}

\begin{EXERCISE}[Shrinking a text by pairing adjacent symbols]\label{pb129}
\newcommand{\Comp}{\mathit{Compress}}%
One of the most powerfull compression techniques is {recompression}. In this
technique there are two crucial operations: shrinking unary runs and {\it
pairing letters}.

A unary run is a maximal occurrence of a factor of length at least 2 that is
a repetition of the same letter. The first phase of the recompression
technique consists in shrinking each unary run into a single letter.

The second phase is to apply the operation $\Comp(x,L,R)$, where $(L,R)$ is a
partition of the alphabet $\mathbf{A}$ of letters, $L\cup R=\mathbf{A}$ and
$L\cap R=\emptyset$. The compressed word $\Comp(x,L,R)$ results from $x$ by
substituting a single letter (identifier of a pair of letters) for each
occurrence of its 2-letter factors $ab$, whenever $a\in L$ and $b\in R$.

\smallskip
The \MC{Pairing problem} consists in computing a partition $(L,R)$ of the
alphabet of $x$ for which $|\Comp(x,L,R)|\le \frac{3}{4}|x|$.

\paragraph{Example}
Consider the word $\sa{abcacbabcbac}$. 
Let $L=\{\sa{a},\sa{c}\}$, $R=\{\sa{b}\}$. Then, substituting $\sa{d}$ for
$\sa{ab}$ and $\sa{e}$ for $\sa{cb}$ produces the word $\sa{dcaedeac}$ of
length $8<\frac{3}{4}12=9$.
On the contrary, setting $L=\{\sa{a}\}$, $R=\{\sa{b}\}$ and substituting
$\sa{d}$ for $\sa{ab}$ in the word $\sa{aaabbb}$ containing two unary runs
produces the word $\sa{aadbb}$ of length $5>\frac{3}{4}6=4.5$, which does not meet the above bound.

\begin{QUESTION}
Let $x$, $|x|\geq 2$, be a word over an integer alphabet containing no unary
runs. Show how to compute in linear time a partition $(L,R)$ of the alphabet
of $x$ for which $|\Comp(x,L,R)|\le \frac{3}{4}|x|$.
\end{QUESTION}

\begin{SOLUTION}
A solution to the pairing problem reduces to the following question.

\paragraph{\bf $\frac{1}{4}$-cut problem} Let $G=(V,{E})$ be a
directed multigraph without self-loops; the goal is to compute a partition
$(L,R)$ of the set $V$ of vertices for which at least $\frac{1}{4}|{E}|$ arcs
lead from $L$ to $R$.

\begin{LEMME}
The $\frac{1}{4}$-cut problem can be solved in time $O(|V|+|{E}|)$.
\end{LEMME}

\begin{PREUVE}
For $A,B\subseteq V$, let ${E}(A,B)$ be the set of arcs leading from $A$ to
$B$ and let $\deg(A,B) = |{E}(A,B)|$.

We use the following algorithm.

\medskip
\begin{algo}{Partition}{G=(V,E) \mbox{ a directed multigraph}}
  \SET{(M,L,R)}{(V,\emptyset,\emptyset)}
  \DOWHILE{M \mbox{ not empty}}
    \ACT"{Let $v\in M$}
    \IF{2\deg(v,R)+\deg(v,M) \geq 2\deg(L,v)+\deg(M,v)}
      \SET{L}{L\cup\{v\}}
    \ELSE
      \SET{R}{R\cup\{v\}}
    \FI
    \SET{M}{M\setminus\{v\}}
  \OD
\RETURN{(L,R)}
\end{algo}

\medskip
To show the result we consider the {potential} expression
$${\mathbf P}=4\deg(L,R) + 2\deg(L,M)+2\deg(M,R)+\deg(M,M)$$
and prove that its value cannot decrease throughout a run of the algorithm.
Let us denote
$$a=\deg(v,R),\; b=\deg(v,M),\; c=\deg(L,v),\; d=\deg(M,v)$$
and consider the effect of moving $v$ from $M$ to $L$ on the four terms
of~${\mathbf P}$:
\begin{itemize}
  \item $\deg(L,R)$ increases by $a$;
  \item $\deg(L,M)$ increases by $b$ and decreases by $c$;
  \item $\deg(M,R)$ decreases by $a$;
  \item $\deg(M,M)$ decreases by $b$ and decreases by $d$.
\end{itemize}
Overall, ${\mathbf P}$ increases by
  $4a + 2(b - c) + -2a - (b + d)  = 2a + b - 2c - d$.
Then, this quantity is non-negative when the algorithm decides to move $v$ to
$L$.

Similarly, if $v$ is moved from $M$ to $R$, ${\mathbf P}$ does not decrease
using a similar argument.

Upon the end of the algorithm, we have ${\mathbf P}=4\deg(L,R)$ due to
$M=\emptyset$, while initially ${\mathbf P} = \deg(M,M)=|{E}|$ due to $M=V$.
Since ${\mathbf P}$ is nondecreasing, we conclude that $4\deg(L,R)\geq
|{E}|$, which proves that $\deg(L,R)\geq \frac{1}{4}|{E}|$ as claimed.

\paragraph{\bf Running time}
To meet the expected running time the graph $G$ is first preprocessed in
linear time to compute the input and output degrees of vertices $v\in V$.
Then, each iteration of the algorithm can be implemented in time
$O(1+\deg(v,V)+\deg(V,v))$, which yields a total running time of
$O(|V|+|{E}|)$ as claimed.
\end{PREUVE}

\paragraph{\bf Reduction of the pairing problem to $\frac{1}{4}$-cut problem}
Let $x$ be a word of length at least $2$ with no unary runs. Let $V=\alph(x)$
be the set of letters occurring in $x$ and let $E$ be the set of edges
$a\rightarrow b$, where $ab$ is a factor of $x$ and $a\ne b$. The number of
edges from $a$ to $b$, that is, the output degree of $a$, is the number of
occurrences of $ab$ in $x$. Now the pairing problem reduces to the
$\frac{1}{4}$-cut in this graph and is solved by Algorithm \Algo{Partition}
that computes a desired partition.

\paragraph{Example}
Consider again the word $\sa{abcacbabcbac}$ of length $12$.
\begin{wrapfigure}{r}{0.27 \textwidth}
\begin{tikzpicture}[scale=.7, node distance=1cm, auto, >=latex']
\tikzstyle{grey} = [circle,draw,thin,fill=lightgray,minimum size=1pt]
\tikzstyle{white} = [circle,draw,thin,minimum size=1pt]
\path[->] node[grey] (q0) at (0,2) {\sa{a}};
\path[->] node[grey] (q1) at (3,2) {\sa{b}}
    (q0) edge [out=25,in=155] node {\small{2}} (q1)
    (q1) edge [out=190,in=-10,above=0mm] node {\small{2}} (q0);
\path[->] node[grey] (q2) at (1.5,0) {\sa{c}}
    (q2) edge [out=160,in=-90,left=0mm] node {\small{1}} (q0)
    (q0) edge [out=-40,in=120,left=0mm] node {\small{2}} (q2)
    (q1) edge [out=-90,in=20,right=0mm] node {\small{2}} (q2)
    (q2) edge [out=60,in=-140,right=0mm] node {\small{2}} (q1);
\end{tikzpicture}
\end{wrapfigure}
Let us run \Algo{Partition} on its associated graph (edges are labelled by
the number of occurrences of their corresponding length-2 factor) and start
with $M=\{\sa{a},\sa{b},\sa{c}\}$, $L=\{\}$, $R=\{\}$.
\begin{itemize}
\item[Node $v=\sa{a}$:] $2\times0+4 \geq 2\times0+3$ gives
    $M=\{\sa{b},\sa{c}\}$, $L=\{\sa{a}\}$, $R=\{\}$.

\item[Node $v=\sa{b}$:] $2\times0+2 \not\geq 2\times2+2$ gives
    $M=\{\sa{c}\}$, $L=\{\sa{a}\}$, $R=\{\sa{b}\}$.

\item[Node $v=\sa{c}$:] $2\times2+1 \geq 2\times2+0$ gives $M=\{\}$,
    $L=\{\sa{a},\sa{c}\}$, $R=\{\sa{b}\}$.
\end{itemize}
Then, factors $\sa{ab}$ and $\sa{cb}$ are replaced by new letters as seen
above.
\end{SOLUTION}

\begin{NOTES}
The recompression technique and the pairing problem can be found in
\cite{Jez16}. This technique was successfully applied to many problems, especially to word equations. The newly created letters correspond to fragments of growing sizes. For recompressing a text, the process of 
pairing letters is iterated while receiving new letters.
The whole process of creating new letters results globally in only a linear number of such letters, since meanwhile the size of the word decreases geometrically.

The recompression technique is technically very complicated, and details depend on the 
particular problem it is applied to.
\end{NOTES}
\vfill
\begin{center}
\begin{tikzpicture}[scale=.20, node distance=1mm, auto, >=latex', on grid, thick]
\draw[opacity=0.6,fill=red] (-3.5,3.6) -- (1.7,6) -- (3.3,5.2) -- (-1.8,2.8)
 -- (-1.8,-0.5) -- (-3.5,0.6) -- (-3.5,3.6);
\draw[opacity=0.6,fill=green] (-1.1,-4.1) -- (-6,-0.9) -- (-6,0.9) -- (-1.1,-2.2)
 -- (1.8,-0.8) -- (1.7,-2.7) -- (-1.1,-4.1);
\draw[opacity=0.6,fill=yellow] (2.6,-5.1) -- (2.9,0.8) -- (0,2.5) -- (1.7,3.3)
 -- (4.4,1.7) -- (4.1,-4.2) -- (2.6,-5.1) ;

\draw (-1.8,-0.5) -- (0,0.3);
\draw (1.8,-0.8) -- (0,0.3);
\draw (0,2.5) -- (0,0.3);
\end{tikzpicture}
\end{center}
\vfill \mbox{}

\end{EXERCISE}
\begin{EXERCISE}[Yet another application of Suffix trees]\label{pb139}
\newcommand{\dif}{\mathit{dif}}%
In this problem, we show how the Suffix tree of a word can be used in three different ways to solve an example problem.
For a string $x$, let $\Sub[k]$ denote the number of (distinct) nonempty factors of $x$ having an occurrence whose position starts in the interval $[0\dd k]$. For simplicity,  assume that $x$ ends with a unique symbol.

\begin{QUESTION}
Show how to compute the table $\Sub[0\dd n-1]$ in linear time.
\end{QUESTION}

\AIDE{Use a suffix tree}

\begin{SOLUTION}
To compute the table $\Sub$, it is enough to compute, for each position $k>0$, the number $\dif[k]$ of factors that start at $k$ but not before. 

\smallskip
\begin{algo}{TableSub}{x \mbox{ word of length } n}
\SET{\dif}{\Call{TableDif}{x}}
\DOFORI{k}{0}{n-1}
  \ACT{\Sub[k] \leftarrow \mbox{ if } k=0 \mbox{ then }\dif[k] \mbox{ else }\Sub[k-1]+\dif[k]}
\OD
\RETURN{\Sub}
\end{algo}

Computing the table $\dif$ can be done in various ways.

\smallskip
\begin{tikzpicture}[scale=.4, node distance=1cm, auto, >=latex'] 
\tikzstyle{white} = [circle,draw,thin] 
\path[->] node[white] [fill=lightgray] (q0) at (0,5) {\tiny 0}
    (-1,5) edge node {} (q0);
\draw node[white] [fill=lightgray] (l0) at (12,11) {\tiny 6};
\draw node[white] (l1) at (6,9) {\tiny 3};
\draw node[white] (l2) at (8,7) {\tiny 4};
\draw node[white] (l3) at (2,5) {\tiny 1};
\draw node[white] [fill=lightgray] (l4) at (10,3) {\tiny 5};
\draw node[white] (l5) at (4,1) {\tiny 2};
\draw node[white] [fill=lightgray] (q1) at (2,8) {\tiny 1};
\draw node[white] [fill=lightgray] (q2) at (4,10) {\tiny 2};
\draw node[white] [fill=lightgray] (q3) at (2,2) {\tiny 1};

\path[->] (q0) edge node {\small{\sa{a}}} (q1);
\path[->] (q1) edge node {\small{\sa{b}}} (q2);
\path[->] (q1) edge [out=-20,in=180] node [below] {\small{\sa{ab\$}}} (l2);
\path[->] (q2) edge [out=20,in=180] node {\small{\sa{aab\$}}} (l0);
\path[->] (q2) edge [below] node {\small{\sa{\$}}} (l1);
\path[->] (q0) edge node {\small{\sa{\$}}} (l3);
\path[->] (q0) edge [left] node {\small{\sa{b}}} (q3);
\path[->] (q3) edge [out=20,in=180] node {\small{\sa{aab\$}}} (l4);
\path[->] (q3) edge [below] node {\small{\sa{\$}}} (l5);

\draw node (t1) at (14,12.5) {$k$};
\draw node (t2) at (16,12.4) {$\dif[k]$};
\draw node (t3) at (19.4,12.4) {$\Sub[k]$};
\draw node (pos0) at (14,11) {$0$};
\draw node (dif0) at (16,11) {$6$};
\draw node (sub0) at (19.4,11) {$6$};
\draw node (pos3) at (14,9) {$3$};
\draw node (dif3) at (16,9) {$.$};
\draw node (sub3) at (19.4,9) {$.$};
\draw node (pos2) at (14,7) {$2$};
\draw node (dif2) at (16,7) {\bf ?};
\draw node (sub2) at (19.4,7) {\bf ?};
\draw node (pos5) at (14,5) {$5$};
\draw node (dif5) at (16,5) {$.$};
\draw node (sub5) at (19.4,5) {$.$};
\draw node (pos1) at (14,3) {$1$};
\draw node (dif1) at (16,3) {$5$};
\draw node (sub1) at (19.4,3) {$11=6+5$};
\draw node (pos4) at (14,1) {$4$};
\draw node (dif4) at (16,1) {$.$};
\draw node (sub4) at (19.4,1) {$.$};
\end{tikzpicture}

\medskip
Let $\aSTree(x)$ be the Suffix tree of $x=x[0\dd n-1]$ (see Notes). Recall that a node of $\aSTree(x)$ is (or can be identified with) a factor $u$ of $x$. In the picture, each branching node (explicit node or fork) $u$ displays $|u|$. 
The weight of an edge $u\rightarrow v$ in $\aSTree(x)$ is the absolute difference between the lengths of its end-nodes, that is, $|v|-|u|$. Each leaf is a non-empty suffix $v$ and is labeled by its starting position on $x$, that is, $|x|-|v|$.

\paragraph{Algorithm 1}
The picture illustrates a step in a run of Algorithm \Algo{TableDif1}, just before processing the suffix $\texttt{aab\$}$ of the word $\texttt{abaab\$}$.
All nodes on the two longest branches are marked following the computation of $\dif[0]$ and of $\dif[1]$. Since the parent of leaf 2 is marked, $\dif[2]=|\texttt{ab\$}|=3$ and $\Sub[2]=11+3=14$.

\begin{algo}{TableDif1}{x \mbox{ word of length } n}
\ACT"{unmark all nodes of $\aSTree(x)$}
\DOFORI{k}{0}{n-1}
  \ACT"{from leaf $k$, go bottom-up until meeting a marked node}
  \ACT"{mark all visited nodes}
  \SET{\dif[k]}{\mbox{sum of weights of visited edges}}
\OD
\RETURN{\dif}
\end{algo}

\paragraph{\bf Algorithm 2}
Instead of running through all suffixes (with variable $k$), the algorithm below processes nodes in any order. But to do so, it recovers suffixes with the value $\mathit{min}(v)$, minimum leaf in the subtree rooted at node $v$.
The algorithm is as follows.

\begin{algo}{TableDif2}{x \mbox{ word of length } n}
\DOFORI{k}{0}{n-1}
  \SET{\dif[k]}{0}
\OD
\ACT"{compute bottom-up $\mathit{min}(v)$ for each node $v$ of $\aSTree(x)$}
\DOFOR"{each non-root node $v$}
  \SET{(k,u)}{(\mathit{min}(v),\mbox{parent of }v)}
  \SET{\dif[k]}{\dif[k]+|v|-|u|}
\OD
\RETURN{\dif}
\end{algo}

\paragraph{Algorithm 3}
The table $\dif$ can be computed during the construction of $\aSTree(x)$ by McCreight algorithm, which combined features in algorithms 1 and 2.
Indeed, this algorithm adds exactly one edge at each iteration on suffix $k$. The weight of the edge is then added to $\dif[k]$.

\end{SOLUTION}
\begin{NOTES}
The Suffix tree of a word is described in \cite[Chapter 1]{CLR21cup} and in references cited in its notes.
McCreight algorithm for its construction is given in \cite{CR02jos} and in \cite[Section 5.2]{CHL07cup}.
\end{NOTES}
\end{EXERCISE}
\begin{EXERCISE}[Two longest subsequence problems]\label{pb140}
\newcommand{\MinSub}{\mathsf{MinSub}}%
\newcommand{\LPS}{\mathsf{LPS}}%
\newcommand{\LMPS}{\mathsf{LMPS}}%
\newcommand{\LCS}{\mathsf{LCS}}%
There are many problems related to subsequences with 
specific properties (see the notes).
We consider two simple problems of this type: for a word $x$, compute its lexicographically
smallest subsequence of a given length $k$, 
and a longest palindromic subsequence. 

\paragraph{The $\MinSub$ problem} 
For a word $x$ drawn from an ordered alphabet, $\MinSub(x,k)$ is defined as the lexicographically
smallest subsequence of a given length $k$, $k\leq |x|$. 
For example, $\MinSub(\sa{bbbbbaeeecffddd},5)=\sa{acddd}$.
Note the subsequence may have several occurrences in $x$.

\begin{QUESTION}
For a word $x$ of length $n$, design an algorithm that computes $\MinSub(x,k)$ in time $O(n)$.
\end{QUESTION}


\paragraph{The $\LPS$ problem}
In this second problem, the goal is to compute a longest palindromic subsequence
$\LPS(x)$ of the word $x$.
For example, $\sa{abba}$ and $\sa{dccd}$ are possible answers for $\LPS(\sa{dcabcdba})$.

\begin{QUESTION}
 Compute a longest palindromic subsequence of a word of length $n$ in time $O(n^2)$. 
\end{QUESTION}


\begin{SOLUTION}
The solution the the first problem is known as a folklore due to its simplicity, which is its main interest.
Algorithm \Algo{MinSuf} is a modification of a very simple algorithm computing a
lexicographically minimal subsequence. 

It uses a stack handled with standard operations $top$, $pop$ and $push$.

\begin{algo}{MinSub}{x \mbox{ word of length } n, \mbox{ integer } k \leq n}
  \SET{\mathit{rest}}{|x|-k}
  \SET{S}{\mbox{empty stack}}
  \DOFORLETTERSIN{a}{x}
    \DOWHILE{S \mbox{ non empty and } a<top(S) \mbox{ and } \mathit{rest}>0}
      \ACT{pop(S)}
      \SET{\mathit{rest}}{\mathit{rest}-1}    
    \OD
    \ACT{push(a,S)}
  \OD
  \RETURN{S}
\end{algo}

The variable $\mathit{rest}$ takes care of the required length $k$. 
If there are not enough unread letters, the algorithm stops comparisons and adds to the stack the remaining unread symbols. 
In particular, if $k=|x|$ the algorithm pushes to the stack the whole word $x$. 

Correctness and complexity of the algorithm are straightforward.

\paragraph{Example}
$\MinSub(\sa{baddbccega}, 7)=\sa{abccega}$, because after reading the subsequence $\sa{ab}$ all remaining letters should be appended to get a word of length 7.

\paragraph{Solution to the second question}
To compute an $\LPS(x)$, the naive approach is to take $\LCS(x,x^\mathrm{R})$. However, it does not work. For example, $\sa{abcd}$ (as one of possible answers) is an $\LCS(\sa{dcabcdba},(\sa{dcabcdba})^\mathrm{R})$ but is not a palindrome.

The problem $\LMPS$ (longest mutually palindromic subsequences) refines the above approach.
$\LMPS(w)$ returns two longest subsequences $u$ and $v$  (possibly the same) of 
$w$ that satisfy $u=v^\mathrm{R}$, together with their {locations}.
In other words we look for the longest word $y$ such that $y$ and $y^R$
occur (it could be the same occurrence if $y$ is a palindrome) as subsequences  of a given word $w$.

Formally, $\LMPS(w)$ is a pair $(\alpha,\gamma)$ of longest increasing sequences of positions on $w$
for which $w[\alpha]$ is the reverse of $w[\gamma]$. It does not guarantee that $w[\alpha]$ is a palindrome.

\paragraph{Example} For $w=\sa{dcabcdba}$, 
$[(0,1,6,7), (2,3,4,5)]$ is a possible value of $\LMPS(w)$. We have $w[\alpha]=\sa{dcba}$ and $w[\gamma]=\sa{abcd}$ and each word is the reverse of the other. However none of them is a palindrome, 
nevertheless $w$ has a palindromic subsequence of length $4$, namely $\sa{abba}$.

\paragraph{Reduction of $\LPS$ to $\LMPS$}
Let $(\alpha,\gamma)=\LMPS(x)$ and $u=x[\alpha]$.
Then, it can be derived a palindromic subsequence of length $|u|$ of $x$.

\begin{PREUVE}
We consider only the case of odd $|u|$ since the even case is similar.
Let  $u$ and $v$ be mutually symmetric subsequences of $x$.
Then, for the ``middle'' letter $c$, let $u=zcy$, $v=y^\mathrm{R}cz^\mathrm{R}$, where $|z|=|y|$. Let $p$ be the  position of letter $c$ on $u$ and $q$ its position on $v$. 
If $p\leq q$ then $y^\mathrm{R}$ is to the left of $y$ and we get a palindromic subsequence $y^\mathrm{R}y$. If $p>q$ then we have a palindromic subsequence $xx^\mathrm{R}$.
\end{PREUVE}

\paragraph{Reduction  $\LMPS \Rightarrow\LCS$}
For two words $u$ and $v$ of length $n$, let $\LCS(u,v)=(\alpha,\beta)$,
where $\alpha$ and $\beta$ are longest increasing sequences of positions on $u$ and on $v$ respectively, for which $u[\alpha]=v[\beta]$.
It is well known that this problem can be solved in $O(n^2)$ time.
The, we compute $(\alpha,\beta)=\LCS(w,w^\mathrm{R})$. 
This gives a solution $(\alpha,\gamma)=(\alpha,\beta')$ to $\LMPS(x)$,
where $\beta'$ results from $\beta$ by numbering positions from the end.
\end{SOLUTION}

\begin{NOTES}
There are other algorithmic problems related to subsequences having other specific properties: 
the longest palindromic subsequence, the longest subsequence that is a square or that is a highly periodic subsequence or that is a Lyndon word,
see \cite{Kosowski}, \cite{BannaiIKKP22}, \cite{DBLP:journals/ipl/BannaiIK23} and \cite{InoueIB20}.
\end{NOTES}
\end{EXERCISE}

\begin{EXERCISE}[Two problems on Run-Length Encoded words]\label{pb141}
\newcommand{\RLE}{\mathrm{RLE}}%
\newcommand{\occur}{\mathit{Occ}}
\newcommand{\LIST}{\mathbf{F}}%
\newcommand{\VAL}{\mathbf{L}}%
\newcommand{\maxgap}{\mathit{maxgap}}%
The run-length encoding of a binary word
$x\in\sa{1}\{\sa{0},\sa{1}\}^*$ is 
$$\RLE(x)= \sa{1}^{p_0}\sa{0}^{p_1}\cdots \sa{1}^{p_{s-2}}\sa{0}^{p_{s-1}},$$ 
where
$s-2\geq0$, $p_i>0$ for $i=0,\dots,s-2$ and
$p_{s-1}\geq 0$. The value $|\RLE(x)|$ denotes the size 
of the compressed version of $x$. Note the length $|x|$ can be 
exponential w.r.t. $\RLE(x)$.

A cover of a non-empty word $x$ is one of its factors whose occurrences cover all positions on $x$. We refer to \cite[Problem 45]{CLR21cup} where 
a list-oriented computation of covers in standard strings is presented
using the prefix table of the word (see \cite[Problem 22]{CLR21cup}).
Our algorithm here follows similar lines, except that
now it operates on sparse sets of (well chosen) positions.

\begin{QUESTION}
Let $x$ be a word whose $\RLE(x)$ is of size $n$. 
Show how to compute in linear time $O(n)$ the length of the shortest cover of $x$, assuming the cost of each arithmetic operation is a constant.
\end{QUESTION}

\AIDE{Extend the list-based algorithm for shortest covers in \cite[Problem 45]{CLR21cup} and the {\it sparse} prefix table}

\begin{QUESTION}
Let a pattern $x$ and a text $y$ given in $\RLE$-form of total size $n$.
Show how to check if $x$ occurs in $y$ in $O(n)$ time.
\end{QUESTION}

\begin{SOLUTION}
Denote by $\occur(u,x)$  the sorted list of starting positions of occurrences of 
a word $u$ in $x$.
Assume that $x$ is a non-unary word (otherwise the solution is trivial)
and $\alpha=\sa{1}^k\sa{0}$ is a prefix of $x$, for $k>0$. 

\paragraph{Observation}
Both $|\occur(\alpha,x)|\le n$ and $\occur(\alpha,x)$ can be computed in time $O(n)$.

\medskip
We use the ``sparse'' prefix table $\tpref$ (\cite[Problem 45]{CLR21cup}) defined (only)  for positions in $\occur(\alpha,x)$:
$\tpref(i)$ is the length of the longest prefix of $x$ 
that has an occurrence starting at position $i$.

Let $$\VAL \,=\, \{\ell \mid \tpref(i)=\ell \ 
\mbox{for some}\ i\in \occur(\alpha,x)\}$$
and, for each length $\ell\in \VAL$, $\ell\geq 0$, 
let $$\tpref^{-1}(\ell) = \{i \mid \tpref(i)=\ell\}.$$ 

Assume each set $\occur(\alpha,x)$, $\tpref^{-1}(\ell)$ and $\VAL$ is represented as a linear ascending double-linked list. 
For $\ell\in \VAL$ denote by $prev(\ell)$ its predecessor in $\VAL$.

Then, Algorithm \Algo{ShortestCover1}
computes the length of the shortest cover of its input.
In fact, it is almost the same solution as in \cite[Problem 45]{CLR21cup},
except that only positions in $\occur(\alpha,x)$ are considered.

\begin{algo}{ShortestCover1}{x \mbox{ non-empty word}}
  \ACT{\mbox{compute the sparse prefix table of }x}
  \ACT{\mbox{compute } \VAL \mbox{ and } \occur(\alpha,x)}
  \ACT{\mbox{compute the sets } \tpref^{-1}(\ell) \mbox{ for } \ell \in \VAL}
  \SET{\LIST}{\occur(\alpha,x)} 
  \COM{the sets $\VAL$ and $\LIST$ are represented as increasing lists}
  \DOFOR{\ell\in \VAL}  
    \IF{\ell\ne \min(\VAL)}
      \ACT{\mbox{remove elements of } \tpref^{-1}(prev(\ell)) \mbox{ from } \LIST}
    \FI
    \ACT{\mbox{update } \maxgap(\LIST)}
    \IF{\maxgap(\LIST)\leq\ell}
      \RETURN{\ell}
    \FI
  \OD
  \RETURN{\mbox{null}}
\end{algo}

We explain now how to compute the sparse prefix table $\tpref$.
We can encode each block consisting of a maximal $k$-repetition of 
the same letter $a$ as a single composite letter $(a,k)$.
We discard the first block and the last block.
The resulting word (consisting of encoded blocks)   $x'$ is  of length $n-2$.

\paragraph{Example} Let $x=1^3\, 2^3\,3^3\,1^3\,2^3\,1^2\,3^3\,1^4\,2^2\,3^2\,1^2\,2^2\,1^5\,2^2$.
The set $\mathbf{F}$ consists of positions indicated by arrows in 
the figure below. We get $\alpha=1^32$ and 

\smallskip\noindent
$x'=(2,3)\,(3,3)\, (1,3)\, (2,3)\, (1,2)\, (3,3)\, (1,4)\, (2,2)\, (3,2)\, (1,2)\, (2,2)\, (1,5).$

\vspace{-2mm}
\centerline{\includegraphics[width=11cm]{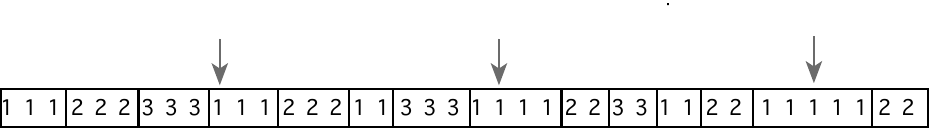}}

\medskip
We compute the additional (full) prefix table $\tpref'$ for $x'$ in $O(n)$ time
using an algorithm for standard strings. Using the table $\tpref'$
it is easy to compute $\tpref(i)$ for each individual position $i\in \occur(\alpha,x)$ in the text $x$ in $O(1)$ time, which is done globally in $O(n)$ time.

\paragraph{Solution to the second question}
Let $z$ be the word $x\#y\#$, where $\#$ does not occur in $xy$. 
Then, the $\RLE$-encoding of $z$ is of size $O(n)$.
After computing the prefix table $\tpref$ of $z$ for the special positions on $z$,
it can be checked $\tpref(i)=|x|$ for each position $i\in \occur(a^kb,x)$ inside $y$, where $a$ and $b$ are letters and $a^kb$ is a prefix of $x$. Which solves the question. 
\end{SOLUTION}


\end{EXERCISE}

\begin{EXERCISE}[Maximal Number of (distinct) Subsequences]\label{pb142}
\newcommand{\subs}{\mathit{subs}}%
For a word $x\in\{\sa{a},\sa{b}\}^*$, let $\SMot(x)$ denote the set of subsequences occurring in $x$, including the empty word, and let $\subs(x)=|\SMot(x)|$.

\begin{QUESTION}
For a word $x\in\{\sa{a},\sa{b}\}^*$, design an efficient (polynomial time) algorithm computing 
$\subs(x)$.
\end{QUESTION}

\begin{QUESTION}
What is a compact formula for the maximal number $S(n)$ of (distinct) subsequences of a binary word of length $n$.
\end{QUESTION}


\begin{SOLUTION}
The present solution uses the subsequence automaton of $x$ (see \cite[Problem 51]{CLR21cup}). Discarding labels of edges, the automaton is a directed acyclic graph with one 
initial node (source). The number of distinct paths from the source can be computed efficiently using the so called topological sorting. This number gives the number of (distinct) subsequences $\subs(x)$.

\paragraph{Solution to the second question}
Let $F_k$ be the $k$-th Fibonacci number ($F_0=0$, $F_1=1$ and $F_n=F_{n-1}+F_{n-2}$ for $n\geq 2$).

\paragraph{Fact 1} $S(n) = F_{n+3}-1$.

\begin{PREUVE}
The proof is by induction. It works for $n=1$ and for $n=2$ considering the word $\sa{ab}$. 
Let $n>2$. For any word $abu$ of length $n$, where $a,b$ are letters, the following inequality
holds true $$\subs(abu)\leq \subs(bu)+\subs(u)+1\leq S(n-1)+S(n-2)+1.$$
Then, using the induction hypothesis, it follows 
$$\subs(abu)\leq (F_{n+2}-1) + (F_{n+1}-1) + 1 = F_{n+3}-1.$$
(The additional unit is for the empty word.)

To conclude, note that if letters $a,b$ are distinct, the first inequality becomes an equality and the rest follows.
\end{PREUVE}



For the word $x=\sa{abab}$, $\subs(x)=12$, that is $F_{4+3}$, since
$\SMot(x)$ is $\{\mv,\sa{a},\sa{ab},\sa{aba},\sa{abab},\sa{aab},\sa{abb},\sa{aa},\sa{b},\sa{ba},\sa{bab},\sa{bb}\}$.
\end{SOLUTION}

\begin{NOTES}
The problem is from \cite{DBLP:journals/combinatorics/FlaxmanHS04}.
\end{NOTES}
\end{EXERCISE}
\begin{EXERCISE}[Avoiding Grasshopper repetitions]\label{pb143}
\newcommand{\fil}{\mathit{fill\_gaps}}%
The problem deals with grasshopper subsequences of words. A grasshopper subsequence of a word $x$ is a word of the form $x[i_1]x[i_2]\cdots x[i_k]$, where $i_t$ is a position on $x$ satisfying $i_{t+1}\in \{i_t+1,i_t+2\}$, for each $t$, $0<t<k$. We can imagine a grasshopper jumping to the right by one or two positions. 


The goal of the problem is related to long words that avoid grasshopper squares and grasshopper cubes over alphabets of size 3 and 6, respectively.

For example, $\sa{abbab}$ contains a grasshopper cube, namely $\sa{bbb}$,
while $\sa{bbaabbaa}$ avoids grasshopper cubes.

Let $A=\{\sa{a},\sa{b},\sa{c}\}$ and $A'=\{\sa{a'},\sa{b'},\sa{c'}\}$ whose elements are called ``primed'' letters.
For a word $v \in A^*$, the word $\Phi(v)$ over the alphabet $A\cup A'$ is defined
using the coding (morphism):
$$\sa{a} \rightarrow \sa{aa'},\; \sa{b} \rightarrow \sa{bb'},\;
\sa{c} \rightarrow \sa{cc'}.$$
For example, $\Phi(\sa{abc})=\sa{aa'bb'cc'}$.

\begin{QUESTION}
Let $x\in A^+$, $y=\Phi(x)$ and $z$ be a grasshopper square in $y$.
Show how to compute in time $O(|z|)$ a (standard) square $v$ in $x$ of length at least $|z|/2$.
\end{QUESTION}

\begin{QUESTION}
For a given integer $n>0$, build a word of length $n$ over a 6-letter alphabet
that avoids grasshopper cubes.
\end{QUESTION}

\begin{SOLUTION}
For a symbol $s\in A\cup A'$, let us define $unprime(s)$ by 
$\sa{a} \rightarrow \sa{a}$, $\sa{a'} \rightarrow \sa{a}$, $\sa{b} \rightarrow \sa{b}$, $\sa{b'} \rightarrow \sa{b}$, $\sa{c} \rightarrow \sa{c}$, $\sa{c'} \rightarrow \sa{c}.$


Algorithm \Algo{RecoverSquare} below constructs a required square $v$ in $x$.
From the grasshopper square $z=z[0\dd k-1]$ in $y=\Phi(x)$
let us denote by $\fil(z)$ the shortest factor of $y$ that contains $z$
and is in $(AA')^+$. In other words $\Phi^{-1}(\fil(z))$
is well defined. In fact, $\fil(z)$ results by filling ``gaps'' created by 
jumps with letters, and eventually expanding the starting and ending part
of $z$; the resulting string should start with a symbol in $A$
and end with a symbol in $A'$. This is what the algorithm computes up to line~\ref{algo-143-line7}
to get the word $v$. 
Additionally, if $v$ is of odd length its last letter is removed to get the output.

\paragraph{Example}
Let $x=\sa{abacba}$. The word $\Phi(x)$ is $\sa{aa'bb'aa'cc'bb'aa'}$ and contains the grasshopper square $z=\sa{a'baa'ba}$. 
\\
Assume we start  the algorithm with $z=\sa{a'baa'ba}$. 
After executing statement in line 6
we get $v=\sa{ababa}$.   The removal of the last symbol of $v$ is necessary and is done in line 9. Ultimately the algorithm produces the square $v=\sa{abab}$ in $x$.

\begin{algo}{RecoverSquare}{x\in A^+, z \mbox{ grasshopper square in }\Phi(x)}
\SET{(k,v,i)}{(|z|,\mv,0)}
\DOWHILE{i<k}
  \IF{z[i]\in A \mbox{ and } z[i+1]\in A'}
    \SET{(s,i)}{(z[i],i+2)}
  \ELSE
    \SET{(s,i)}{(unprime(z[i]),i+1)}
  \FI
\SET{v}{v\cdot s}
\OD
\COM{\ $v=\Phi^{-1}(\fil(z))$} \label{algo-143-line7}
\IF{|v|\ \mbox{is odd}}
\SET{v}{(v \mbox{ without its last symbol})}
\FI
\RETURN{v \mbox{ (square in }x)}
\end{algo}

The correctness of Algorithm \Algo{RecoverSquare} follows from the fact
that odd positions on $y$ point to primed symbols and even
positions to unprimed symbols. Hence, due to limited jumps
(one or two steps), whenever we have a factor $cd'$, for $c\in A$ and $d'\in A'$, we know that $d'=c'$ and the factor becomes $cc'$ that decodes to $c$.

\paragraph{Solution to the second question}
Use any sufficiently long square-free word $w$ over the 3-letter alphabet
$\{\sa{a},\sa{b},\sa{c}\}$ and compute $\Phi(w)$.
The solution to the previous question guarantees that $\Phi(w)$
avoids grasshopper cubes.
\end{SOLUTION}

\begin{NOTES}
For a given integer $n>0$ there is a word of length $n$ over
a 3-letter alphabet that avoids grasshopper cubes.
To see it, let $v$ be a cube-free word over the alphabet $\{\sa{a},\sa{b}\}$.
We create the required word $w$ over the alphabet $\{\sa{a},\sa{b},\sa{c}\}$
by applying to $v$ the coding defined by:
$\sa{a} \rightarrow \sa{c}^2\sa{a},\; \sa{b} \rightarrow \sa{c}^2\sa{b}$.
The correctness can be proved similarly to the above solution for squares,
see \cite{DebskiPW16}.

Our presentation is a version of constructions in \cite{DebskiPW16}.
Computer experiments show that 5 letters are not enough to avoid
grasshopper squares: every word of length 23 over a 5-letter alphabet
contains a grasshopper square.
Besides, two letters are not enough to avoid grasshopper cubes. 
Hence, the numbers 3 and 6 here are minimal.
\end{NOTES}
\end{EXERCISE}
\begin{EXERCISE}[Counting unbordered words and relatives]\label{pb144}
\newcommand{\vv}{\mathbf{v}}%
\newcommand{\un}{\mathbf{u}}%
\newcommand{\tv}{\mathbf{t}}%
We assume, for simplicity, that the alphabet is binary.
A word is called {\it unbordered} if it has no proper border.
For example the word $ababb$  is unbordered, as well as the empty word.
Denote by $\un(n)$
the number of unbordered binary words of length $n$.
We have:
\[\un(0),\un(1),\un(3),\un(4),\ldots = \, 1, 2, 2, 4, 6, 12, 20, 40, 74,\ldots.\]

\paragraph{\bf Observation}
A word $w$ is unbordered if and only if it has no border of length at most
$|w|/2$.

\medskip
There are $2^{n-k}$ 
binary words with border of size $k$, for $k\le n/2$. Due to  the observation we have 
an exponential lower bound on $\un(n)$
\[\un(n)\geq 2^n-2^{n-1}-2^{n-2}-\cdots - 2^{\lceil n/2\rceil}\, \geq 2^{n/2}.\]

Denote by $\vv(n)$ and $\tv(n)$ the number of binary words of length $n$
without nontrivial prefix palindrome of even length and odd length,
respectively.

\begin{QUESTION}
Describe  algorithms computing $\un(n),\, \vv(n)$ and $\tv(n)$ in $O(n)$
time
\end{QUESTION}

\begin{SOLUTION}
We can use recurrences:
\[(*)\ \  
 \un(2n+1)=2 \cdot \un(2n),\ \ \un(2n)=2 \cdot \un(2n - 1) - \un(n)
\]
The first equality follows from the fact that a $(2n+1)$-length word is
unbordered if and only if it is unbordered after removing the ``middle'' letter.
There are two possible letters, hence we have the coefficient 2.

Similarly, an even length word $w=a\,w_1b\,w_2$ of length $2n$,
where $|w_1|=|w_2|=n-1$, is
unbordered if and only if $aw_1w_2$ (of length $2n-1$) is unbordered
and $aw_1,\,bw_2$ are not equal unbordered words
($\un(n)$ possibilities to exclude).

\medskip
The computation of $\vv(n)$ is easy due to equality
{$\vv(n)=\un(n)$ for each $n$.}
We prove it the following operation.
For two words $x=a_1a_2\cdots a_m,\, y=b_1b_2\cdots b_m$ denote
$x\,\otimes\, y\,=\, a_1b_1a_2b_2\cdots a_mb_m.$

\medskip\noindent Now we construct a bijection $\mathbf{F}$ in the following way.

\medskip
If $w=uav,\, |u|=|v|, |a|\le 1$ then $\mathbf{F}(w)=u\otimes v^R\,a $.

\medskip\noindent {\bf Example.} Let $w=abcd\,\circ\circ\circ\circ \, \bullet\,\star\star\star\star\,abcd$.
We have
\[
\mathbf{F}(w)\,=\, adbccbda\, \circ\star\circ\star\circ\star\circ\star\, \bullet.\]
Observe that $w$ has a border $abcd$ and $\mathbf{F}(w)$ has prefix
palindrome $ adbc\,cbda$.

\medskip\noindent  It is easy to see
that 

\smallskip
\centerline{
$w$ is bordered $\Leftrightarrow$ $\mathbf{F}(w)$ has  nontrivial even prefix palindrome}

\bigskip\noindent {\bf Computing $\tv(n)$}. The numbers $\tv(n), \vv(n)$
are ``almost'' the same.
The computation of $\tv(n)$ is easy due to equality
\[(**)\ \ 
\vv(2n+1)=\tv(2n+1),\ \ \tv(2n)=2\cdot \tv(2n-1)
\]

\medskip\noindent We justify the first  equality using simple
algebraic trick. 
\\
Denote by $\oplus$ the operation of addition modulo 2.
For a word \\ $w=a_1a_2\cdots a_m$ define \[\mathbf{F'}(w)=b_1b_2\cdots b_{m-1},\
\mbox{where}\ b_i=a_i\oplus a_{i+1}\ \mbox{for}\ i<m.\]

\medskip\noindent {\bf Observation.} A word 
$x$ is a nontrivial  odd palindrome  if and only if
$\mathbf{F'}(x)$ is a nontrivial  even palindrome.

\medskip Due to the observation we have a mapping $\mathbf{F'}$ of the  set of
length-$(2n+1)$ words without odd palindromes onto the set of length-$2n$
words without even palindromes.
 $\mathbf{F'}$ is not bijection, however it is a ``2-bijection'':
$$|\mathbf{F'}(y)|=2\  \mbox{for each word}\ y,\ \mbox{such that}\ |y|\geq 2.$$
Consequently,
$$\tv(2n+1)=2\cdot \vv(2n)=\vv(2n+1),$$ due to Equation~$(*)$.

The second equality in Equation~$(**)$
follows from the fact that for each length-$2n$  word $w$ we  can create two
length-$(2n+1)$ words $w_1,w_2$ by inserting 0 or 1 in the middle. The word $w$
has no nontrivial prefix palindrome of odd length if and only if $w_1,w_2$ have the
same property.
 \end{SOLUTION}

\begin{NOTES}
There are other relations between unbordered words and palindromes.
 \emph{Prime palstars} are even nonempty palindromes  which are not a concatenation of smaller even nonempty palindromes. It is known that the number of prime palstars of length $2n$ equals $\un(n)$, see \cite{DBLP:journals/ipl/RampersadSW11}.
Another problem concerns the number $A_3(n)$  of ternary words without {\bf any} nontrivial
palindromic prefix, then we have a recurrence similar to $(*)$:
$$ A_3(n)= 3\,A_3(n-1)-A_3(\lceil n/2 \rceil).$$
The relation between length-$n$ unbordered words
and words without even palindromic prefix is from
\cite{GabricS21}.
The cardinalities of  the sets of length-$n$
unbordered words with fixed ``weight'' have been
investigated in  \cite{HarjuN05a}.
The weight of a binary word is the number of ones. Let $U(n,k)$ denote the number of length-$n$ unbordered binary words of weight $k$.\\
If $0<k<n$ then
$$U(n,k)=U(n-1,k)+U(n-1,k-1)-\alpha(n,k)\cdot U(n/2,k/2),$$
where $\alpha(n,k)=1$  if 
both $n,k$ are even, otherwise  $\alpha(n,k)=0$.\\
Interestingly, a different natural sequence of numbers of $n$-length sequences  of  +1 and -1, not summing together to  zero, looks
 initially the same  as $\un$: $1, 2, 2, 4, 6, 12, 20, 40$.
Afterwards it differs from $\un$.
\end{NOTES}
\end{EXERCISE}
\begin{EXERCISE}[Cartesian Tree Pattern-Matching]\label{pb145}
%
\def\CT{\mathbf{CTree}}%
\def\PD{\mathsf{PD}}%
\def\ctbord{\mathsf{CTBord}}%
\def\DQ{\mathsf{Q}}%
In the problem we consider words drawn from a linear-sortable alphabet
$\Sigma$ of integers.
Let $x=x[0\dd m-1]$ be a word of length $m$. 
The Cartesian Tree $\CT(x)$ of $x$ is a binary tree in which:
\begin{itemize}
\item
the root is the position $i$ of the minimal element
 $x[i]$ (if there are several occurrences of the minimal element, its leftmost position is chosen);
\item
the left subtree of the root is $\CT(x[0\dd i-1])$;
\item
the right subtree of the root is $\CT(x[i+1\dd m-1])$.
\end{itemize}

\noindent The {\it Cartesian tree pattern-matching} problem
is naturally defined as follows: given a pattern $x$ and a text
$y$ of length $m$ and $n$ respectively, find all factors of $y$ that have the same Cartesian tree as $x$.

\paragraph{Example} Let
 $x = \sa{3}\; \sa{1}\; \sa{6}\; \sa{4}\; \sa{8}\; \sa{6}\; \sa{7}\; \sa{5}\; \sa{9},$ and 
 $$y = \sa{10}\; \sa{12}\; \sa{16}\; \underline{\sa{15}\; \sa{6}\; \sa{14}\; \sa{9}\; \sa{12}\; \sa{11}\;
  \sa{14}\; \sa{9}\; \sa{17}}\; \sa{12}\; \sa{10}\; \sa{12}$$ 
The underlined factor $u=\sa{15}\; \sa{6}\; \sa{14}\; \sa{9}\; \sa{12}\; \sa{11}\; \sa{14}\; \sa{9}\; \sa{17}$ of $y$ has the same Cartesian tree as $x$: $\CT(u)=\CT(x)$.

\tikzstyle{general} = [circle, draw]
\medskip
\begin{center}
\begin{tikzpicture}[scale=0.35,node distance=1.5cm, auto,>=latex']
\path[-] node[general] (q1) at (0,6) {\scriptsize\bf 0};

\path[-] node[general] (q2) at (2,8) {\scriptsize\bf 1}
              (q1) edge node {} (q2);

\path[-] node[general] (q3) at (4,4) {\scriptsize\bf 2};

\path[-] node[general] (q4) at (6,6) {\scriptsize\bf 3}
              (q2) edge node {} (q4)
              (q3) edge node {} (q4);

\path[-] node[general] (q5) at (8,0) {\scriptsize\bf 4};

\path[-] node[general] (q6) at (10,2) {\scriptsize\bf 5}
              (q5) edge node {} (q6);

\path[-] node[general] (q7) at (12,0) {\scriptsize\bf 6}
              (q7) edge node {} (q6);

\path[-] node[general] (q8) at (14,4) {\scriptsize\bf 7}
              (q8) edge node {} (q4)
              (q8) edge node {} (q6);

\path[-] node[general] (q9) at (16,2) {\scriptsize\bf 8}
              (q8) edge node {} (q9);
\end{tikzpicture}
\end{center}

\medskip\noindent
 
\begin{QUESTION}
Design an online linear time and space algorithm that
builds the Cartesian tree of a word $x\in\Sigma^*$.
\end{QUESTION}

\AIDE{Consider only  nodes on the rightmost paths of the tree}

\begin{QUESTION}
Design a linear-time algorithm for the Cartesian tree pattern-matching related to a pattern $x$ and a text $y$ in $\Sigma^*$.
\end{QUESTION}

\AIDE{Find a linear representation of Cartesian trees and design a notion of border table (see \cite[Problems 19 and 26]{CLR21cup}) adequate to the problem}

\begin{SOLUTION}
The algorithm considers the right path of the tree (starting from the root and always going right) as a stack of positions.
The Cartesian tree of $x[0]$ consists thus of a single root node $0$ with a stack
 containing this element only.
Then, given the Cartesian tree of a prefix $x[0\dd i]$ of $x$, with $0\le i < |x|-1$,
 the Cartesian tree of $x[0\dd i+1]$
 is obtained by popping from the stack all positions $j\le i$ for which $x[j] > x[i+1]$.
If no such element, $i+1$ is a leaf and inserted as the right child of
 the top element of the stack.
Otherwise, let $k$ be the last popped element from the stack for which $x[k] > x[i+1]$.
There are two cases:
\begin{enumerate}
\item
The stack is not empty.
Let $k'$ be its top element ($x[k'] < x[i+1]$).
Then $i+1$ is inserted as the right child of $k'$ and $k$ becomes the left child of $i+1$.
\item
The stack is empty.
Then $i+1$ is inserted at the root of the Cartesian tree of $x[0\dd i+1]$ and $k$ 
 becomes the left child of $i+1$.
\end{enumerate}
Eventually, $i+1$ is pushed on the stack.

In all cases, the new element $i+1$ is always the last element of the right path (thus the top element of the stack).

Since each index $j$ can be popped only once from the stack, the whole online process takes linear worst-case time.

\paragraph{Solution to Cartesian tree pattern-matching}
The present solution is based on the notion of a Parent-Distance array $\PD_w$ of a word $w$, which is defined as follows for $0\leq i <|w|$:

\smallskip
$\PD_w[i] =
\begin{cases}
  i - \max_{0 \leq j < i}\{{j \mid w[j] \leq w[i]} \} & \mbox{if such } j \mbox{ exists},\cr
  0 & \mbox{otherwise}.
\end{cases}$

The Parent-Distance representation has a one-to-one mapping to the Cartesian tree.

Below is the Parent-Distance representation of
$x = \sa{3}\; \sa{1}\; \sa{6}\; \sa{4}\; \sa{8}\; \sa{6}\; \sa{7}\; \sa{5}\; \sa{9}$.

\begin{center}
\begin{tabular}{|l||c|c|c|c|c|c|c|c|c|}
\hline
$i$ & 0 & 1 & 2 & 3 & 4 & 5 & 6 & 7 & 8\\
\hline
\hline
$x[i]$ & \sa{3} & \sa{1} & \sa{6} & \sa{4} & \sa{8} & \sa{6} & \sa{7} & \sa{5} & \sa{9}\\
\hline
\hline
$\PD_x[i]$ & 0 & 0 & 1 & 2 & 1 & 2 & 1 & 4 & 1\\
\hline
\end{tabular}
\end{center}

The Parent-Distance representation of a word $w$ 
 can be computed in time $O(|w|)$ time using an algorithm similar to the one
 for building the Cartesian tree of $w$.
Given the Parent-Distance representation of $w$, 
 the Parent-Distance of a factor $w[i\dd j]$ of $w$ satisfies:

\smallskip
$\PD_{w[i\dd j]}[k] =
\begin{cases}
0 & \mbox{if } \PD_w[i+k-1] \geq k,\cr
\PD_w[i+k-1] & \mbox{otherwise}.
\end{cases}$

Then, the Cartesian border table of $w$ is defined in the following way:
$\ctbord[0] = -1$ and, for $1 \leq k < i$,
\[\ctbord[i] = \max\{k \mid \CT(w[0\dd k]) = \CT(w[i-k+1\dd i]) 
\]
Below is the Cartesian border table of
$x = \sa{3}\; \sa{1}\; \sa{6}\; \sa{4}\; \sa{8}\; \sa{6}\; \sa{7}\; \sa{5}\; \sa{9}$.

\begin{center}
\begin{tabular}{|l||c|c|c|c|c|c|c|c|c|}
\hline
$i$ & 0 & 1 & 2 & 3 & 4 & 5 & 6 & 7 & 8\\
\hline
\hline
$x[i]$ & 3 & 1 & 6 & 4 & 8 & 6 & 7 & 5 & 9\\
\hline
\hline
$\ctbord[i]$ & -1 & 0 & 0 & 1 & 2 & 3 & 4 & 1 & 2\\
\hline
\end{tabular}
\end{center}

Algorithm \Algo{CTMatch} answers the second question.
It first builds the Parent-Distance representation of $x$ and $y$ and
 builds the Cartesian border table of $x$.
It uses a deque $Q$ to represent the right path of the Cartesian tree of substring
 of $y$ that matches the Cartesian tree of a prefix of $x$.
 
 \begin{algo}{CTMatch}{x,y\mbox{ non-empty words}}
  \SET{\PD_x,\PD_y}{\mbox{Parent-Distance representations of }x \mbox{ and } y}
  \SET{\ctbord}{\mbox{Cartesian border table of }x}
  \SET{i}{-1}
  \SET{\DQ}{\mbox{empty stack}}
  \DOFORI{j}{0}{|y|-1}
    \ACT"{delete elements $(v,k)$ from back of $\DQ$ with $v > y[j]$}
    \DOWHILE{i > -1 \mbox{ and } \PD_{y[j-i\dd j]}[i+1] = \PD_x[i+1]}
      \SET{i}{\ctbord[i]}
      \ACT"{delete elements $(v,k)$ from front of $\DQ$ with $k < j-i$}
    \OD
    \ACT"{add $(y[j],j)$ at the back of $\DQ$}
    \SET{i}{i+1}
    \IF{i=|x|-1}
      \ACT"{\textbf{output:} match at position $j-|x|+1$}
      \SET{i}{\ctbord[i]}
      \ACT"{delete elements $(v,k)$ from front of $\DQ$ with $k < j-i$}
    \FI
  \OD
\end{algo}

\end{SOLUTION}

\begin{NOTES}
Cartesian trees have been introduced by Vuillemin~\cite{Vuillemin80}. More information in 
\url{https://en.wikipedia.org/wiki/Cartesian_tree}
with various applications of them.
Multiple pattern Cartesian tree matching and a suffix tree for Cartesian tree matching is considered in \cite{ParkALP19}.
Fast practical solutions are presented in \cite{SongGRFLP21}.

An obvious application of this type of matching is to detect analogue ups and downs behaviour in time series without processing their absolute values.

There is a strong connection between Cartesian trees and (right) Lyndon trees. Indeed,
the Lyndon tree of a word is a Cartesian tree built from the lexicographic rank of its suffixes (see \cite{HohlwegR03,CrochemoreR20}). The notion of Lyndon tree is essential tool to deal with repetitions in words (see \cite{BannaiIINTT17}).
\end{NOTES}
\end{EXERCISE}
 
\begin{EXERCISE}[List-Constrained Square-Free Strings]\label{pb146}
\newcommand{\pop}{\mathrm{pop}}%
\newcommand{\append}{\mathrm{append}}%
\newcommand{\HalfSquare}{\frac{1}{2}\mathrm{square}}%
\newcommand{\RR}{\mathbf{C}}%
Let $L$ be a list of finite alphabets $(L_1,L_2,\ldots,L_n)$.
A word $a_1a_2\cdots a_n$ is said to be $L$-constrained if 
$a_i\in L_i$ for each $i$, $1\le i \le n$.
The aim is to find $L$-constrained square-free words of length $n$.

\paragraph{Example} For the list
$L=(\{\sa{a},\sa{b},\sa{c},\sa{e}\},
\{\sa{b},\sa{c},\sa{d},\sa{e}\},
\{\sa{a},\sa{c},\sa{d},\sa{e}\},
\{\sa{c},\sa{a},\sa{b},\sa{e}\},$
$\{\sa{a},\sa{b},\sa{c}\},
\{\sa{b},\sa{c},\sa{d}\},
\{\sa{a},\sa{b},\sa{c},\sa{d},\sa{e}\},
\{\sa{a},\sa{c},\sa{d},\sa{e}\},
\{\sa{c},\sa{a},\sa{b},\sa{d}\})$,
among many others, the word $\sa{abcabdbca}$ 
is an $L$-constrained square-free word.

For simplicity, assume from now on that each $L_i$ is of size 5.
The constructed word $u$ is treated as a stack: adding a symbol at the end 
corresponds to a {\it push} operation and removing the last symbol corresponds
to a {\it pop} operation.
Let $\pop^k$ be the sequence of $k$ pop operations.
Let also $\HalfSquare(u)$ be the maximal half-length of the suffix of $u$ that is a square. 
Let $\RR=\{1,2,3,4,5\}^{8n}$ and $symbol_j(t)$ denote the $t$-th symbol on the list $L_j$. Informally speaking,
each element  $c \in \RR$ is treated as a ``control sequence''. 
During the $i$-th iteration of Algorithm \Algo{H} below, the letter 
$symbol_j(c[i])$ is inserted at the $j$-th position of $u$ by pushing it onto the stack. 
The following function \Algo{H} runs a naive backtracking way controlled by the sequence $c\in \RR$. The result is $(u,\beta)$, where $u$ is a square-free word and $\beta$ is an auxiliary value. We have $|u|\le n$.

\begin{algo}{H}{c\in \RR}
  \SET{(u,i)}{(\mbox{empty stack},1)}
  \DOWHILE{i\le 8n\ \mbox{and}\ |u|< n}
     \SET{j}{|u|+1}
    \ACT"{push $symbol_j(c[i])$}
    \IF{u \mbox{ contains a suffix square}}
      \SET{k}{\HalfSquare(u)}
      \SET{u}{pop^k(u)}
    \FI
    \SET{i}{i+1}
  \OD
  \SET{\beta}{\mbox{the sequence of executed push and pop operations}}
  \ACT"{($\beta$ is the sequence of symbols ``push'' and ``pop'')}
  \RETURN{(u,\beta)}
\end{algo}

\begin{QUESTION}
Show constructively that there exists an $L$-constrained square-free word of length $n=|L|$ if each set $L_i$ of $L$ is of size $5$.
\end{QUESTION}

We say that $c\in \RR$ is successful if $\Algo{H}(c)=(u,\beta$, where
$|u|=n$.
Our algorithm is to compute the function $\Algo{H}(c)$ for  all possible $c\in \RR$ and choose any $c$ for which $\Algo{H}(c)$ is successful.
Then we return $u$, where $\Algo{H}(c)=(u,\beta)$.
It is enough to show that such $c$ exists.

\paragraph{Observation 1}
If $\Algo{H}(c)=(u,\beta)$ with $|u|<n$ then 
$\beta$ contains $8n$ symbols ``push'' and $8n-|u|$ symbols
``pop''.

\smallskip
$\Algo{H}(c)$ records the computation history: 
both the sequence of moves of the stack (pops and pushes) and the word $u$ as final content of the stack, with $|u|\le n$.
This is sufficient to reconstruct the word $c$ if $|u|<n$.

\begin{SOLUTION}
If $n\leq 5$ there is obviously an $L$-constrained
square-free word of length $n$. Hence, we assume later $n\geq 6$.
It is enough  to show that for at least one $c\in \RR$
the algorithm is successful, in other words $\Algo{H}(c)=(u,\beta)$, where $|u|=n$.
Define
$V=\{\Algo{H}(c)\mid c\in \RR\}.$
The following fact says that in the unsuccessful case, that is, $|u|<n$, from $(u,\beta)$, the sequence of symbols pushed onto the stack can be recovered by
reversing the algorithm.
Hence, $(u,\beta)$ uniquely determines the sequence $c=\Algo{H}^{-1}(u,\beta)$. If, for each $c\in \RR$, $\Algo{H}(c)=(u,\beta)$ with $|u|<n$ then the function 
$\Algo{H}$ is a one-to-one mapping. This implies the following fact

\paragraph{Observation 2}  Assume that for each $c\in \RR$
the algorithm is unsuccessful. Then
$|V|\geq |\RR|$


\smallskip
Now we show that our algorithm is successful. The proof is by contradiction. 
Assume the algorithm is unsuccessful for each $c$.

There are at most $2\cdot 4^{ 8n}$ sequences consisting of $8n$ push operations and at most $8n$ pop operations. The number of possible values of $u$ is at most $2\cdot 5^n$.
So, $|V|\le 4\cdot 5^n\cdot 4^{8n}$. Besides, $|\RR|=5^{8n}$.
Together, this gives  $|V|\le 4\cdot 5^n\cdot 4^{8n}<5^{8n}=|\RR|\ \mbox{for}\ n> 5.$

Therefore, the unsuccessful assumption, due to Observation2, leads to a contradiction, and proves that the algorithm is successful for at least one $c$.
\end{SOLUTION}

\begin{NOTES}
Our presentation is a deterministic version of the probabilistic algorithm
from~\cite{GrytczukKM13}, where list elements of size 4 are shown to work
similarly as for size 5. But the proof needs certain properties of Catalan numbers. The above algorithm has a pessimistic exponential time. However, by choosing randomly a control sequence $c$, it is  claimed in \cite{GrytczukKM13} that it
gives a randomized linear-time algorithm.

It is conjectured that there are also list-constrained square-free words when list elements are of size 3. The conjecture was confirmed by Matthieu Rosenfeld~\cite{Rosenfeld21}
for the case where all list elements of size 3 are subsets of the same alphabet of size 4.

There are other square-free problems on ``special'' words, for example, Abelian square-free words with 4 letters, see
\cite{DBLP:conf/icalp/Keranen92}, or circular square-free words with 3 letters of the length not in \{5, 7, 9, 10, 14, 17\}, see \cite{DBLP:journals/combinatorics/Shur10}.
\end{NOTES}
\end{EXERCISE}

\begin{EXERCISE}[Superstrings of shapes of permutations]\label{pb147}
\newcommand{\shape}{\mathsf{shape}}%
\newcommand{\SHAPES}{\mathsf{SHAPES}}%
\newcommand{\Extend}{\mathsf{Extend}}%
Two words $u$ and $v$ of the same length
are said to be {\it order-equivalent}\INDEX{order equivalent}, written
 $u\approx v$, if
 $u[i] <  u[j]\Longleftrightarrow v[i] < v[j]$
 for all pairs of positions $i,j$ on the words.
For a word $u$  of length $n$ with all letters distinct we define 
 $\shape(u)$ as the $n$-permutation of $\{1,2,\ldots,n\}$
 order-equivalent to $u$. For example $\shape(2,5,4)=(1,3,2)$.
Define 
$$\SHAPES_n(w)=\{\shape(u)\,:\, u\ \mbox{is a factor of}\ w\ \mbox{of length}\ n\}$$
For $n>2$ there is no word containing exactly once each $n$-permutation,
 but surprisingly 
 in order-preserving case such a word exists for each $n$.
 
A word  of size $n!+n-1$  containing shapes of all $n$-permutations is
 called a {\it universal word}, it is a superstring of shapes of all $n$-permutations.
Obviously $n!+n-1$ is the smallest length of such word.

\paragraph{\bf  Example}
The word 
 $3\,  5\,  1\,   0\,  5\,  1\,  2\, 3$ of length $3!+2$  is 3-universal:
The sequence of shapes of its  factors of length 3 is:
$$(2,3,1)\rightarrow (3,2,1)\rightarrow (2,1,3)\rightarrow (1,3,2)\rightarrow (3,1,2)\rightarrow (1,2,3).$$
%

\begin{QUESTION}
Construct, for a given $n$,  a universal
  word of size $n!+n-1$ (shortest possible).
\end{QUESTION}

\AIDE{Use a construction similar to that for linear
 de Bruijn words}

\begin{SOLUTION} 
Denote by $suf(w)$/$pref(w)$ the suffix/prefix of length $n-1$
of the word $w$.
We construct a graph $G_n$. Its nodes are $(n-1)$-permutations and edges correspond to 
$n$-permutations.
The  edge corresponding to the $n$-permutation 
$\pi$ is defined as 
$$\shape(pref_{n-1}(\pi))\, \stackrel{k}{\rightarrow}\, \shape(suf(\pi)),$$
where the label $k$ is the last element of $\pi$, see the figure.
For example the 5-permutation $(3,5,4,1,2)$ corresponds to the edge   $(2,4,3,1) \stackrel{2}{\rightarrow}\, (4,3,1,2)$. 

\medskip\noindent 
Let $lift(\alpha,a)$ be the operation of 
adding 1 to  each  element of $\alpha$ equal or larger than $a$. 
For example $lift((3,2,4,1),\,3)=(4,2,5,1)$.

\paragraph{Observation}  Assume $\alpha$ is a sequence of integers,
$suf(\alpha)$ consists  of distinct integers
and $\alpha'$ results by replacing $suf(\alpha)$ with 
$lift(suf(\alpha),a)$. 
Then
$\SHAPES(\alpha')=\SHAPES(\alpha)$.

%

\medskip\noindent
We describe the following operation for $1\le k\le n$ and a sequence $\alpha$ of distinct integers
of length at least $n-1$. 

\medskip
\begin{minipage}{0.8\textwidth}
 \noindent 
 $\Extend_n(\alpha,k)$:

\smallskip
 \quad $\beta := suf(\alpha)$

 \quad if $k<n$ then 
  $a$ := $k$-th smallest element of $\beta$

 \quad\quad  else $a$ := $\max\, (\beta)\,+\, 1 $
 
 \quad $\alpha:=lift(\alpha,a)$ (now $a\notin \alpha$)

 \quad $\alpha := \alpha\cdot a$

 \quad return $\alpha$
\end{minipage}

\paragraph{\bf Example} Let $\alpha\,=\, (4,6,7,5,1,7,6,4,3,1).$
Then
\[\Extend_4(\alpha,2)=(5,7,8,6,1,8,7,5,4,1,3),\] 
\[\Extend_4(\alpha,4)=(4,7,8,6,1,8,7,4,3,1,5).\]
Denote by $\mathsf{EulerCycle}(G_n,\alpha)$ the 
sequence of labels of any Euler cycle
of $G_n$ starting in the node $\alpha$ (an $(n-1)$-permutation).

\begin{algo}{Superstring}{n,\ \mbox{positive integer }}
   \SET{\alpha}{(1,2,\ldots,n-1)}
  \ACT"{$C$ := $\mathsf{EulerCycle}(G_n,\alpha)$\ \ 
  (Assume $\,C\,=\, c_1c_2\cdots c_{n!}\,$)}
  \DOFORI{i}{1}{n!}               
     \SET{\alpha}{\Extend_n(\alpha,c_i)}
  \OD
  \RETURN{\alpha}
\end{algo}

\begin{center}
\begin{figure}
\centering
\includegraphics[width=3.7cm]{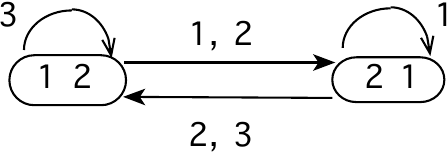}\ \
\includegraphics[width=5.8cm]{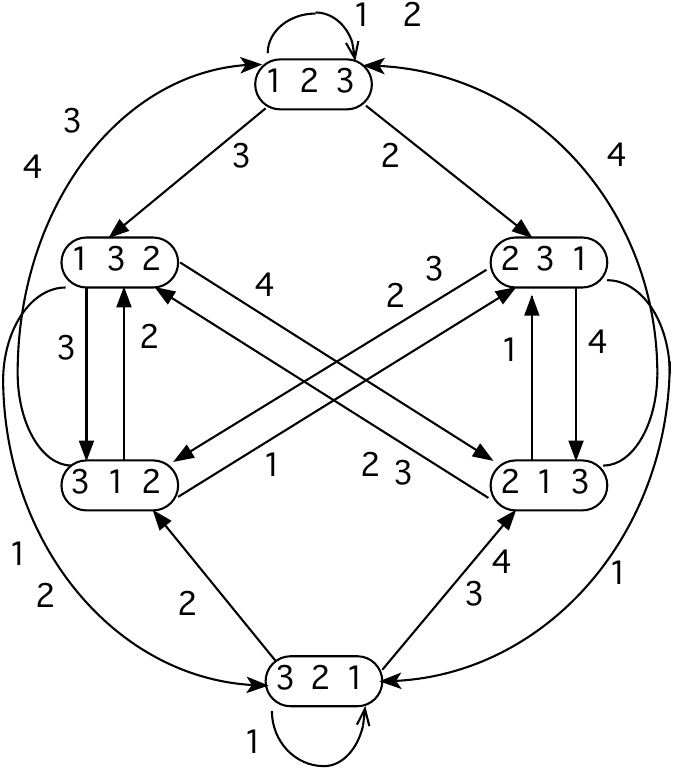}
\end{figure}
\end{center}

\noindent {\bf Example.} 
$\mathsf{EulerCycle}(G_3,(1\,2))\;=\; 1\,2\,1\,1\,2\,2\,3\,3$.
The algorithm returns a universal word
 $\alpha=78613245$. We can reduce the alphabet and get $56413245$, with the same sequence of length-3 shapes.

\medskip It is easy to see that $G_n$ is Eulerian. The computed $\alpha$ 
contains all permutations as shapes. It follows from the following fact. 

\medskip\noindent {\bf Observation.} 
Assume the edges of the Euler cycle given by labels $C=c_1c_2\dd c_{n!}$ correspond to sequence of 
permutation $\pi_1,\pi_2,\dd \pi_{n!}$.
If $\SHAPES(\alpha)\,=\,\{\pi_1,\pi_2,\dd \pi_{i-1}\}$ then 
$$\SHAPES_n(\Extend(\alpha,c_i))\,=\,\{\pi_1,\pi_2,\dd \pi_{i-1},\pi_i\}$$
\end{SOLUTION}

\begin{NOTES}
The resulting universal word produced in the algorithm 
presented here uses huge amount of letters, however the most interesting is that universal words exist at all.
The algorithm  is a version of that in \cite{GaoKSZ19}, where 
it was also given construction of cyclic universal words. 

There exist universal words with  only $n+1$ 
letters, but the construction is too complex to present it here,
we refer to \cite{DBLP:journals/dm/Johnson09b}.
\end{NOTES}
\end{EXERCISE}
\begin{EXERCISE}[Linearly generated words and primitive  polynomials]\label{pb148}
\newcommand{\LS}{\mathsf{LShift}}%
We consider sequences  of length-$n$ binary non-unary  words 
(each containing at least one nonzero bit).
There are $N=2^n-1$ such word. By $\oplus$
 denote operation {\it xor} on bits. 
%
\newcommand{\LFSR}{\mathsf{LFSR}}
\newcommand{\GEN}{\mathsf{GEN}}
Let $\alpha=(a_0,a_1,\ldots,a_{n-1})$ be a (control) sequence of bits.
The {\bf LFSR-sequence} associated with $\alpha$, denoted by $\LFSR(\alpha)$,
 is the sequence $b_1b_2b_3\cdots b_{N+n-1}$  of bits, such that
  for $n<k<N+n-2$.
\[\hspace*{0.5cm}
 b_1b_2\cdots b_n=0^{n-1}1,\]
\[\hspace*{0.5cm}
 b_{k+1}\, =\,a_0\cdot  b_{k-n+1}\, \oplus\,  a_1\cdot  b_{k-n+2}\, \oplus\,  a_2\cdot  b_{k-n+3} \, \oplus \,
\cdots \, \oplus\, a_{n-1}\cdot  b_{k} 
\]
For example for $\alpha=11010$ and $n=5$ the recurrence is
$$b_{k+1}\, =\,  b_{k-4}\, \oplus\,  b_{k-3}\, \oplus\, b_{k-1}.$$
%
We fix the starting prefix $0^{n-1}1$ for simplicity.
Observe that $N+n-2$ is the smallest length of a binary sequence 
containing each length-$n$ nonzero word.
Denote by $\GEN(\alpha)$ the sequence of consecutive length-$n$ factors in $\LFSR(\alpha)$.
Observe that $\GEN(\alpha)$ is of length $N$. 
%
%

\paragraph{Example} We have:\
$ \LFSR(110)\,=\,0\,0\,1\,0\,1\,1\,1\,0\,0,$
$$\GEN(110)\,=\, 
001,\  010,\ 101,\ 011,\ 111,\ 110,\ 100.$$ 
The polynomial related to $\LFSR(\alpha)$, where $\alpha=a_0a_1\cdots a_{n-1}$  is 
$$W_{\alpha}\,=\,x^n+a_{n-1}x^{n-1}+a_{n-2}x^{n-2}+\cdots+a_1x^1+a_0$$
$W_{\alpha}$ is called a  generating polynomial for
$\LFSR(\alpha)$. 
If all words in $\GEN(\alpha)$ are different then the sequence $\LFSR(\alpha)$
 is called a {\it PN sequence} (pseudo-noise) sequence.
 The polynomial $P$ is called {\it primitive}
if all polynomials $x^i \ \mbox{mod}\  P(x)$ are distinct, for $1\le i \le 2^n-1$
(it is maximal number of nonzero binary polynomials of degree smaller than $n$).
It is known, and surprising, that  $\LFSR$-sequences corresponding  to primitive polynomials are
PN sequences. In this way construction of PN sequences is reduced to construction of 
primitive polynomials, which is  easier since one can use  algebraic tools.

\begin{QUESTION}
Compute the $m$-th word of $\GEN(\alpha)$ in $O(n^3\log m)$ time,
using matrix multiplications.
\end{QUESTION}

\begin{QUESTION}  
Improve time complexity of computing the $m$-th word of $\GEN(\alpha)$ to $O(n\log n\cdot \log m+n^2)$ time, 
using polynomial multiplications.
\end{QUESTION}

\begin{SOLUTION}
Each binary polynomial $V(x)=a_{n-1}x^{n-1}+a_{n-2}x^{n-2}+\cdots+a_0$ of degree
at most $n-1$ (some of $a_i$ could equal zero)
can be represented as the word $string(V)\,=\,a_{n-1}a_{n-2}\cdots a_0$
The length of $string(V) $ equals $n$.

\paragraph{Example} For $n=5$ and polynomials $P$ 
of degree less than $5$ each $string(P)$ is of length $5$, we have
$$string(x^2+1)=00101, \ string(x)=00010,\ string(x^3)=01000.$$
\medskip\noindent
We create the $n\times n$ matrix $A$.
The $i$-th column, for $1\le i\le n$, 
 equals $string(x^i \ \mbox{mod}\ W_{\alpha})$ written bottom-up.

In particular the $n$-th column is the sequence $a_0,a_1,\ldots,a_{n-1}$ read top-down,
 and the $i$-th column, for $i<n$ is the sequence $0^i10^{n-i-1}$, read also 
 top-down. 

Then $\GEN(\alpha)$ is the 
 sequence of the first rows of $A^1, A^2,A^3,\ldots$ 
Consequently, the required result equals the first row of $A^m$.

\medskip\noindent{\bf Fast computation of $A^m$}
The computation of $A^m$ 
 can be done in time $O(n^3\log m$ by first computing
 all powers $A^t$, where $t$ is a power of 2 not exceeding $2^n$.

\medskip\noindent {\bf Example} Consider $\alpha=a_0a_1a_2a_3a_4a_5=10100$ and 
$$W_{\alpha}= x^5+a_4x^4+a_3x^3+a_2x^2+a_1x^1+a_0=x^5+x^2+1,$$
Then the consecutive first  
rows of $A^i$, for $i=1,2,3,\ldots$ give  the sequence  $\GEN(\alpha)$. 
The first 6 powers of $A=A^1$ are:

{\small
\vspace*{0.5cm}
\begin{center}
\begin{tabular}{ccccc}
$A^1\,=\,$&
\begin{tabular}{|c|c|c|c|c|}
\hline
 {\bf 0} & {\bf 0} & {\bf 0} & {\bf 0} & {\bf 1} \\
\hline
 1 & 0 & 0 & 0 & 0  \\
\hline
 0 & 1 & 0 & 0 & 1\\
\hline
 0 & 0 & 1 & 0 & 0\\
\hline
 0 & 0 & 0 & 1 & 0\\
\hline
\end{tabular}
&
\hspace*{0.5cm} $A^2\,=\,$
&
\begin{tabular}{|c|c|c|c|c|}
\hline
{\bf 0} & {\bf 0} & {\bf 0} & {\bf 1} & {\bf 0} \\
\hline
0 & 0 & 0 & 0 & 1  \\
\hline
1 & 0 & 0 & 1 & 0\\
\hline
0 & 1 & 0 & 0 & 1\\
\hline
0 & 0 & 1 & 0 & 0\\
\hline
\end{tabular}
\end{tabular}
\end{center}

\begin{center}
\begin{tabular}{ccccc}
$A^3\,=\,$
&
\begin{tabular}{|c|c|c|c|c|}
\hline
{\bf 0} & {\bf 0} & {\bf 1} & {\bf 0} & {\bf 0} \\
\hline
0 & 0 & 0 & 1 & 0  \\
\hline
0 & 0 & 1 & 0 & 1\\
\hline
1 & 0 & 0 & 1 & 0\\
\hline
0 & 1 & 0 & 0 & 1\\
\hline
\end{tabular}
&
\hspace*{0.5cm} $A^4\,=\,$
&
\begin{tabular}{|c|c|c|c|c|}
\hline
{\bf 0} & {\bf 1} & {\bf 0} & {\bf 0} & {\bf 1} \\
\hline
0 & 0 & 1 & 0 & 0  \\
\hline
0 & 1 & 0 & 1 & 1\\
\hline
0 & 0 & 1 & 0 & 1\\
\hline
1 & 0 & 0 & 1 & 0\\
\hline
\end{tabular}
\end{tabular}
\end{center}

\begin{center}
\begin{tabular}{ccccc}
$A^5\,=\,$
&
\begin{tabular}{|c|c|c|c|c|}
\hline
{\bf 1} & {\bf 0} & {\bf 0} & {\bf 1} & {\bf 0} \\
\hline
0 & 1 & 0 & 0 & 1  \\
\hline
1 & 0 & 1 & 1 & 0\\
\hline
0 & 1 & 0 & 1 & 1\\
\hline
0 & 0 & 1 & 0 & 1\\
\hline
\end{tabular}   
&
\hspace*{0.5cm} $A^6\,=\,$
&
\begin{tabular}{|c|c|c|c|c|}
\hline
{\bf 0} & {\bf 0} & {\bf 1} & {\bf 0} & {\bf 1} \\
\hline
1 & 0 & 0 & 1 & 0  \\
\hline
0 & 1 & 1 & 0 & 0\\
\hline 
1 & 0 & 1 & 1 & 0\\
\hline
0 & 1 & 0 & 1 & 1\\
\hline
\end{tabular}   
\end{tabular}
\end{center}
}
\end{SOLUTION}

\begin{SOLUTION}
We go now to the second question, using the following observation.

\smallskip
\paragraph{\bf Observation}
 \begin{itemize}
     \item
 The columns (read
  left-to-right) of the matrix $A^m$ correspond to\\
 $string(x^m),\, string(x^{m+1}),\,\dd\,string(x^{m+n-1}).$
 \item
The sequence $\GEN(\alpha)$ consists of  first rows of consecutive matrices.
 \end{itemize}

\medskip\noindent
 In the example the first 6  words of $\GEN(10100)$ are
 \[00001,\ 00010,\
 00100,\ 01001,\ 10010,\ 00101.\]
%
%
The columns of $A^6$ correspond to:
\[x^6 \bmod  W_{\alpha}\,=\, x+x^3,\ \ \
x^7 \bmod  W_{\alpha}\,=\, x^2+x^4,\]
\[ x^8 \bmod  W_{\alpha}\,=\, 1+x^2+x^4,\  \ \ x^9 \bmod  W_{\alpha}\,=\, x+x^3+x^4\] 
\[x^{10} \bmod  W_{\alpha}\,=\, 1+x^4.\]
Instead of using matrix multiplication we can compute $x^m \bmod W_{\alpha}$,
by computing first $x^{2^i}$ for all $i$'s such that  $2^i\le m$.
The cost $O(n^3)$ of matrix multiplication is reduced to
$O(n\log n)$ time of polynomial multiplication using {\sf FFT}.
We need $O(\log m)$ such multiplications.
Once we know $x^m$ we can compute all columns of $A^m$, since they correspond to $x^m,x^{m+1},\dd x^{m+n-1}$, it  needs  $O(n^2)$ time.
Altogether we need $O(n^2)$ time.
\end{SOLUTION}

\paragraph{Primitive trinomials}
We consider now  trinomials (polynomials with exactly three
nonzero coefficients), althought the next problem applies also to
general polynomials (but the answer is more difficult).
A very partial list of primitive trinomials is:

$x^3+x+1,\, x^4+x+1,\, x^5+x^2+1,\, x^6+x+1,\,
 x^7+x+1,\, $

$x^9 + x^4 + 1,\, x^{10} + x^3 +
1,\, x^{11}+x^2+1,\, x^{15}+x+1,\,
x^{100}+x^{37}+1,\,$

$x^{900}+x+1,\, x^{74 207 281}+x^{9999621}+1,\, x^{6972593}+x^{3037958}+1.
$

\medskip
\begin{QUESTION}
Assume $W(x)=x^n+x^k+1$ is a primitive binary trinomial  of degree $n$.
 Prove that $\LFSR(W)$ is  a simple PN sequence
\end{QUESTION}

\begin{SOLUTION} The main trick is to use cyclic shifts of words $v_i$.
Denote by $\LS_k(w)$ the cyclic left shift of $w$
by $k$ positions (suffix of size $k$ is moved to the front of $w$).
For example $\LS_3(abcde)= cdeab$.

\paragraph{\bf Observation} Assume $P=a_{n-1}x^{n-1}+a_{n-2}x^{n-2}...+a_0$
then\\ $x\cdot P(x) \ \mbox{mod}\ W(x))$ equals
$$ a_{n-2}x^{n-1}+a_{n-3}x^{n-3}\dd +(a_{k-1}\oplus a_{n-1})x^k\dd +
a_0x+a_{n-1}$$
If $string(P)=a_{n-1}a_{n-2}\dd a_0$ then $string(x\cdot P\ \mbox{mod}\ W)
$ results by applying $\LS_{n-1}$ and adding $a_{n-1}$ to $(n-k)$-th bit.
For $k=3$ we have $$100101 \rightarrow 000011,\ 01111\rightarrow 11110$$

\paragraph{\bf Fact}  If $W$ is a primitive trinomial then
$\LFSR(W)$ generates a PN sequence.

\begin{PREUVE}
Let $W_i=
 x^i \ \mbox{mod}\ W(x)) $  and $v_i=string(W_i)$.
Let $w_i=\LS_k(v_i)$.
The function $\LS_k(w)$ is a bijection between consecutive
words $v_i$ and $w_i$. Hence we have $2^n-1$ distinct $w_i$,
since we have $2^n-1$ distinct $v_i$, due to primitivity of
the polynomial $W$.
\end{PREUVE}

\medskip
The construction is demonstrated in the table below for $W(x)=x^4+x^3+1$.
Top sequence presents here the binary representations of all 15 polynomials
$x^1,x^2,x^3,x^4...x^{15}$
modulo $x^4+x^3+1$, where  the polynomial $W(x)=a_1x^3+a_2x^2+a_3x^1+a_4$
is represented by $(a_1,a_2,a_3,a_4)$. Bottom sequence represents a  PN sequence -
the
words given by the bijection $\LS_3$
applied  to the words in the top sequence.

\medskip
{\small

0\underline{010} 0\underline{100} 1\underline{000} 1\underline{001} 1\underline{011} 1\underline{111} 0\underline{111} 1\underline{110} 0\underline{101} 1\underline{010} 1\underline{101}

0\underline{011} 0\underline{110} 1\underline{100} 0\underline{001}

\medskip
 \underline{010}0 \underline{100}0 \underline{000}1 \underline{001}1 \underline{011}1
\underline{111}1 \underline{111}0  \underline{110}1
\underline{101}0 \underline{010}1 \underline{101}1

\underline{011}0 \underline{110}0 \underline{100}1 \underline{001}0
}

\end{SOLUTION}

\begin{NOTES}
\noindent Using the observation it is relatively easy, though tedious,
to show that primitive polynomials generate PN-sequences. It is enough to show
that the top row of $A^i$ determines the whole matrix $A^i$. Then, if a
polynomial is primitive,  the
values of the first column correspond to powers of $x$, which are different due to polynomial primitivity, so all the first rows are distinct.

Hence if the polynomial $W_{\alpha}$ is primitive then
$\LFSR$ generates $2^n-1$ distinct words.
The proof of this fact is nontrivial

If we know any $n$ consecutive values of $\GEN(\alpha)$ then
we can compute $\alpha$ in polynomial time using Berlekamp-Massey algorithm, see \cite{Berlekamp}.
Linear feedback shift register  and PN sequences were introduced by Solomon Golomb.
\end{NOTES}
\end{EXERCISE}
\begin{EXERCISE}[An application of  linearly generated words]\label{pb149}
\newcommand{\Cyc}{\mathsf{CycF}}%
\newcommand{\PN}{\mathsf{LFSR'}}%
\newcommand{\LFSR}{\mathsf{LFSR}}%
\newcommand{\GEN}{\mathsf{GEN}}%
In this problem we  want to decompose each binary de Bruijn graph $G_{n+1}$, 
disregarding two loops, into two
edge-disjoint simple cycle. The word ``simple'' means that nodes 
do not repeat on  the cycle. It is easy to see that each cycle should 
be of length $2^n-1$. LFSR-sequences provide
a surprisingly simple algorithm.

Assume  the alphabet is $\{0,1\}$.
Denote by $\Cyc_m(w)$ the set of all
cyclic length-$m$ factors of a word $w$. A word of length $2^{n}$ is a de Bruijn word  of rank $n$ if $\Cyc_n(w)=2^n$.
We say that 
a word $w$ is a semi-deBruijn word of rank $n$  if $|w|=2^n-1$
and $\Cyc_n(w)=2^n-1$. Two semi-deBruijn words $u,w$ are {\it orthogonal}
if $\Cyc_{n+1}(w)\, \cap \Cyc_{n+1}(u)
\,=\,  \emptyset.$ 
We are interested in finding two orthogonal semi-deBruijn words $u,w$.

%

\smallskip
A  simple cycle of length $2^n-1$ in  $G_n$ 
is called a {\it semi-Hamiltonian} cycle. Orthogonal semi-deBruijn words correspond to such cycles.
The figure shows the decomposition of the graph $G_5$ without loops 
 into two edge-disjoint semi-Hamiltonian cycles.

\smallskip
\noindent
\begin{tikzpicture}[scale=.50]
\tikzset{font=\small} 
\path[->] node (o1) at (3,12) {0};
\path[->] node (o2) at (0,10.5) {1};
\path[->] node (o3) at (6,10.5) {8};
\path[->] node (o4) at (1.5,9) {2};
\path[->] node (o5) at (4.5,9) {4};
\path[->] node (o6) at (3,7.5) {9};
\path[->] node (o7) at (0,6) {3};
\path[->] node (o8) at (1.5,6) {5};
\path[->] node (o9) at (4.5,6) {10};
\path[->] node (o10) at (6,6) {12};
\path[->] node (o11) at (3,4.5) {6};
\path[->] node (o12) at (1.5,3) {11};
\path[->] node (o13) at (4.5,3) {13};
\path[->] node (o14) at (0,1.5) {7};
\path[->] node (o15) at (6,1.5) {14};
\path[->] node (o16) at (3,0) {15};

\path[->,dotted]
    (o1) edge node {} (o2)
    (o2) edge node {} (o4)
    (o3) edge node {} (o1)
    (o4) edge node {} (o8)
    (o5) edge node {} (o6)
    (o6) edge node {} (o7)
    (o7) edge node {} (o11)
    (o7) edge node {} (o14)
    (o8) edge[bend right] node {} (o9)
    (o10) edge node {} (o3)
    (o11) edge node {} (o13)
    (o12) edge node {} (o14)
    (o13) edge node {} (o12)
    (o14) edge node {} (o15)
    (o15) edge node {} (o10)
    (o9) edge node {} (o5)
;

\path[->, color=blue]
    (o2) edge node {} (o7)
    (o3) edge node {} (o2)
    (o4) edge node {} (o5)
    (o5) edge node {} (o3)
    (o6) edge node {} (o4)
    (o7) edge node {} (o14)
    (o9) edge[bend right] node {} (o8)
    (o8) edge node {} (o12)
    (o10) edge node {} (o6)
    (o11) edge node {} (o10)
    (o12) edge node {} (o11)
    (o13) edge node {} (o9)
    (o14) edge node {} (o16)
    (o15) edge node {} (o13)
    (o16) edge node {} (o15)
;
\end{tikzpicture}
\hspace*{2cm} 
\begin{tikzpicture}[scale=.50]
\tikzset{font=\small} 
\path[->] node (o1) at (3,12) {0};
\path[->] node (o2) at (0,10.5) {1};
\path[->] node (o3) at (6,10.5) {8};
\path[->] node (o4) at (1.5,9) {2};
\path[->] node (o5) at (4.5,9) {4};
\path[->] node (o6) at (3,7.5) {9};
\path[->] node (o7) at (0,6) {3};
\path[->] node (o8) at (1.5,6) {5};
\path[->] node (o9) at (4.5,6) {10};
\path[->] node (o10) at (6,6) {12};
\path[->] node (o11) at (3,4.5) {6};
\path[->] node (o12) at (1.5,3) {11};
\path[->] node (o13) at (4.5,3) {13};
\path[->] node (o14) at (0,1.5) {7};
\path[->] node (o15) at (6,1.5) {14};
\path[->] node (o16) at (3,0) {15};

\path[->,color=red]
    (o1) edge node {} (o2)
    (o2) edge node {} (o4)
    (o3) edge node {} (o1)
    (o4) edge node {} (o8)
    (o5) edge node {} (o6)
    (o6) edge node {} (o7)
    (o7) edge node {} (o11)
    (o8) edge[bend right] node {} (o9)
    (o10) edge node {} (o3)
    (o11) edge node {} (o13)
    (o12) edge node {} (o14)
    (o13) edge node {} (o12)
    (o14) edge node {} (o15)
    (o15) edge node {} (o10)
    (o9) edge node {} (o5)
;

\path[->,dotted]
    (o2) edge node {} (o7)
    (o3) edge node {} (o2)
    (o4) edge node {} (o5)
    (o5) edge node {} (o3)
    (o6) edge node {} (o4)
    (o7) edge node {} (o14)
    (o9) edge[bend right] node {} (o8)
    (o8) edge node {} (o12)
    (o10) edge node {} (o6)
    (o11) edge node {} (o10)
    (o12) edge node {} (o11)
    (o13) edge node {} (o9)
    (o14) edge node {} (o16)
    (o15) edge node {} (o13)
    (o16) edge node {} (o15)
;
\end{tikzpicture}


We refer to problems \cite[Problem 18]{CLR21cup}, \cite[Problem 69]{CLR21cup}  for the
 formal definition of de Bruijn graph $G_{n+1}$.
The nodes of $G_{n+1}$ are  words of length $n$, edges correspond to words of 
length $n+1$. The word $a_1a_2\cdots a_{n+1}$ corresponds to the edge 
$$a_1a_2\cdots a_n \stackrel{a_{n+1}}{\rightarrow}  a_2a_3\cdots a_{n+1}$$
We discard two loops in the graph and ask to compute two semi-Hamiltonian
cycles covering all non-loop edges. 
In the figure each node $i$ corresponds to the 4-bit binary representation of $i$.

\begin{QUESTION}
Assume you have a primitive polynomial $W(x)$  over $Z_2$ of degree $n$.
Compute two  orthogonal semi-deBruijn words $u,w$. Equivalently, compute two
edge-disjoint semi-Hamiltonian cycles in $G_{n+1}$ covering all non-loop edges.
\end{QUESTION}

\AIDE{Use LFSR-sequences}

\begin{SOLUTION}
Let $\alpha$ be the sequence 
 of coefficients of $W(x)$, without coefficient at $x^n$, in the order
 of increasing powers of $x^i$:
$\alpha=(a_0,a_1,\ldots,a_{n-1})$. 
For a binary word $w$ denote by $\overline{w}$ 
its bitwise negation, for example $\overline{0011}=1100$.
Using LFSR the construction is as follows.

\paragraph{\bf Algorithm} 

$w:=\LFSR(\alpha)$;

$u:=\overline{w}$;

remove the last $n-1$ letters in $u$ and in $w$;

{\bf return} $u,w$

\medskip\noindent
{\bf Example.}
We have $LFSR(1001)\,=\,000111101011001\underline{000}.$ The algorithm deletes the underlined fragment and returns
\[
w\,=\, 000111101011001,\ \ u\,=\, 111000010100110\] 
Observe that $w$ is the word corresponding to the the cycle indicated 
on the left in the figure. $u$ is  the negation of $w$ and corresponds to  the remaining cycle.
The words $u,w$ are requied words.
The nodes of the cycle in $G_6$ related to $w$ are:
\[
 0001\rightarrow  0011\rightarrow  0111\rightarrow  
1111\rightarrow  1110\rightarrow   1101\rightarrow  
1010\rightarrow  0101\]
\[
\rightarrow1011\rightarrow 
0110\rightarrow  1100\rightarrow  1001\rightarrow  0010\rightarrow 
0100\rightarrow  1000 \]
 This cycle, when nodes are written as decimal numbers is (see the figure)
\[1,3, 7, 15 14, 13, 10, 5, 11, 6, 12, 9, 2 4, 8,\]

\medskip\noindent
Correctness of the algorithm follows directly from the following fact.

\medskip\noindent {\bf Fact.} If $\alpha$ corresponds to a primitive 
polynomial then the word 
$\LFSR(\alpha)$ and its bitwise negation  have no
common factor of length $n+1$.

\begin{PREUVE} We use the following property of $\alpha$.

\medskip\noindent{\bf Claim.}
The number of $1$'s in  $\alpha$ is even.
The number of 1's in $\alpha_n$ cannot be odd, otherwise $\LFSR(\alpha)$
would loop at $1^n$

\medskip\noindent
The proof  is now by contradiction.

\medskip\noindent
Assume $(n+1)$-length word $vs$, where $s$ is a letter,  is both
 in $\LFSR(\alpha)$ and $ \overline{\LFSR(\alpha)}$.\\
Then $v\,s,\, \overline{v}\,\overline{s} \in \LFSR(\alpha)$.
However the factors $v$ and $\overline{v}$ should be  followed by 
the same letter, which follows from the claim and definition of $\LFSR$
- a contradiction.
\end{PREUVE}
\end{SOLUTION}

\begin{NOTES} 
The problem is related to the number of edge-disjoint  simple 
cycles in de Bruijn graph. 
It is known that {\it  maximal} number of  such cycles
in de Bruijn graph of rank $n$  equals the number of conjugate (cyclic) classes
of binary words of length $n$. It was a difficult problem known as Golomb's 
conjecture.
We were interested here in {\it minimal} 
number of simple cycles containing all edges of de Bruijn graph. 
The considered problems is related to finding so called double helices.
A double helix is a Hamiltonian cycle in de Bruijn such that after its deletion 
the remaining graph consists of one simple cycle and two loops.
The notion of double helix was motivated by
some problems in genetics.
If we have two edge-disjoint semi-Hamiltonian cycles then it is easy to convert 
one of them into a double helix.
Double helices were considered in the context of primitive polynomials in \cite{mcconnell2013debruijn}. 
The  algorithm presented here is  algebraic and depends heavily on primitive polynomials.
A different combinatorial construction (without use of primitive polynomials) 
 of double helices  was given
 in~\cite{RepkeR18}, where double helices correspond to so called complementary cycles: two Hamiltonian cycles of de Bruijn graph $G_{n+1}$ which are edge-disjoint except
 4 edges which are necessarily contained in each Hamiltonian cycle of $G_{n+1}$.
 Such two cycles can be trivially converted to 
 edge-disjoint semi-Hamiltonian cycles. The algorithm presented here gives always words $u,w$ 
 which are negation of each other, the algorithm in~\cite{RepkeR18} would give 
 in many cases the words $u,w$ not having this property.
\end{NOTES}
\end{EXERCISE}

\begin{EXERCISE}[Testing idempotent equivalence  of words]\label{pb131}
We consider a relation between words of $A^*$, for a finite alphabet $A$, that identifies a square $uu$ to its root $u$.
More precisely, any factor $u$ occurring in a word $x\in A^*$ can be replaced by $uu$, and any occurrence of $uu$ can be replaced by $u$. 
Two words are idempotent equivalent if one can be transformed in the other using 
such replacements.
This defines an equivalence relation $\approx$ between words of $A^*$.
For example, $\sa{aababa}\approx \sa{aba}$ since $\underline{\sa{aa}}\sa{baba} \approx \underline{\sa{abab}}\sa{a} \approx \sa{aba}$, and obviously  $\sa{a}^{10}\approx \sa{a}^{111}$.
A nontrivial example is
$\sa{bacbcabc} \approx \sa{bacabc}$.
For a given alphabet the number of equivalence classes is finite, but grows considerably fast.
For alphabet sizes $1,2,3,4,5$ the number of equivalence classes are respectively $1,2,7,160,332381$.
The goal of the problem is to design an efficient algorithm for testing the $\approx$-equivalence of two words. To do so, with each $x\in
A^*$ is associated a (characteristic) quadruple $\Psi(x)=(p,a,b,q)$, where
$a,b\in A$, $pa$ is a shortest prefix and $bq$ is a shortest suffix of $x$ for which $\alph(pa)=\alph(bq)=\alph(x)$ (Recall that $\alph(u)$ is the set of letters occurring in $u$). For example, $\Psi(\sa{ababbbcbcbc})=(\sa{ababbb},\sa{c},\sa{a},\sa{bbbcbcbc}).$
The sought algorithm is based on the following result (see Notes for reference).

\begin{LEMME}[Equivalence Criterion]\label{t:equiv_crit}
Let $x,y\in A^*$ be two words and their quadruples $\Psi(x)=(p,a,b,q)$ and $\Psi(y)=(p',a',b',q')$.
Then,
$x \approx y$ iff
$p \approx p'$, $a=a'$, $b=b'$ and $q \approx  q'$.
\end{LEMME}
For example $\sa{bacbcabc} \approx \sa{bacabc}$ since
$\Psi(\sa{bacbcabc})=(\sa{ba},\sa{c},\sa{a},\sa{bc})= \Psi(\sa{bacabc})$.

\begin{QUESTION}
Assuming $A$ is an integer alphabet (sortable in linear time),
show how to check if $x\approx y$ in $(n\cdot |A|)$ time,
where $n=|x|+|y|$.
\end{QUESTION}

\begin{SOLUTION}
We assume $\alph(x)=\alph(y)$ since otherwise $x,y$ are certainly not equivalent.

First, we change the problem to testing the equivalence of two factors $x,y$ of the same word $z=x\$y$, where \$ is new symbol. Observe that $\Psi(z)=(x,\$,\$,y)$.

Next, we restrict the set of factors of $z$ as follows.
Let $\Rang(u)=|alph(u)|$ the rank of a word $u$.
We say that a proper factor $z[i\dd j]$ of $z$ is {\em essential} if
$$\Rang(z[i\dd j])+1=\Rang(z[i\dd j+1])
\mbox{ or } \Rang(z[i\dd j])+1=\Rang(z[i-1\dd j]).$$
Denote by $E$ and $E_k$ the set of all and
of rank $k$ essential factors of $z$ respectively.
Note that $x$ and $y$ are essential factors of $z$.

\paragraph{\bf Data structure}
Our main data structure to answer the question consists of collections of tables of two types: for each $k<|A|$, when $z[i\dd j]\in E_k$
$$\mathit{RIGHT}_k[i]=j \mbox{ and } \mathit{LEFT}_k[j]=i.$$
The tables are used to compute the quadruple of each $z[r\dd s]\in E_{k+1}$: $$\Psi(z[r\dd s])=(z[r\dd r'],z[r'+1],z[s'-1],
z[s'\dd s]),$$
where $r'=\mathit{RIGHT}_k[r], s'=\mathit{LEFT}_k[s]$.

Note that all the tables and quadruples can be computed in total time $O(n\cdot |A|)$ since there are only $O(n\cdot |A|)$ essential factors.

\paragraph{Sketch of algorithm}
It is based on a dynamic programming technique to compute,
for each essential factor $u$, an identifier $\mathit{ID}(u)$ of
its equivalence class. It is an integer in the range $[1\dd n]$ that must satisfy the condition: if $u,v\in E_k$
$$(*)\  u\approx v \Longleftrightarrow \mathit{ID}(u)=\mathit{ID}(v).$$
The main step implements the Equivalence Criterion in Lemma~\ref{t:equiv_crit}.

\begin{algo}{Equivalence}{x,y \mbox{ words in } A^+}
  \SET{z}{x\$y}
  \ACT"{compute all tables $\mathit{RIGHT}$ and $\mathit{LEFT}$}
  \DOFOR"{all $u\in E_1$}           \label{alg-approx-equi-2}
    \ACT"{compute $\mathit{ID}(u)$} \label{alg-approx-equi-3}
  \OD
  \DOFORI{k}{2}{|A|}
    \DOFOR"{all $u\in E_k$}
      \SET{\Psi(u)}{\mbox{quadruple}\ (p,a,b,q)\ \mbox{corresponding to}\ u}
    \OD
    \ACT"{radix-sort all quadruples and give the same $\mathit{ID}$}
    \EXT{to words having the same quadruple $\Psi$}
  \OD
  \RETURN{\mathit{ID}(x)=\mathit{ID}(y)}
\end{algo}

The computation at lines~\ref{alg-approx-equi-2}-\ref{alg-approx-equi-3} is straightforward because there factors of $z$ have length $1$.
The whole computation has the required running time, mostly because both each radix-sort works in time $O(|E_k|)$ and we have $|E_k|=O(n)$ for each $k$.






\end{SOLUTION}

\begin{NOTES}
The Equivalence Criterion Lemma is stated in \cite{Lothaire83}.
The above computation is similar to the computation of the Dictionary of Basic factors of a word (see \cite[Problem 66]{CLR21cup} or \cite{CR94oup}).
The present algorithm is a version of the one given in   \cite{RadoszewskiR10}.
\end{NOTES}
\end{EXERCISE}


\small
\bibliographystyle{abbrv}
\bibliography{clr}
\normalsize
\end{document}